\documentclass[aps,prb,twocolumn,showpacs,eqsecnum,superscriptaddress,floatfix,10pt]{revtex4-1}

\usepackage{graphicx}
\usepackage{epstopdf}
\usepackage{color}\usepackage{enumerate}
\usepackage{amsmath}\usepackage{amssymb}\usepackage{stmaryrd}\usepackage{wasysym}
\usepackage{multirow}
\usepackage{float}
\usepackage{longtable,booktabs}
\usepackage{hyperref}\usepackage{url}
\usepackage{subfigure}%
\usepackage{dsfont}

\newcommand{\dslash}{{\mathchar'57\mkern-10mu\partial}}

\begin{document}

\title{Coupled wire model of symmetric Majorana surfaces of topological superconductors}

\author{Sharmistha Sahoo}
\author{Zhao Zhang}
\author{Jeffrey C. Y. Teo}\email{jteo@virginia.edu}
\affiliation{Department of Physics, University of Virginia, Virginia 22904, USA}

\date{\today}

\begin{abstract} Time reversal symmetric topological superconductors in three spatial dimensions carry gapless surface Majorana fermions. They are robust against any time reversal symmetric single-body perturbation weaker than the bulk energy gap. We mimic the massless surface Majorana's by coupled wire models in two spatial dimensions. We introduce explicit many-body interwire interactions that preserve time reversal symmetry and give energy gaps to all low energy degrees of freedom. We show the gapped models generically carry non-trivial topological order and support anyonic excitations. 
\end{abstract}

\maketitle

\section{Introduction}

Topological superconductors (TSC) are electronic phases of matter with finite excitation energy gaps that are not continuously connected to a conventional BCS $s$-wave superconductor. In particular BCS superconductors in three dimensions can have non-trivial topologies protected by time reversal symmetry.\cite{SchnyderRyuFurusakiLudwig08,Kitaevtable08,QiHughesRaghuZhang09} There is a bulk integral quantity $N$ of the mean-field system, known as chirality, that cannot change upon any adiabatic evolution unless the energy gap is closed or time reversal symmetry (TRS) is broken. TSC also exhibits unique physical signature along its surface. Despite there is a bulk energy gap, the surface of a TSC hosts $N$ gapless Majorana (real) fermion modes that are robust, in the single-body mean-field framework, to all symmetry and bulk gap preserving perturbations. The superfluid $^3$He-B\cite{Volovik:book,Aoki2005,Choi2006,Murakawa2009} and perhaps superconducting Cu$_x$Bi$_2$Se$_3$\cite{FuBerg2010,SasakiKrienerSegawaYadaTanakaSatoAndo11} are candidates of TSC.

The $\mathbb{Z}$ classification of TSC -- or class DIII band theories according to the Altland-Zirnbauer classification\cite{AltlandZirnbauer97} -- relies heavily on the single-body BCS description of the electronic structure. It has recently been shown that under strong many-body interaction, the surface state of sixteen copies of a TSC can be gapped without breaking time reversal symmetry or introducing surface topological order. This reduces the integer classification of TSC into $\mathbb{Z}_{16}$.\cite{LukaszChenVishwanath,MetlitskiFidkowskiChenVishwanath14,WangSenthil14,Senthil2014,Kapustin2014e,Qi_AxionTFT,Witten15} This suggests the many-body extension allows a continuous path that connects sixteen copies of a TSC to a trivial $s$-wave superconductor in three dimensions without breaking symmetry or closing the bulk gap. In fact, the surface Majorana modes of {\em any} TSC can be gapped without breaking symmetries. However, there would generically be a residue topological order, unless $N$ is a multiple of 16, that allows non-trivial anyonic excitations to live on the surface.\cite{LukaszChenVishwanath,MetlitskiFidkowskiChenVishwanath14} As a result, these 3D bulk systems are still topologically distinct from a trivial state.

Similar phenomena were also seen in topological insulators\cite{FuKaneMele3D,Roy07,MooreBalents07,QiHughesZhang08} in three dimensions and topological superconductors\cite{Kitaevchain} in one dimension. Many-body interactions allow the surface Dirac mode of a topological insulator to acquire an energy gap without breaking time reversal or charge conservation symmetries. However a non-trivial surface topological order would be left behind.\cite{WangPotterSenthilgapTI13, MetlitskiKaneFisher13b, ChenFidkowskiVishwanath14, BondersonNayakQi13} This indicates the bulk insulator still carries a non-trivial $\mathbb{Z}_2$ symmetry protected topology (SPT) even in the many-body framework. On the other hand, the $\mathbb{Z}$ classification of time reversal symmetric BDI superconductors in one dimension breaks down to $\mathbb{Z}_8$ in the presence of strong interaction.\cite{FidkowskiKitaev10,FidkowskiKitaev11,TurnerPollmannBerg11,ChenGuWen11}

The topological order of a gapped symmetric surface of a topological insulator or superconductor was deduced mainly using vortex condensation or other topological field theory techniques. They do not specify the microscopic many-body surface gapping interactions that give rise to these exotic surface states. A pioneer work that addressed this issue was done by Fidkowski and Kitaev in Ref.\onlinecite{FidkowskiKitaev10} where they constructed explicit time reversal symmetric 4-fermion interactions that give an energy gap to eight boundary Majorana zero modes of a 1D TSC. Another insightful work was published by Mross, Essin and Alicea in Ref.\onlinecite{MrossEssinAlicea15} where they mimicked the surface Dirac mode of a topological insulator using a coupled wire model and wrote down explicit symmetric gapping interactions that lead to different gapped or gapless surface states.

Sliding Luttinger liquids\cite{OHernLubenskyToner99,EmeryFradkinKivelsonLubensky00,VishwanathCarpentier01,SondhiYang01,MukhopadhyayKaneLubensky01} and coupled wire constructions\cite{KaneMukhopadhyayLubensky02} are immensely powerful in building two dimensional topological phases. They model 2D systems by arrays of coupled 1D chains, where interaction effects are more controlled and better understood. This theoretical technique has been frequently used in the study of fractional quantum Hall states\cite{KaneMukhopadhyayLubensky02,TeoKaneCouplewires,KlinovajaLoss14,MengStanoKlinovajaLoss14,SagiOregSternHalperin15}, anyon models\cite{OregSelaStern14,StoudenmireClarkeMongAlicea15}, spin liquids\cite{MengNeupertGreiterThomale15,GorohovskyPereiraSela15}, (fractional) topological insulators\cite{NeupertChamonMudryThomale14,KlinovajaTserkovnyak14,SagiOreg14,SagiOreg15,SantosHuangGefenGutman15} and superconductors\cite{mongg2,SeroussiBergOreg14}.

In this article, we imitate the surface Majorana modes of a 3D topological superconductor using a coupled Majorana wire model, construct explicit 4-fermion interactions that lead to a finite excitation energy gap, and study the residue surface topological order.

\subsection{Summary of results}
\begin{figure}[htbp]
\centering\includegraphics[width=0.4\textwidth]{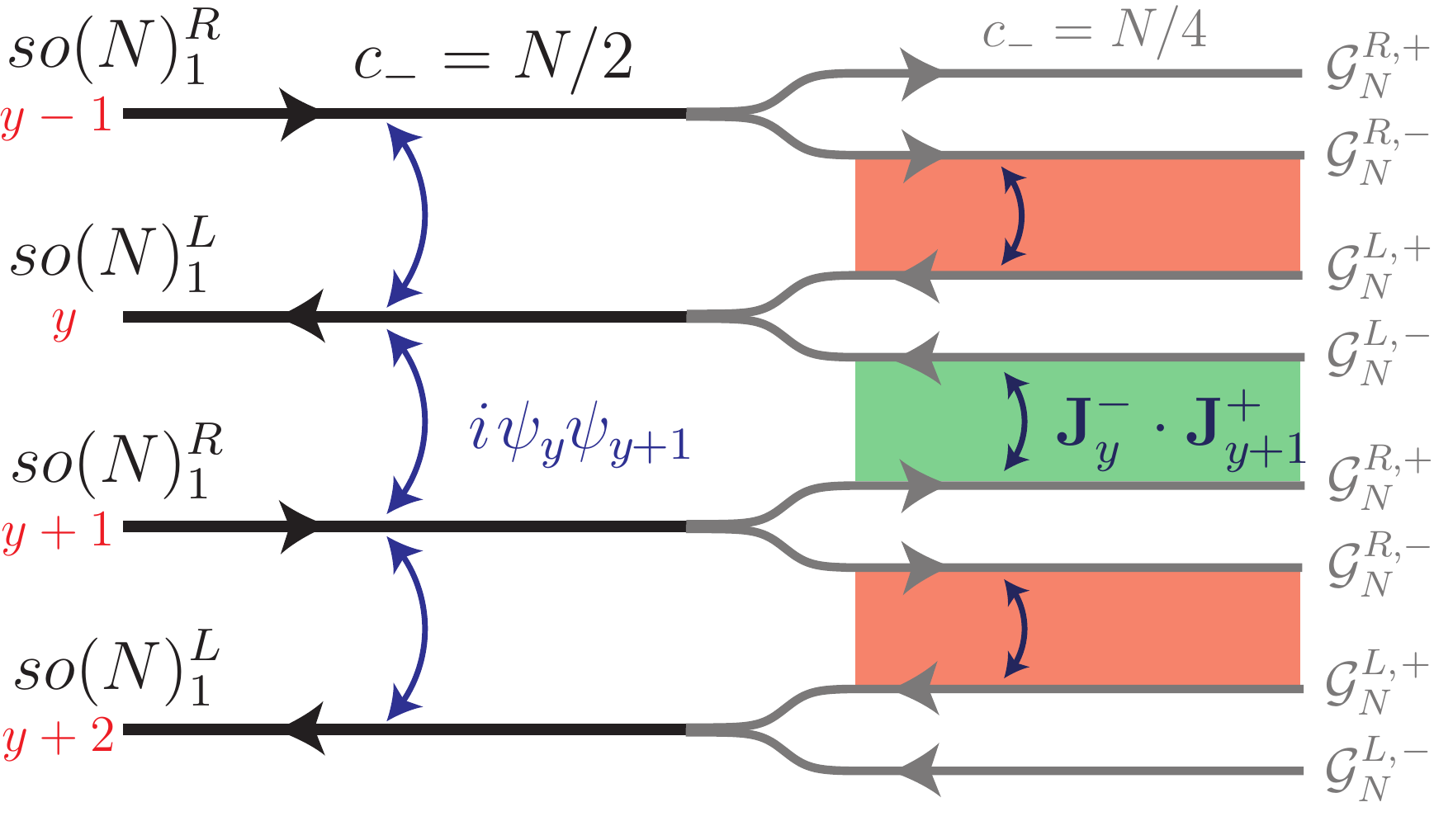}
\caption{(Left) Coupled wire model \eqref{NMajcone} of $N$ gapless surface Majorana cones. (Right) Fractionalization \eqref{fractioneq} and couple wires construction \eqref{bcint} of gapped anomalous and topological surface state.}\label{fig:couplewires}
\end{figure}

We consider a 2D array of chiral Majorana wires, each of which carries $N$ Majorana fermion channels that propagate in a single direction. The chiralities of wires alternate so that adjacent wires counter-propagate and Majorana's can backscatter to their neighbors through electron tunneling (see figure~\ref{fig:couplewires}). When the interwire backscattering is uniform, the 2D system is gapless. In the long wavelength continuum limit, the energy spectrum is linear in both $k_x$ and $k_y$ directions and the model gives $N$ Majorana cones.

If the $N$ massless Majorana species decouple, each of them is protected by a non-local ``antiferromagnetic" time reversal symmetry (TRS) that translates all Majorana's to the next wire while reversing the propagating direction. This symmetry requires uniform interwire backscattering and forbids a fermion mass. However, this TRS is qualitatively different from the conventional one, which is local. For instance a gap can be introduced when $N$ is even by inter-species single-body backscatterings which preserves symmetry.\cite{MrossEssinAlicea15} Thus the surface classification becomes $\mathbb{Z}_2$ instead of $\mathbb{Z}$ even in the single-body framework. Despite the discrepancy, the coupled wire model does bare resemblance to the original problem of topological superconductor surfaces, especially when the number of chiral species $N$ is odd.

A major result of this work is to construct many-body gapping potentials that freeze out all low energy degrees of freedom while preserving the non-local time reversal symmetry. This is achieved by {\em fractionalizing} or {\em bipartitioning} the Majorana channels on each wire into a pair of independent sectors without interfering each other. They can then be backscattered to adjacent wires in opposite directions. When there are even Majorana species, the decomposition is obvious as $N=2r=r+r$ and one can simply separate the first $r$ Majorana's from the remaining $r$. The fractionalization in the odd case is more involved but are well-known in the conformal field theory (CFT) community. Firstly the emergent rotation symmetry $\psi^a\to O^a_b\psi^b$ among the fermion species corresponds to a chiral $so(N)$ current algebra at level 1, also known as an affine Kac-Moody algebra or Wess-Zumino-Witten (WZW) theory, along each wire.\cite{bigyellowbook} Low energy excitations along the chiral wires, referred as primary fields, are irreducible representations of the $so(N)_1$ algebra. Fractionalization of the WZW CFT is also known as level-rank duality\cite{bigyellowbook,NaculichRiggsSchnitzer90,MlawerNaculichRiggsSchnitzer91} or conformal embedding\cite{bigyellowbook,BaisEnglertTaorminaZizzi87,SchellekensWarner86,BaisBouwknegt87} \begin{align}so(N)_1\supseteq\mathcal{G}^+_N\times\mathcal{G}^-_N\end{align} where the two $\mathcal{G}_N$'s are mutually commuting subalgebras of $so(N)$. For example $so(9)_1$ can be decomposed into the tensor product $so(3)_3\otimes so(3)_3$ as $9=3\times3$. This splits each wire into a pair of fractional channels (see figure~\ref{fig:couplewires}). For instance the chiral central charge, $c_-=N/2$, which loosely speaking counts the degrees of freedom and characterizes the heat current\cite{KaneFisher97, Cappelli01, Kitaev06, Luttinger64} running along each wire, also decomposes so that each $\mathcal{G}_N$ channel carries $c_-=N/4$.

The many-body gapping interaction are given by interwire current-current backscattering (see also figure~\ref{fig:couplewires}) \begin{align}\mathcal{H}_{\mathrm{int}}=u\sum_y{\bf J}_{\mathcal{G}^-_N}^y\cdot{\bf J}_{\mathcal{G}^+_N}^{y+1}\label{Hintintro}\end{align} where ${\bf J}_{\mathcal{G}^\pm_N}^y$ are the $\mathcal{G}^\pm_N$ currents operators along wire $y$. All current operators are certain combinations of fermion bilinears, and the backscattering interaction therefore consists of 4-fermion terms. This Hamiltonian is exactly solvable. It preserves the ``antiferromagnetic" time reversal symmetry and opens up an excitation energy gap.

The symmetric gapped surface generically carries a non-trivial $G_N$ topological order. \begin{align}G_N=\left\{\begin{array}{*{20}l}SO(r)_1,&\mbox{for $N=2r$}\\SO(3)_3\boxtimes_bSO(r)_1,&\mbox{for $N=9+2r$}\end{array}\right.\label{GNdefintro}\end{align} where both $N$ and $r$ can be extended to negative integers. It can be inferred, using the bulk-boundary correspondence\cite{Wenedgereview,MooreRead,ReadMoore,Kitaev06}, from the $(1+1)$D $\mathcal{G}_N$ CFT living along the interface that separates the TR symmetric topological gapped domain and a TR breaking trivial gapped domain. The anyon structure\cite{Wilczekbook,Fradkinbook,Wenbook,Kitaev06}, which encodes the quasiparticle types together with their statistics and fusion properties, follows a 32-fold periodicity in the sense that $G_N\cong G_{N+32}$. Moreover, these thirty-two topological states exhibit a natural $\mathbb{Z}_{32}$ {\em relative} tensor product structure, $G_{N_1}\boxtimes_bG_{N_2}\cong G_{N_1+N_2}$, where certain set $b$ of non-trivial bosons are condensed\cite{BaisSlingerlandCondensation} under the tensor product. 

It is important to clarify at this point that the coupled wire construction is a $(2+1)$D model where the interaction \eqref{Hintintro} is built out of local bosonic current operators ${\bf J}$. Under this interpretation, the coupled $SO(N)_1$ wire model is bosonic and Majorana fermions are treated as anyonic excitations that carries a quasiparticle string. There is a $\mathbb{Z}_2$ gauge degree of freedom that couples to the fermions, $\psi^a\to-\psi^a$, and there are deconfined $\pi$-fluxes (or $hc/2e$-fluxes), which are anyonic excitations that are non-local with fermions. 

When $N$ is a multiple of four, the $G_N$ topological order is Abelian with four distinct anyon types $1,\psi,s_+,s_-$, where $s_\pm$ are $\pi$-fluxes with opposite fermion parities. When $N$ is 2 mod 4, the $G_N$ state resembles an Ising topological order with anyons $1,\psi,\sigma$. When $N$ is odd, the topological state has 7 anyon types, $1,\alpha_\pm,\gamma_\pm,\beta,f$, and has a structure similar to $SO(3)_3$ (or equivalently $SU(2)_6$). All these anyon theories contain $\pi$-fluxes, which should be absent on the surface of a fermionic topological superconductor. In Ref.\onlinecite{LukaszChenVishwanath}, the surface topological order of a $N=1$ fermionic TSC only contains 4 quasiparticles $1,\gamma_\pm,f$ instead of 7. The additional $\pi$-fluxes in our coupled wire model could become confined by re-introducing single-body interwire fermion backscattering. In this case, the thirty two bosonic topological states reduces down to two fermionic ones, (1) a trivial state containing $1,\psi$ similar to copies of $p_x+ip_y$ superconductors when $N$ is even, or (2) a non-trivial fermionic $SO(3)_3$ state with anyons $1,\gamma_\pm,f$ when $N$ is odd. This $\mathbb{Z}_2$ classification, instead of $\mathbb{Z}_{16}$, is a natural consequence of the ``antiferromagnetic" time reversal symmetry. In particular there is no reason to expect the result would match that of Ref.\onlinecite{LukaszChenVishwanath,MetlitskiFidkowskiChenVishwanath14} when $N$ is even.

We will introduce the single-body coupled Majorana wire model at the beginning of section~\ref{sec:couplewire}. A review on the $so(N)_1$ WZW CFT will be given in section~\ref{sec:SO(N)} and \ref{sec:bosonization} as well as in appendix~\ref{sec:so(N)app}, \ref{sec:kleinfactors} and \ref{sec:kleinfactors2}. In section~\ref{sec:gapping}, we will construct time reversal symmetry 4-fermion interactions that will open up an excitation energy gap. The discussion will be decomposed into the even and odd $N$ cases in section~\ref{sec:even} and \ref{sec:odd} respectively. In the even case, the gapping Hamiltonian will match the $O(r)$ Gross-Neveu model\cite{GrossNeveu,ZamolodchikovZamolodchikov78,Witten78,ShankarWitten78} and we will show an energy gap in section~\ref{sec:GNmodel} by (partially) bosonizing the problem. The gapping potential for the odd case will rely on a conformal embedding and relate to the Zamolodchikov and Fateev $\mathbb{Z}_6$ parafermion CFT\cite{FateevZamolodchikov82,ZamolodchikovFateev85}. This will be discussed and reviewed in section~\ref{sec:conformalembedding}, \ref{sec:Z6parafermions} as well as in appendix~\ref{sec:appZ6parafermion}. The symmetric gapping interactions will correspond to non-trivial surface topological orders. This will be discussed in section~\ref{sec:topologicalorder} where we will present the class of 32-fold periodic topological $G_N$ states. In section~\ref{sec:otherpossibilities}, we will describe alternative gapping interactions that would lead to even more possibilities. Lastly, we will conclude the article in section~\ref{sec:conclusion} where we will also discuss some possible future exploration.

\section{Couple wire construction of surface Majorana cones}\label{sec:couplewire}
A time reversal symmetric BCS superconductor is described by a Bogoliubov - de Gennes (BdG) Hamiltonians $H_{BdG}({\bf k})$. Symmetries require $TH_{BdG}({\bf k})T^{-1}=H_{BdG}(-{\bf k})$ and $CH_{BdG}({\bf k})C^{-1}=-H_{BdG}(-{\bf k})$ where $T$ and $C$ are the antiunitary time reversal and particle-hole operators. When the symmetries square to $C^2=-T^2=1$, the BdG theory belongs to the symmetry class DIII according to the Altland-Zirnbauer classification\cite{AltlandZirnbauer97} and theories in three dimensions with finite excitation energy gaps are topologically classified by integers\cite{SchnyderRyuFurusakiLudwig08,Kitaevtable08,QiHughesRaghuZhang09}. Superconducting $^3$He in the B-phase\cite{Volovik:book,Aoki2005,Choi2006,Murakawa2009} and certain doped topological insulators\cite{FuBerg2010,SasakiKrienerSegawaYadaTanakaSatoAndo11} were suggested to carry non-trivial topologies.

Topological superconductors host protected gapless surface Majorana modes. The simplest version is a single Majorana cone, which is the spectrum of a massless two-component real fermion $\mathcal{H}_\pm=iv\boldsymbol\psi^T\dslash_\pm\boldsymbol\psi$, where $\dslash_\pm=\partial_y\tau_x\pm\partial_x\tau_z$ and the Pauli matrices $\tau_x,\tau_y,\tau_z$ act on the surface real fermion $\boldsymbol\psi=(\psi_R,\psi_L)$. 
Majorana fermions are hermitian $\psi_j^\dagger=\psi_j$ and obey the anti-commutation relation $\{\psi_j({\bf r}),\psi_{j'}({\bf r}')\}=2\delta_{jj'}\delta({\bf r}-{\bf r}')$. Time reversal switches the components $\mathcal{T}(\alpha_1\psi_L+\alpha_2\psi_R)\mathcal{T}^{-1}=\alpha_2^\ast\psi_L-\alpha_1^\ast\psi_R$ so that $\mathcal{T}^2=-1$. The sign in the Hamiltonian $\mathcal{H}_\pm$ determines its {\em chirality}. A general surface state could consist of multiple copies of Majorana cones with different chiralities \begin{align}\mathcal{H}_c=\sum_{a=1}^{N_R}iv_a\boldsymbol\psi_a^T\dslash_+\boldsymbol\psi_a+\sum_{b=1}^{N_L}iv_b\boldsymbol\psi_b^T\dslash_-\boldsymbol\psi_b.\label{Hc}\end{align} Fermions $\boldsymbol\psi_a$ and $\boldsymbol\psi_b$ with opposite chiralities can annihilate each other by the time reversal symmetric mass term $im\boldsymbol\psi_a^T\tau_z\boldsymbol\psi_b$. Quadratic terms among fermions of the same chirality would however either break time reversal or only move the gapless Majorana cones away from zero momentum without destroying them. The net surface chirality $N=N_R-N_L$ is thus a robust topological signature that distinguishes and characterizes 3D bulk topological superconductors. It cannot be altered by any time reversal symmetric two-body perturbations that are not strong enough to close the bulk excitation energy gap. 

Recent theoretical studies suggest many-body interactions can remove these gapless surface degrees of freedom. To construct explicit gapping terms, we turn to an anisotropic description of surface Majorana fermions using an array of coupled fermion wires (see figure~\ref{fig:couplewires}). The horizontal wires are labeled according to their vertical position $y=\ldots,-2,-1,0,1,2,\ldots$ and each carries $N$ chiral (real) Majorana fermions $\boldsymbol\psi_y=(\psi^1_y,\ldots,\psi^N_y)$ which propagate only to the right (or left) if $y$ is even (resp.~odd). The number of flavors $N$ here is going to be identified with the net chirality of the surface Majorana cone. Time reversal symmetry is non-local in this model as it relates fermions on adjacent wires that propagate in opposite directions, \begin{align}\mathcal{T}\left(\sum_{a=1}^N\alpha_a\psi^a_y\right)\mathcal{T}^{-1}=(-1)^y\sum_{a=1}^N\alpha_a^\ast\psi^a_{y+1}.\label{TR}\end{align} Similar to the symmetry of an antiferrormagnet, here time reversal on the single-fermion Hilbert space squares to a primitive translation up to a sign, $\mathcal{T}^2=-\hat{t}_y$ for $\hat{t}_y$ the vertical lattice translation $y\to y+2$ that relates nearest co-propagating wires. In the many-body Hilbert space, \begin{align}\mathcal{T}^2=(-1)^F\hat{t}_y\end{align} where $(-1)^F$ is the fermion parity operator whose sign depends on the eveness or oddness of fermion number.

\begin{figure}[htbp]
\centering\includegraphics[width=0.2\textwidth]{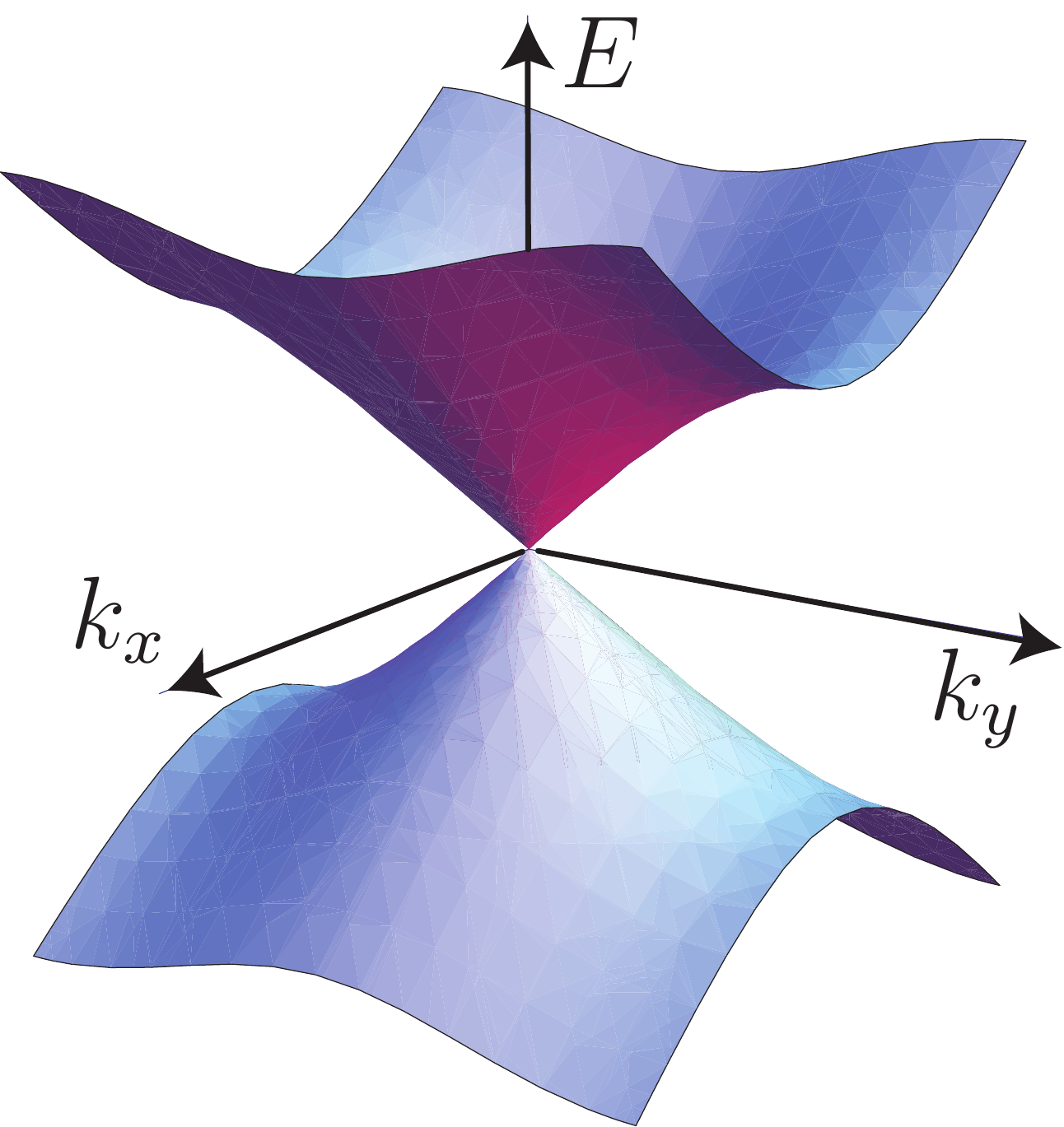}
\caption{The energy spectrum of the coupled Majorana wire model \eqref{NMajcone}}\label{fig:Majoranacone}
\end{figure}

We mimic $N$ copies of surface Majorana cones by the coupled wire Hamiltonian \begin{align}\mathcal{H}_0=\sum_{y=-\infty}^\infty iv_{\mathsf{x}}(-1)^y\boldsymbol\psi_y^T\partial_x\boldsymbol\psi_y+iv_{\mathsf{y}}\boldsymbol\psi_y^T\boldsymbol\psi_{y+1}\label{NMajcone}\end{align} where the $N$-component Majorana fermion $\boldsymbol\psi$ disperses linearly (for small $k_y$) with velocities $v_{\mathsf{x}},v_{\mathsf{y}}$ along the horizontal and vertical axes (see figure~\ref{fig:Majoranacone}). By applying \eqref{TR}, we see $\mathcal{T}\mathcal{H}_0\mathcal{T}^{-1}=\mathcal{H}_0$ and the coupled wire model is therefore time reversal symmetric. Moreover, $\mathcal{H}_0$ has continuous translation symmetry along $x$ and discrete translation along $y\to y+2$. The alternating sign in the first term of \eqref{NMajcone} specifies the propagating directions of the wires. Projecting to the $k_x=0$ zero modes along the wires, the second term in \eqref{NMajcone} effectively becomes a 1D Kitaev Majorana chain\cite{Kitaevchain} which has a linear spectrum for small $k_y$. More explicitly, by using the Nambu basis $\boldsymbol\xi_{\bf k}=(c_{\bf k}^a,{c_{-{\bf k}}^a}^\dagger)^T$ for $c^a_{\bf k}=\sum_{xy}e^{i(k_xx+k_yy)}c^a_y(x)$ the Fourier transform of the Dirac fermion $c^a_y(x)=(\psi^a_{2y-1}(x)+i\psi^a_{2y}(x))/2$, the coupled wire Hamiltonian \eqref{NMajcone} can be expressed as $\mathcal{H}_0=\sum_{\bf k}\boldsymbol\xi_{\bf k}^\dagger H_{\mathrm{BdG}}^0({\bf k})\boldsymbol\xi_{\bf k}$, where the BdG Hamiltonian is given by\begin{align}H_{\mathrm{BdG}}^0({\bf k})=2v_{\mathsf{x}}k_x\tau_x+v_{\mathsf{y}}\left[-\sin k_y\tau_y+(1-\cos k_y)\tau_z\right]\label{NMajconeBdG}\end{align} for $-\infty<k_x<\infty$ and $-\pi\leq k_y\leq\pi$. It has a linear spectrum near zero energy and momentum as shown in figure~\ref{fig:Majoranacone}.

We notice in passing that if the time reversal operation in \eqref{TR} was defined without the alternating sign $(-1)^y$, it would sqaure to a different sign $\mathcal{T}^2=+\hat{t}_y$ in the single-fermion Hilbert space and the vertical term in \eqref{NMajcone} would need to be modified into $\sum_yiv_{\mathsf{y}}(-1)^y\boldsymbol\psi_y^T\boldsymbol\psi_{y+1}$ in order to preserve the symmetry. This would correspond to an alternating Majorana chain in the $y$-direction, where the gapless Majorana cone would be positioned at $k_y=\pi$ instead of 0 and would still be protected by Kramers theorem as $T_{k_y=\pi}^2=e^{ik_y}=-1$. This scenario is actually equivalent and related to the original by a gauge transformation $(\psi_{4y},\psi_{4y+1},\psi_{4y+2},\psi_{4y+1+3})\to(\psi_{4y},\psi_{4y+1},-\psi_{4y+2},-\psi_{4y+1+3})$, and therefore the sign of $\mathcal{T}^2$ is unimportant in this problem. Nevertheless we will stick with previous convention defined in \eqref{TR} in the following discussions.

The chirality $N$ of the coupled Majorana wire model \eqref{NMajcone} is set by the {\em chiral central charge} $c_-=N/2$ along each wire. This quantity is defined by the difference of central charges\cite{bigyellowbook} between right and left moving modes, and determines the energy (thermal) current $I_T\approx c_-\frac{\pi^2k_B^2}{6h}T^2$ flowing along the wire in low temperature\cite{KaneFisher97, Cappelli01, Kitaev06, Luttinger64}. In general, a Majorana wire carrying $N_R$ right moving fermions and $N_L$ left moving ones has the kinetic Hamiltonian $\mathcal{H}=iv_{\mathsf{x}}\boldsymbol\psi^T\dslash_x\boldsymbol\psi$, where $\dslash_x=[\openone_{N_R}\oplus(-\openone_{N_L})]\partial_x$ acts on the $(N_R+N_L)$-component real fermion $\boldsymbol\psi$. In \eqref{NMajcone} we consider the simplest case when $(N_R,N_L)=(N,0)$ for $y$ even or $(0,N)$ for $y$ odd.

\begin{figure}[htbp]
\centering\includegraphics[width=0.4\textwidth]{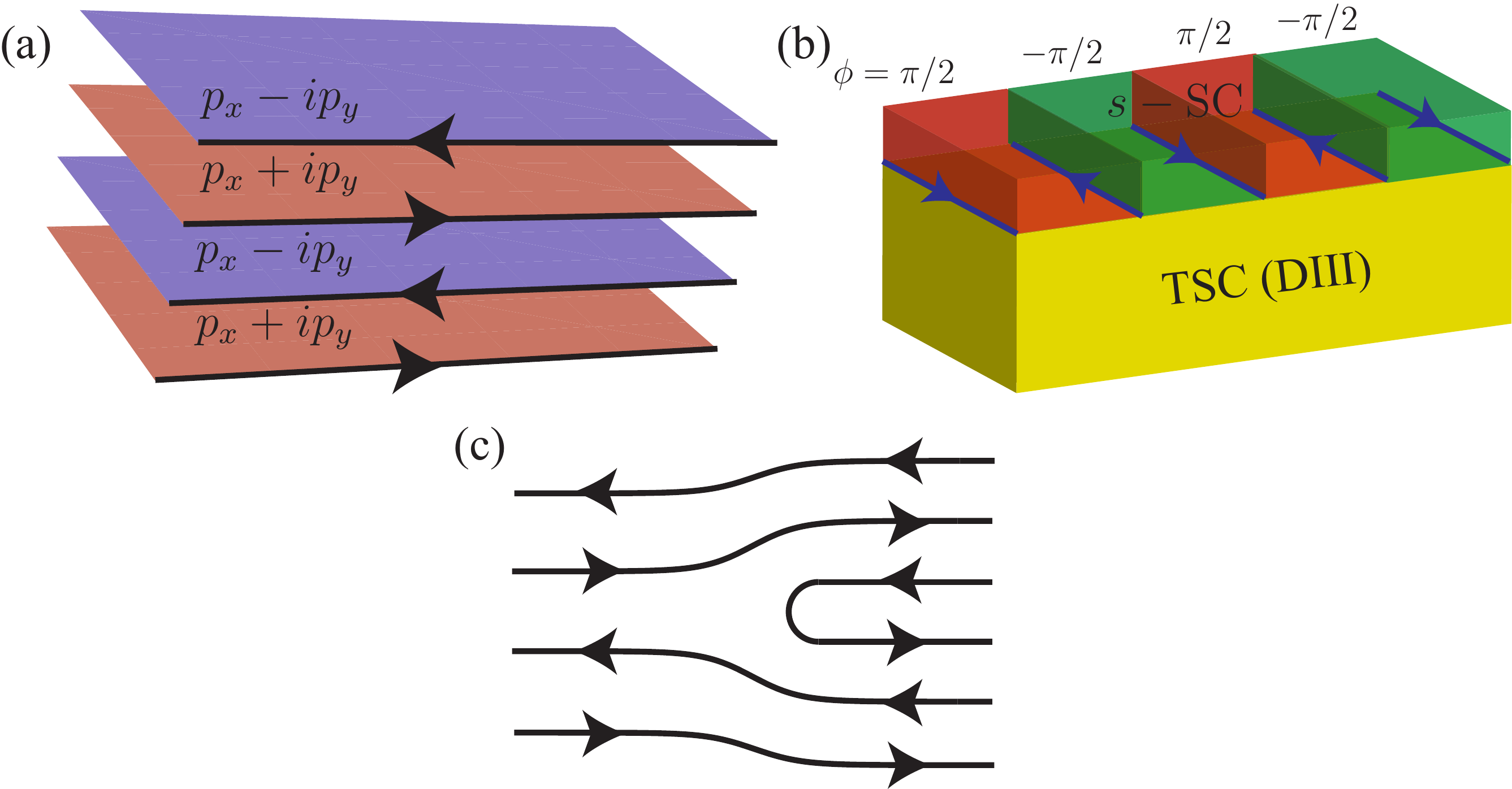}
\caption{Coupled Majorana wire model on the surface of (a) a stack of alternating $p_x\pm ip_y$ superconductors, and (b) a class DIII topological superconductor (TSC) with alternating TR breaking surface domains. (c) A dislocation.}\label{fig:dislocation}
\end{figure}

A chiral 1D system violates fermion doubling\cite{NielsenNinomiya83} and can only be realized as an anomalous edge of a gapped 2D bulk\cite{Volovik92,ReadGreen,Kitaev06}. The coupled Majorana wire model, \eqref{NMajcone} or figure~\ref{fig:couplewires}, must therefore also be holographic and living on the surface of a 3D bulk superconductor. This can be modeled by a stack of alternating layers of spinless $p_x\pm ip_y$ superconductors (see figure~\ref{fig:dislocation}(a)). The interwire backscattering in \eqref{NMajcone} can be generated by bulk interlayer electron tunneling and pairing that are not competing with the intralayer $p+ip$ pairing. Time reversal \eqref{TR} extends to the three dimensional bulk by relating fermions on adjacent layers. The coupled Majorana wire model can also live on the surface of a 3D class DIII topological superconductor where each chiral Majorana mode is bound between adjacent domains with opposite time reveral breaking phases $\phi=\pm\pi/2$ (see figure~\ref{fig:dislocation}(b)).\cite{TeoKane,Qi_AxionTFT} The discrete translation order along the $y$-axis perpendicular to the wire direction can be melted by proliferating dislocations (see figure~\ref{fig:dislocation}(c)). With continuous translation symmetry restored, time reversal symmetry becomes local with $T^2=-1$ and the coupled Majorana wire model \eqref{NMajcone} recovers the surface Majorana cone \eqref{Hc} in the continuum limit for small $k_y$.

The non-local time reversal symmetry \eqref{TR} actually provides a weaker topological protection to gapless surface Majorana's than a conventional local one. For instance in section~\ref{sec:gapping}, we will show that the $N=2$ coupled Majorana wire model can be gapped by single-body backscattering terms without breaking time reversal, leaving behind a surface with trivial topological order. 
This reduced robustness stems from the half-translation component in the antiferrormagnetic time reversal. In the BdG description \eqref{NMajconeBdG}, the time reversal operator takes the momentum dependent form \begin{align}T_{\bf k}=\left(\frac{1+e^{ik_y}}{2}\tau_y+i\frac{1-e^{ik_y}}{2}\tau_z\right)\mathcal{K}\label{TRBdG}\end{align} for $\mathcal{K}$ the complex conjugation operator. It commutes with the BdG Hamiltonian $T_{\bf k}H_{\mathrm{BdG}}^0({\bf k})=H_{\mathrm{BdG}}^0(-{\bf k})T_{\bf k}$ as well as the particle-hole (PH) $CT_{\bf k}=T_{-{\bf k}}C$, for $C=\tau_x\mathcal{K}$ the PH operator. In the continuum limit or for small $k_y$, $T\simeq\tau_y\mathcal{K}$ agrees with the conventional local time reversal operator and protects a zero energy Majorana Kramers' doublet. The BdG Hamiltonian has a chiral symmetry $\Pi_{\bf k}H_{\mathrm{BdG}}^0({\bf k})=-H_{\mathrm{BdG}}^0({\bf k})\Pi_{\bf k}$, for $\Pi_{\bf k}=iCT_{\bf k}$ the chiral operator. It can be used to assign the chirality of a Majorana cone by an integral winding number \begin{align}n=\frac{1}{2\pi i}\oint_{\mathcal{C}_\varepsilon({\bf k}_0)}\mathrm{Tr}\left[h({\bf k})^{-1}\nabla_{\bf k}h({\bf k})\right]\cdot d{\bf l}\end{align} {\em locally} around a loop $\mathcal{C}_\varepsilon({\bf k}_0)$ $\varepsilon$ away from the zero mode at ${\bf k}_0$. Here $h({\bf k})$ is the elliptic operator \begin{align}h({\bf k})=P_{\bf k}^+H_{\mathrm{BdG}}^0({\bf k})P_{\bf k}^-\end{align} for $P_{\bf k}^\pm=(P_{\bf k}^\pm)^2$ the two {\em local} projectors diagonalizing the chiral operator $\Pi_{\bf k}=e^{-ik_y/2}(P_{\bf k}^+-P_{\bf k}^-)$. However, as time reversal squares to $T_{\bf k}T_{-{\bf k}}=-e^{ik_y}$, which is the eigenvalue of the primitive translation $-\hat{t}_y$ at momentum ${\bf k}$, so does the {\em non-symmorphic} chiral operator $\Pi_{\bf k}^2=e^{-ik_y}$. The two chiral branches $\Pi_{\bf k}=\pm e^{-ik_y/2}$ switch across the Brillouin zone when $k_y\to k_y+2\pi$. As a result, a {\em global} winding number can only be defined modulo 2.

\subsection{The \texorpdfstring{$so(N)_1$}{so(N)} current algebra}\label{sec:SO(N)}
We notice the couple Majorana wire model \eqref{NMajcone} has a $SO(N)$ symmetry that rotates the $N$-component Majorana fermion $\psi_y^a\to O^a_b\psi_y^b$. 
Consequently, there is a chiral $so(N)$ Wess-Zumino-Witten (WZW) theory\cite{WessZumino71,WittenWZW} or affine Kac-Moody algebra at level 1 along each wire. Here we review some relevant features of the $so(N)_1$ algebra, which are well-known and can be found in standard texts on conformal field theory (CFT) such as Ref.\onlinecite{bigyellowbook}.

The $so(N)_1$ currents have the free field representation \begin{align}J^\beta(z)=\frac{i}{2}\boldsymbol\psi(z)^Tt^\beta\boldsymbol\psi(z)=\frac{i}{2}\sum_{ab}\psi^a(z)t^\beta_{ab}\psi^b(z)\label{so(N)current}\end{align} where the $t^\beta$'s are antisymmetric $N\times N$ matrices that generate the $so(N)$ Lie algebra (see appendix~\ref{sec:so(N)app}), $z=e^{\tau+ix}$ is the complex space-time parameter, and \eqref{so(N)current} is normal ordered. The coupled Majorana wire model carries currents that propagate in alternating directions (see figure~\ref{fig:couplewires}) so that $J^\beta_y(z)$ are holomorphic for even $y$ and $J^\beta_y(\overline{z})$ are anti-holomorphic for odd $y$. Focusing on an even wire, from the operator product expansion (OPE) \begin{align}\psi^a(z)\psi^b(w)=\frac{\delta^{ab}}{z-w}+\ldots\label{fermionOPE}\end{align} the $so(N)_1$ currents obey the product expansion \begin{align}J^\beta(z)J^\gamma(w)=\frac{\delta^{\beta\gamma}}{(z-w)^2}+\sum_\delta\frac{if_{\beta\gamma\delta}}{z-w}J^\delta(w)+\ldots\label{so(N)1Jrelation}\end{align} where $f_{\beta\gamma\delta}$ are the structure constants of the $so(N)$ Lie algebra with $\left[t^\beta,t^\gamma\right]=\sum_\delta f_{\beta\gamma\delta}t^\delta$ (see appendix~\ref{sec:so(N)app}). The Sugawara energy momentum tensor (along a single wire) is equivalent to the free fermion one\cite{GoddardNahmOlive85} \begin{align}T(z)=\frac{1}{2(N-1)}{\bf J}(z)\cdot{\bf J}(z)=-\frac{1}{2}\boldsymbol\psi(z)^T\partial_z\boldsymbol\psi(z)\label{SugawaraT}\end{align} for ${\bf J}=(J^\beta)$ the current vector and $\boldsymbol\psi=(\psi^1,\ldots,\psi^N)$ the $N$-component real fermion.
The energy momentum tensor defines a chiral Virasoro algebra and characterizes a chiral CFT. It satisfies the OPE \begin{align}T(z)T(w)=\frac{c_-/2}{(z-w)^4}+\frac{2T(w)}{(z-w)^2}+\frac{\partial_wT(w)}{z-w}+\ldots\end{align} where the chiral central charge $c_-=N/2$, loosely speaking, counts the conformal degrees of freedom on the Majorana wires and is proportional to the energy current\cite{KaneFisher97, Cappelli01, Kitaev06, Luttinger64} and entanglement entropy\cite{AffleckLudwig91,HolzheyLarsenWilczek94,CalabreseCardy04} carried by the wire. 

Excitations of the $N$-component Majorana wire transform acording to the $SO(N)$ symmetry. They decompose into {\em primary fields} and their corresponding descendants. A primary field ${\bf V}_\lambda=(V^1,\ldots,V^d)$ is a simple excitation sector that irreducibly represents the $so(N)_1$ Kac-Moody algebra. \begin{align}J^\beta(z)V^r(w)=-\sum_{s=1}^d\frac{(t^\beta_\lambda)_{rs}}{z-w}V^s(w)+\ldots\label{currentrepOPE}\end{align} where $\lambda$ labels some $d$-dimensional irreducible representation of $so(N)$ and $t^\beta_\lambda$ is the $d\times d$ matrix representing the generator $t^\beta$ of $so(N)$. For example it is straightforward to check by using the definition \eqref{so(N)current} and the OPE \eqref{fermionOPE} that the Majorana fermion $\boldsymbol\psi=(\psi^1,\ldots,\psi^N)$ is primary with respect to the fundamental representation, i.e. \begin{align}J^\beta(z)\psi^a(w)=-\sum_{b=1}^N\frac{t^\beta_{ab}}{z-w}\psi^b(w)+\ldots.\end{align} From \eqref{SugawaraT}, space-time translation of a primary field ${\bf V}_\lambda$ is governed by \begin{align}T(z){\bf V}_\lambda(w)=\frac{h_\lambda}{(z-w)^2}{\bf V}_\lambda(w)+\frac{\partial_w{\bf V}_\lambda(w)}{z-w}+\ldots\label{TVdim}\end{align} where the conformal (scaling) dimension is given by \begin{align}h_\lambda=\frac{\mathcal{Q}_\lambda}{2(N-1)}\end{align} for $-\sum_\beta t^\beta_\lambda t^\beta_\lambda=\mathcal{Q}_\lambda\openone_{d\times d}$ the quadratic Casimir operator. For instance $\mathcal{Q}_\psi$, the quadratic Casimir eigenvalue for the fundamental representation, is $N-1$ (see appendix~\ref{sec:so(N)app}) and therefore the fermion $\boldsymbol\psi$ has conformal dimension $h_\psi=1/2$. This agrees with the OPE \eqref{fermionOPE} by dimension analysis.

There are extra primary fields other than the the trivial vacuum $1$ and the fermion $\psi$. The spinor representations (see appendix~\ref{sec:so(N)app}) $\sigma$, for $N$ odd, or $s_+$ and $s_-$, for $N$ even, also correspond to primary fields of $so(N)_1$. Their conformal dimensions can be read off from their quadratic Casimir values \eqref{spinorQ}, and are \begin{align}h_\sigma=\frac{N}{16},\quad h_{s\pm}=\frac{N}{16}.\label{spinsigmas}\end{align} Unlike the infinite number of irreducible representations of a Lie algebra, the extended affine $so(N)_1$ algebra only has a truncated set of primary fields $\{1,\sigma,\psi\}$, for $N$ odd, or $\{1,s_+,s_-,\psi\}$, for $N$ even. 

These $so(N)_1$ primary fields take more explicit operator forms after bosonization and can be found in appendix~\ref{sec:kleinfactors} and \ref{sec:kleinfactors2}.

\subsection{Bosonizing even Majorana cones}\label{sec:bosonization}

In the case when $N=2r$ is even, the $N$ Majorana (real) fermions on each wire can be paired into $r$ Dirac (complex) fermions and {\em bosonized}\cite{witten1984,Fradkinbook}\begin{align}c_y^j=\frac{\psi_y^{2j-1}+i\psi_y^{2j}}{\sqrt{2}}\sim\frac{1}{\sqrt{l_0}}\exp\left(i\widetilde{\phi}_y^j\right)\label{complexfermion}\end{align} where $\widetilde\phi^1_y,\ldots,\widetilde\phi^r_y$ are real bosons on the $y^{\mathrm{th}}$ wire, and the vertex operator in \eqref{complexfermion} is normal ordered. The bosons obey the equal-time commutation relation 
\begin{align}\left[\widetilde{\phi}_y^j(x),\widetilde{\phi}_{y'}^{j'}(x')\right]=&i\pi(-1)^{\mathrm{max}\{y,y'\}}\Big[\delta_{yy'}\delta^{jj'}\mathrm{sgn}(x'-x)\nonumber\\&+\delta_{yy'}\mathrm{sgn}(j-j')+\mathrm{sgn}(y-y')\Big]\label{ETcomm0}\end{align} where $\mathrm{sgn}(s)=s/|s|=\pm1$ for $s\neq0$ and $\mathrm{sgn}(0)=0$. The first line of \eqref{ETcomm0} is equivalent to the commutation relation between conjugate fields \begin{align}\left[\widetilde{\phi}_y^j(x),\partial_{x'}\widetilde{\phi}_{y'}^{j'}(x')\right]=2\pi i(-1)^y\delta_{yy'}\delta^{jj'}\delta(x-x')\label{ETcomm00}\end{align} and is set by the ``$p\dot{q}$" term of the Lagrangian density \begin{align}\mathcal{L}_0=\frac{1}{2\pi}\sum_{y=-\infty}^\infty\sum_{j=1}^r(-1)^y\partial_x\widetilde{\phi}_y^j\partial_t\widetilde{\phi}_y^j.\label{L0}\end{align} The second line of \eqref{ETcomm0} guarantees the correct anticommutation relations between Dirac fermions along distinct channels. The alternating signs $(-1)^y$ in \eqref{ETcomm00} and \eqref{L0} specify the propagating directions along each wire, $R$ (or $L$) for $y$ even (resp.~odd). Eq.\eqref{ETcomm0} is symmetric under time reversal \eqref{TR}, which sends \begin{align}\mathcal{T}c_y^j\mathcal{T}^{-1}=(-1)^y{c_y^j}^\dagger,\quad\mathcal{T}\widetilde{\phi}^i_y\mathcal{T}^{-1}=\widetilde{\phi}^i_{y+1}+\pi y.\label{bosonTR}\end{align} We notice time reversal, in this convention, flips the fermion parity as it interchanges between the creation and annihilation operators. 

The entire coupled Majorana wire Hamiltonian \eqref{NMajcone}, when $N=2r$ is even, can be turned into a model of coupled boson wires. The total Lagrangian density is a combination \begin{align}\mathcal{L}=\mathcal{L}_0-\mathcal{H}=\mathcal{L}_0-\left(\mathcal{H}_\|+\mathcal{H}_\perp\right)\end{align} where the Hamiltonian density $\mathcal{H}=\mathcal{H}_\|+\mathcal{H}_\perp$ consists of the sliding Luttinger liquid\cite{OHernLubenskyToner99,EmeryFradkinKivelsonLubensky00,VishwanathCarpentier01,SondhiYang01,MukhopadhyayKaneLubensky01} (SLL) component along each wire \begin{align}\mathcal{H}_\|=V_{\mathsf{x}}\sum_{y=-\infty}^{\infty}\sum_{j=1}^r\partial_x\widetilde{\phi}^j_y\partial_x\widetilde{\phi}^j_y\label{SLL}\end{align} and the backscattering component between wires \begin{gather}\mathcal{H}_\perp=-V_{\mathsf{y}}\sum_{y=-\infty}^{\infty}\sum_{j=1}^r(-1)^y\cos\left(2\vartheta_{y+1/2}^j\right)\label{SLLbackscattering}\\2\vartheta_{y+1/2}^j=\widetilde\phi^j_y-\widetilde\phi^j_{y+1}.\label{2Thetadef}\end{gather} The SLL Hamiltonian \eqref{SLL} contains the (normal ordered) kinetic term $i\boldsymbol\psi_y^T\partial_x\boldsymbol\psi_y=i(c_y^\dagger\partial_x c_y+c_y\partial_x c_y^\dagger)$ in \eqref{NMajcone} as well as possible forward scattering terms like the density-density coupling $(c_y^\dagger c_y)(c_y^\dagger c_y)$. 
The interwire backscattering Hamiltonian \eqref{SLLbackscattering} is identical to the second term $i\boldsymbol\psi_y^T\boldsymbol\psi_{y+1}=i(c_y^\dagger c_{y+1}+c_yc_{y+1}^\dagger)$ in \eqref{NMajcone}. This can be derived directly by applying the bosonization \eqref{complexfermion} and the Baker--Campbell--Hausdorff formula $e^{i\widetilde{\phi}_y}e^{-i\widetilde{\phi}_{y+1}}=e^{i(\widetilde{\phi}_y-\widetilde{\phi}_{y+1})+[\widetilde{\phi}_y,\widetilde{\phi}_{y+1}]/2}$. The alternating sign $(-1)^y$ in \eqref{SLLbackscattering} is crucial to preserve time reversal symmetry \eqref{bosonTR}, which relates $\mathcal{T}2\vartheta_{y+1/2}^j\mathcal{T}^{-1}=2\vartheta_{y+3/2}^j-\pi$.

The $r$ sine-Gordon terms in \eqref{SLLbackscattering} between the same pair of adjacent wires mutually commute \begin{align}\left[2\vartheta_{y+1/2}^j(x),2\vartheta_{y+1/2}^{j'}(x')\right]=0\end{align} and share simultaneous eigenvalues. If there was a single pair of counter-propagating wires, these potentials would have pinned $\langle2\vartheta_{y+1/2}^j(x)\rangle=(2n+y)\pi$ between the two wires. However, they compete with the sine-Gordon terms between the next pair of wires due to the non-commuting relation \begin{align}&\left[2\vartheta_{y+1/2}^j(x),2\vartheta_{y+3/2}^{j'}(x')\right]\nonumber\\=&2\pi i(-1)^y\left[\theta(j-j')+\delta^{jj'}\theta(x'-x)\right]
\end{align} where the unit step function $\theta(s)=0$ when $s\leq0$, or 1 when $s>0$. In other words, the vertex operators $e^{i2\vartheta^j_{y+1/2}}$ produces fluctuations to adjacent pairs, \begin{align}&e^{-i2\vartheta^j_{y+1/2}(x)}2\vartheta^j_{y+3/2}(x')e^{i2\vartheta^j_{y+1/2}(x)}\nonumber\\=&2\vartheta^j_{y+3/2}(x')+2\pi(-1)^y\theta(x'-x).\end{align} The uniform backscattering strength $V_{\mathsf{y}}$, as protected by time reversal \eqref{TR}, exactly balances the competing potentials so that the Hamiltonian $\mathcal{H}=\mathcal{H}_\|+\mathcal{H}_\perp$ remains gapless.





\section{Gapping surface Majorana cones}\label{sec:gapping}

The previous section describes the gapless surface Majorana fermions of a 3D topological superconductor using a coupled wire model \eqref{NMajcone}. It consists of an array of chiral wires, each of which carries $N$ flavors of Majorana fermions co-propagating in alternating directions (see figure~\ref{fig:couplewires}). Together with uniform backscattering interactions between adjacent wires, the model captures $N$ surface Majorana cones with linear energy dispersion about zero energy and momentum (see figure~\ref{fig:Majoranacone}). 
In this section we construct explicit fermion interactions that introduce an excitation energy gap to the surface Majorana cones while preserving time reversal symmetry. Generically, this leaves behind a fermionic surface topological order, which will not be discussed until the next section.

\begin{figure}[htbp]
\centering
\includegraphics[width=0.4\textwidth]{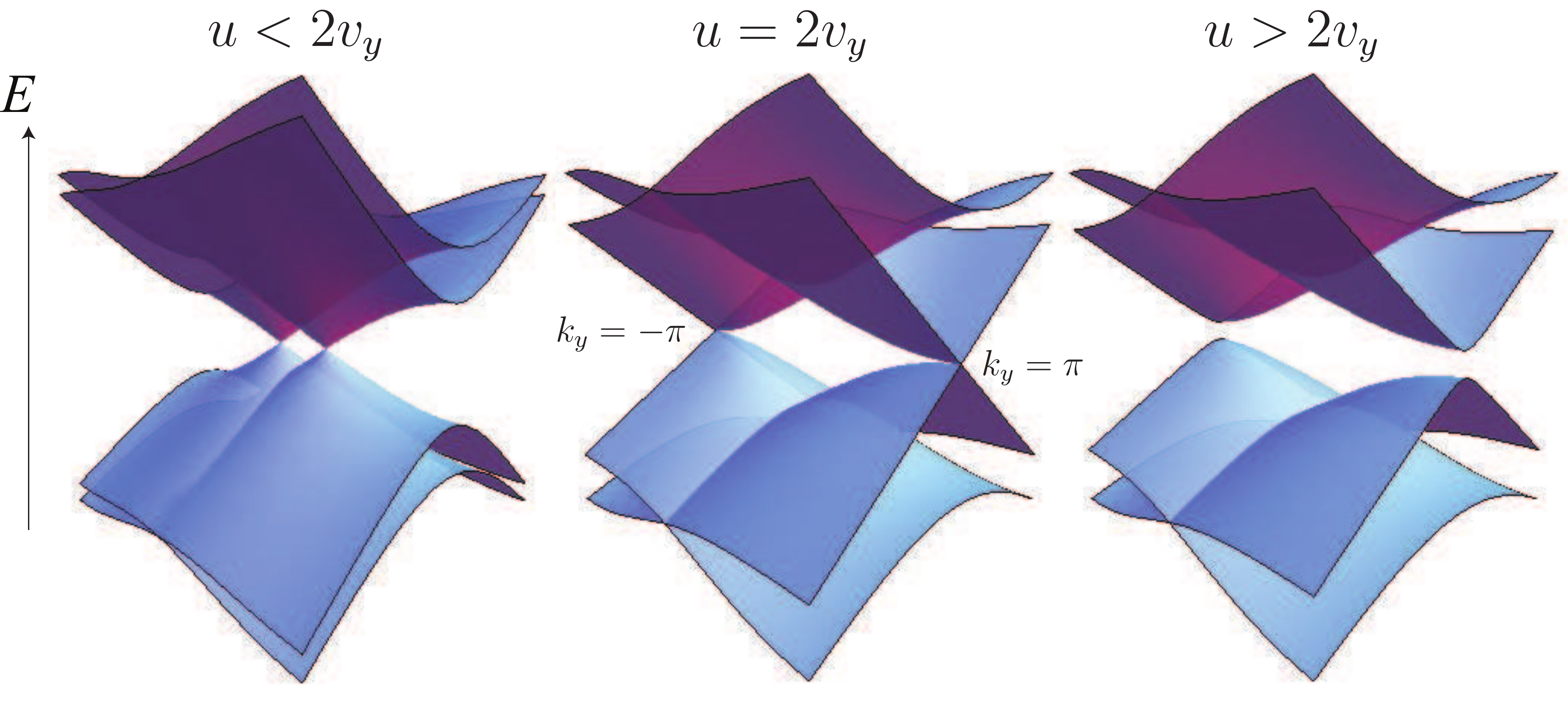}
\caption{Energy spectrum of the $N=2$ coupled Majorana wire model with inter-flavor mixing.}\label{fig:bands}
\end{figure}

We begin with the simplest case when there are $N=2$ chiral Majorana channels along each wire and correspond to two surface Majorana cones. As eluded in section~\ref{sec:couplewire}, due to the non-local nature of time reversal, the coupled wire model can be gapped by single-body backscattering terms without violating the symmetry. Although this cannot be applied to a conventional topological superconductor with local time reversal, this model demonstrates the idea of {\em fractionalization}, which can be generalized to the many-body interacting case and subsequently lead to surface topological order. The Hamiltonian $\mathcal{H}=\mathcal{H}_0+\mathcal{H}_{\mathrm{bc}}$ consists of the original model \eqref{NMajcone} with two fermion flavors $\boldsymbol\psi_y=(\psi_y^1,\psi_y^2)$ and the inter-flavor backscattering \begin{align}\mathcal{H}_{\mathrm{bc}}=iu\sum_{y=-\infty}^\infty\psi^1_y\psi^2_{y+1}\label{HbcN=2}\end{align} which is symmetric under the time reversal \eqref{TR}, $\mathcal{T}:\psi^a_y\to(-1)^y\psi^a_{y+1}$. The BdG Hamiltonian $H_{\mathrm{BdG}}({\bf k})=H_{\mathrm{BdG}}^0({\bf k})+H_{\mathrm{BdG}}^{\mathrm{bc}}({\bf k})$ is the combination of \eqref{NMajconeBdG} and \begin{align}H_{\mathrm{BdG}}^{\mathrm{bc}}({\bf k})=&\frac{u}{2}\left[(1-\cos k_y)\sigma_x\tau_z+(1+\cos k_y)\sigma_y\tau_y\right.\nonumber\\&\;\;\;\left.-\sin k_y(\sigma_y\tau_z+\sigma_x\tau_y)\right]\end{align} which is symmetric under $T_{\bf k}$ in \eqref{TRBdG}. The energy spectrum depends on the relative strength between the two interwire couplings $iv_{\mathsf{y}}(\psi^1_y\psi^1_{y+1}+\psi^1_y\psi^1_{y+1})$ and $iu\psi^1_y\psi^2_{y+1}$ (see figure~\ref{fig:bands}). When $u=0$, the two Majorana cone coincide at zero momentum. A finite $u$ separates the two until they have traveled across the Brillouin zone and annihilate each other at $k_y=\pi$ when $u>2v_{\mathsf{y}}$. Once an energy gap has opened up, the BdG Hamitonian has a unit Chern invariant \begin{align}\mathrm{Ch}=\frac{i}{2\pi}\int_{-\infty}^{\infty}dk_x\int_{-\pi}^\pi dk_y\mathrm{Tr}\left(\mathcal{F}_{\bf k}\right)=1\end{align} where $\mathrm{Tr}\left(\mathcal{F}_{\bf k}\right)=\mathrm{Tr}\left(\langle\partial_{k_y}u^a_{\bf k}|\partial_{k_x}u^b_{\bf k}\rangle-\langle\partial_{k_x}u^a_{\bf k}|\partial_{k_y}u^b_{\bf k}\rangle\right)$ is the Berry curvature constructed from the two occupied eigenstates $u^1_{\bf k},u^2_{\bf k}$ below zero energy of $H_{\mathrm{BdG}}({\bf k})$. The coupled Majorana wire model thus behaves like a chiral $p+ip$ topological superconductor\cite{Volovik:book,ReadGreen}. However the single-body Hamiltonian does not possesses a topological order in the sense that it does not support anyonic excitations. For instance the $\psi\to-\psi$ $\mathbb{Z}_2$ symmetry is global and $\pi$-vortices are not quantum excitations of the model but rather introduced as classical extrinsic defects.

This example relies on a simple decomposition of the degrees of freedom along each wire, $N=2=1+1$. The two Majorana fermions $\psi^1_y,\psi^2_y$ are backscattered independently to adjacent wires in opposite directions. Unlike the intra-flavor couplings $iv_{\mathsf{y}}(\psi^1_y\psi^1_{y+1}+\psi^1_y\psi^1_{y+1})$, inter-flavor terms $iu\psi^1_y\psi^2_{y+1}$ freeze independent degrees of freedom and they are not competing with each other. It is useful to notice that the decomposition breaks the $SO(2)_1$ symmetry described in section~\ref{sec:SO(N)}, and as a result the $so(2r)_1$ CFT along each wire splits into a pair of chiral Ising CFT's.

\begin{figure}[htbp]
\centering
\includegraphics[width=0.3\textwidth]{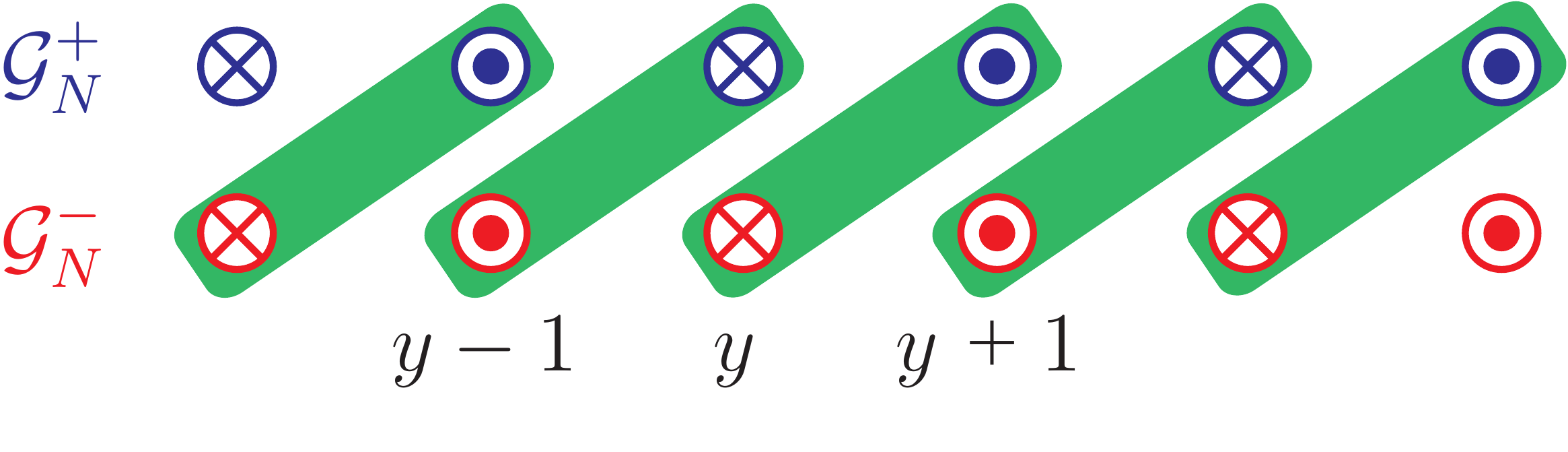}
\caption{Interwire gapping terms \eqref{bcint} (green rectangular boxes) between chiral fractional $\mathcal{G}_N^{R,\pm},\mathcal{G}_N^{L,\pm}$ sectors (resp.~$\otimes,\odot$) in opposite direction.}\label{fig:fractionalization}
\end{figure}

We can now generalize this idea to all $N$, but with many-body interwire interactions. From now on, unless specified otherwise, we turn off all single-body scattering terms. For instance, the vertical velocity now vanishes, $v_{\mathsf{y}}=0$, in the kinetic part $\mathcal{H}_0$ of the coupled wire model \eqref{NMajcone}. First we seek a decomposition of the $so(N)_1$ degrees of freedom along each wire (see section~\ref{sec:SO(N)}) into a pair of identical but independent sectors (also see figure~\ref{fig:couplewires})  \begin{align}so(N)_1\supseteq\mathcal{G}_N^+\times\mathcal{G}_N^-\label{fractioneq}\end{align} where $\mathcal{G}_N^\pm$ are the Kac-Moody subalgebras \begin{align}\mathcal{G}^\pm_N=\left\{\begin{array}{*{20}l}so(N/2)_1&\mbox{for $N$ even}\\so(3)_3\times so\left(\frac{N-9}{2}\right)_1&\mbox{for $N$ odd}\end{array}\right.\label{GNfractionalization}\end{align} to be discussed below. This fractionalization has to be complete in the sense that the Sugawara energy-momentum tensor exactly splits into \begin{align}T_{so(N)_1}=T_{\mathcal{G}_N^+}+T_{\mathcal{G}_N^-}.\end{align} In particular the central charge divides \begin{align}c_-\left(so(N)_1\right)=2c_-\left(\mathcal{G}_N\right)=c_-\left(\mathcal{G}_N^+\right)+c_-\left(\mathcal{G}_N^-\right)\end{align} and there are no degrees of freedom left behind. Using the subalgebra current operators ${\bf J}_{\mathcal{G}_N^\pm}$, which are quadratic in $\psi$'s, we construct the four-fermion backscattering interaction \begin{align}\mathcal{H}_{\mathrm{int}}&=u\sum_{y=-\infty}^\infty{\bf J}_{\mathcal{G}_N^-}^y\cdot{\bf J}_{\mathcal{G}_N^+}^{y+1}\label{bcint}\\&=u\sum_{y'=-\infty}^\infty{\bf J}_{\mathcal{G}_N^{L,-}}^{2y'-1}\cdot{\bf J}_{\mathcal{G}_N^{R,+}}^{2y'}+{\bf J}_{\mathcal{G}_N^{R,-}}^{2y'}\cdot{\bf J}_{\mathcal{G}_N^{L,+}}^{2y'+1}\nonumber\end{align} for $u$ positive, and $R,L$ labels the propagating directions of the currents. This is pictorially presented in figure~\ref{fig:couplewires} and \ref{fig:fractionalization}. 

In this section, we design the fractionalization \eqref{fractioneq} of $so(N)_1$ for all $N$ and show that the backscattering interactions \eqref{bcint} open an excitation energy gap without breaking time reversal. In CFT context, \eqref{fractioneq} is also known as a conformal embedding\cite{bigyellowbook,BaisEnglertTaorminaZizzi87,SchellekensWarner86,BaisBouwknegt87}. When $N=2r$ is even, there is an obvious decomposition \begin{align}so(2r)_1\supseteq so(r)_1^+\times so(r)_1^-\end{align} where the ``$+$" sector contains $\psi^1,\ldots,\psi^r$ while the ``$-$" one contains the rest $\psi^{r+1},\ldots,\psi^{2r}$. In section~\ref{sec:even}, we review how the ${\bf J}_{so(r)_1^R}\cdot{\bf J}_{so(r)_1^L}$ interactions contribute an energy gap. This is a direct application of the well-studied $O(N)$ Gross-Neveu problem\cite{GrossNeveu,ZamolodchikovZamolodchikov78,Witten78,ShankarWitten78} in 1D. In the discrete limit, this is related to the Haldane $O(3)$ antiferrormagnetic spin chain\cite{Haldanespinchain1,Haldanespinchain2}, the Affleck - Kennedy - Lieb - Tasaki (AKLT) spin chains\cite{AKLT1,AKLT2} and the $SO(n)$ Heisenberg chain\cite{TuZhangXiang08,TuOrus11,AletCapponiNonneLecheminantMcCulloch11}. When $N$ is odd, the splitting \eqref{fractioneq} is less trivial. We will make use of the level-rank duality\cite{bigyellowbook,NaculichRiggsSchnitzer90,MlawerNaculichRiggsSchnitzer91} \begin{align}so(n^2)_1\supseteq so(n)_n\times so(n)_n\end{align} which comes from the fact that the tensor product $SO(n)\otimes SO(n)$ is a Lie subgroup in $SO(n^2)$. In particular, we will demonstrate the simplest case in section~\ref{sec:odd} when $n=3$. The division of $so(9)_1$ can subsequently be generalized to $so(N)_1$ for all odd $N$ effectively by writing $N=9+2r$. This sets $\mathcal{G}_N^\pm=so(3)_3\times so(r)_1$ in \eqref{fractioneq} and the corresponding interwire backscattering interactions \eqref{bcint}.

\subsection{Gapping even Majorana cones}\label{sec:even}
We begin with the coupled Majorana wire model \eqref{NMajcone} (or figure~\ref{fig:couplewires}) with $N=2r$ chiral fermion channels per wire and corresponds to the same number of gapless Majorana cones. Similar to the previously shown $N=2$ case, the gapless modes can be removed using simple single-body backscattering terms. We however are interested in finding gapping interactions that would support surface topological order as well. In section~\ref{sec:SO(N)} and appendices~\ref{sec:kleinfactors}, \ref{sec:kleinfactors2}, we described the $so(N)_1$ WZW theory, which along the $y^{\mathrm{th}}$ wire is generated by chiral current operators \eqref{so(N)current} \begin{align}J^{(a,b)}_y=(-1)^yi\psi^a_y\psi^b_y.\label{TRsymmcurrentconvention}\end{align} We take the alternating sign convention $(-1)^y$ so that under time reversal, $\mathcal{T}J^{(a,b)}_y\mathcal{T}^{-1}=J^{(a,b)}_{y+1}$. We consider two subsets of generators, $so(r)_1^+$ containing $J^{(a,b)}$ for $1\leq a<b\leq r$, and $so(r)_1^-$ containing $J^{(a,b)}$ for $r+1\leq a<b\leq2r$. As they act on independent fermion sectors, the two sets of operators commute or equivalently their operator product expansions (OPE) are trivial up to non-singular terms. Moreover the Sugawara energy-momentum tensor \eqref{SugawaraT} for $so(N)_1$ completely splits into a sum between 
\begin{align}T_{so(r)_1^+}=-\frac{1}{2}\sum_{a=1}^r\psi^a\partial\psi^a,\quad T_{so(r)_1^-}=-\frac{1}{2}\sum_{a=r+1}^{2r}\psi^a\partial\psi^a.\end{align} This ensures all degrees of freedom in $so(2r)_1$ are generated by tensor products between those in the $so(r)_1^\pm$ sectors. Precisely this means any $so(2r)_1$ primary field is a fusion channel of the OPE of certain primary field pair in $so(r)_1^+$ and $so(r)_1^-$. Thus as long as the gapping terms independently freeze both sectors, they remove all gapless degrees of freedom.

The backscattering interactions \eqref{bcint} couples the $so(r)_1^-$ sector on the $y^{\mathrm{th}}$ wire with the $so(r)_1^+$ sector on the $(y+1)^{\mathrm{th}}$ one. They can explicitly written as \begin{align}\mathcal{H}_{\mathrm{int}}=u\sum_{y=-\infty}^{\infty}\sum_{1\leq a<b\leq r}\psi_y^{r+a}\psi_y^{r+b}\psi_{y+1}^a\psi_{y+1}^b.\label{bcinteven}\end{align} Firstly, the interactions are time reversal symmetric as \eqref{bcinteven} is unchanged by $\psi_y^a\to(-1)^y\psi_{y+1}^a$. Secondly, it breaks the $O(2r)$ symmetry to $O(r)^+\times O(r)^-$. The symmetry breaking can be faciliated by forward scattering within wires that renormalizes the velocities differently between the $so(r)^\pm_1$ sectors. Eq.\eqref{bcinteven} is also a combination allowed by the chiral $O(r)$ symmetry \begin{align}\psi^a_y\to\left(\mathcal{O}^{(-1)^y}\right)^a_b\psi_y^b,\quad\psi^{r+a}_y\to\left(\mathcal{O}^{(-1)^{y+1}}\right)^a_b\psi_y^{r+b}.\end{align} The chiral symmetry only allows cross couplings ${\bf J}_{so(r)^\pm_1}^y\cdot{\bf J}_{so(r)^\mp_1}^{y+1}$ between adjacent wires. Instead of \eqref{bcinteven}, another possibility would be its mirror image with summands $\psi_y^a\psi_y^b\psi_{y+1}^{r+a}\psi_{y+1}^{r+b}$. This competes with the original, but as long as mirror symmetry is broken and their strength is asymmetric, an energy gap will open. In the following we will ignore the mirror image by assuming it is weaker.

Next we notice that the four-fermion interaction \eqref{bcinteven} is marginally relevant when velocity $v_{\mathsf{x}}$ is uniform. The dimensionless coupling strength $u$ follows the renormalization group (RG) flow equation \begin{align}\frac{du}{d\lambda}=+4\pi(r-2)u^2\label{RGeven}\end{align} when length scale renormalizes by $l\to e^\lambda l$. This can be verified by applying the RG formula among marginal operators\cite{Cardybook} \begin{align}\frac{dg_l}{d\lambda}=-2\pi \sum_{mn}C^{mn}_lg_mg_n\label{RGeqn}\end{align} where $C^{mn}_l$ is the fusion coefficient of the OPE $\mathcal{O}_m\mathcal{O}_n=C^{mn}_l\mathcal{O}_l+\ldots$ between operators in the perturbative action $\delta S=\int d\tau dx\sum_mg_m\mathcal{O}_m$. In the current case, the fusion coefficient $\mathcal{O}\mathcal{O}=-2(r-2)\mathcal{O}+\ldots$ can be evaluated simply by applying the Wick's theorem of fermions, for $\mathcal{O}=-\sum_{y,a,b}\psi_y^{r+a}\psi_y^{r+b}\psi_{y+1}^a\psi_{y+1}^b$. The plus sign in \eqref{RGeven} shows the interacting strength grows at weak coupling. To show that the backscattering \eqref{bcinteven} indeed opens up a gap, we first focus on a single coupled pair of counter-propagating $so(r)_1$ channels (see figure~\ref{fig:fractionalization}).

\subsubsection{The \texorpdfstring{$O(r)$}{O(r)} Gross-Neveu model}\label{sec:GNmodel}
Here we concentrate on a particular set of backscattering terms in \eqref{bcinteven} at say an even $y$. We relabel $\psi_y^{r+a}=\psi^a_R$ and $\psi_{y+1}^a=\psi^a_L$, for $a=1,\ldots,r$. The interaction between the $y^{\mathrm{th}}$ and $(y+1)^{\mathrm{th}}$ wire is identical to that of the $O(r)$ Gross-Neveu (GN) model\cite{GrossNeveu,ZamolodchikovZamolodchikov78,Witten78,ShankarWitten78} \begin{align}\mathcal{H}_{\mathrm{GN}}=-\frac{u}{2}\left(\boldsymbol\psi_R\cdot\boldsymbol\psi_L\right)^2\label{GN}\end{align} where the minus sign is from the fermion exchange statistics $\psi_R^a\psi_R^b\psi_L^a\psi_L^b=-\psi_R^a\psi_L^a\psi_R^b\psi_L^b$. This GN model is known to have an excitation energy gap for $r>2$. 

For even $r=2n>2$, the Majorana fermions can be paired into Dirac ones and subsequently bosonized (see section~\ref{sec:bosonization}), $c^j_{R/L}=(\psi^{2j-1}_{R/L}+i\psi^{2j}_{R/L})/\sqrt{2}\sim e^{i\widetilde{\phi}^j_{R/L}}$, for $j=1,\ldots,n$. Using \begin{align}\boldsymbol\psi_R\cdot\boldsymbol\psi_L=\sum_{j=1}^nc^j_R(c^j_L)^\dagger+(c^j_R)^\dagger c^j_L\sim\sum_{j=1}^n\cos\left(2\Theta^j\right)\end{align} for $2\Theta^j=\widetilde{\phi}^j_R-\widetilde{\phi}^j_L$ (also see \eqref{2Thetadef}) are mutually commuting variables, the GN interation \eqref{GN} takes the bosonized form \begin{align}\mathcal{H}_{GN}&\sim 
u\sum_{j=1}^n\partial_x\widetilde{\phi}^j_R\partial_x\widetilde{\phi}^j_L-u\sum_{j_1\neq j_2}\sum_\pm\cos\left(2\Theta^{j_1}\pm2\Theta^{j_2}\right)\nonumber\\&=u\sum_{j=1}^n\partial_x\widetilde{\phi}^j_R\partial_x\widetilde{\phi}^j_L-u\sum_{\boldsymbol\alpha\in\Delta}\cos\left(\boldsymbol\alpha\cdot2\boldsymbol\Theta\right)\label{GNso(2n)}\end{align} 
where $2\boldsymbol\Theta=(2\Theta^1,\ldots,2\Theta^n)$ and $\boldsymbol\alpha$ are roots of $so(2n)$ (see \eqref{approots}).
The first term renormalizes the velocity $V_{\mathsf{x}}$ in \eqref{SLL} as well as the Luttinger parameter. We assume $V_{\mathsf{x}}>>u$ so that the first term can be dropped. The remaining sine-Gordon terms are responsible for gapping out all low energy degrees of freedom. Firstly the angle parameters mutually commute and share simultaneous eigenvalues. The ground state minimizes the energy by uniformly pinning the ground state expectation value (GEV) \begin{align}\left\langle2\Theta^j(x)\right\rangle=\pi m_\psi^j,\quad m_\psi^j\in\mathbb{Z}.\label{VEVpsi}\end{align} 

We notice in passing that the following subset of sine-Gordon terms \begin{align}-u\sum_{I=1}^n\cos\left(\boldsymbol\alpha_I\cdot2\boldsymbol\Theta\right)&=-u\sum_{I=1}^n\cos\left[\sum_{J=1}^nK_{IJ}(\phi^J_R-\phi^J_L)\right]\nonumber\\&=-u\sum_{I=1}^n\cos\left({\bf n}_I^T\mathbb{K}\boldsymbol\Phi\right)\label{nullvectorsubset}\end{align} using the simple roots $\boldsymbol\alpha_I$ in \eqref{simpleroots}, is already enough to remove all low energy degrees of freedom. Here $K_{IJ}$ is the Cartan matrix \eqref{Kso(2r)} of $so(2n)$ that appears in the Lagrangian density \begin{align}\mathcal{L}_0=\frac{1}{2\pi}\partial_x\boldsymbol\Phi^T\mathbb{K}\partial_t\boldsymbol\Phi\end{align} for $\mathbb{K}=K\oplus(-K)$ and $\boldsymbol\Phi=(\boldsymbol\phi_R,\boldsymbol\phi_L)$, and $\phi$ is related to $\widetilde{\phi}$ by the basis transformation \eqref{phiphitilde}. For instance, the $n$ vector coefficients ${\bf n}_J=({\bf e}_J,{\bf e}_J)$ in \eqref{nullvectorsubset} form a null basis \begin{align}{\bf n}_I^T\mathbb{K}{\bf n}_J=0\end{align} and guarantee an energy gap according to Ref.\onlinecite{Haldane95}. The remaining GN terms in \eqref{GNso(2n)} are compatible with \eqref{nullvectorsubset} as they share the same minima. 

There are constraints on the GEV $m_\psi^j$ in \eqref{VEVpsi}. In order to minimize $-\cos(\boldsymbol\alpha\cdot2\boldsymbol\Theta)$ in \eqref{GNso(2n)}, $\langle\boldsymbol\alpha\cdot2\boldsymbol\Theta\rangle$ must be an integer multiple of $2\pi$. This restricts uniform parity among $m_\psi^j$ so that the sign in the fermion backscattering amplitude \begin{align}\left\langle \psi_R^a(x)\psi_L^a(x)\right\rangle&=\left\langle c_R^j(x)c_L^j(x)^\dagger\right\rangle\nonumber\\&\sim\left\langle e^{i2\Theta^j(x)}\right\rangle=(-1)^{m_\psi}.\label{fbackVEV}\end{align} does not depend on fermion flavor $j$. This is not the only non-zero GEV as $\psi$ is not the only primary field in $so(2n)_1$. The backscattering of spinor fields $V_{s_\pm}=e^{i\boldsymbol\varepsilon\cdot\widetilde{\boldsymbol\phi}/2}$ \eqref{sprimaryso(2r)} corresponds to the two GEV's \begin{gather}
\left\langle V_{s_\pm}^R(x)V_{s_\pm}^L(x)^\dagger\right\rangle=\left\langle e^{i\boldsymbol\varepsilon\cdot\boldsymbol\Theta(x)}\right\rangle=e^{i\pi m_{s_\pm}/2}\label{VEVspm2}\end{gather} where $\boldsymbol\varepsilon=(\varepsilon_1,\ldots,\varepsilon_n)$ for $\varepsilon_j=\pm1$, and the overall sign $\prod_j\varepsilon_j$ is positive for the even spinor field $s_+$, or negative for $s_-$. Here the GEV \eqref{VEVspm2} does not depend on the choice of $\boldsymbol\varepsilon$. This is because given $\boldsymbol\varepsilon$ and $\boldsymbol\varepsilon'$ with the same overall parity $\prod\varepsilon_j=\prod\varepsilon'_j$, $\boldsymbol\varepsilon\cdot\boldsymbol\Theta$ and $\boldsymbol\varepsilon'\cdot\boldsymbol\Theta$ differ by some combination of $\boldsymbol\alpha\cdot2\boldsymbol\Theta$, which takes expectation value in $2\pi\mathbb{Z}$.

There are extra constraints between $m_\psi$ and $m_{s_\pm}$ from the fusion rules of the primary fields of $so(2n)_1$ (see \eqref{so(2r)fusion1} and \eqref{so(2r)fusion2}). Firstly, $s_\pm\times\psi=s_\mp$ requires \begin{align}m_{s_+}\equiv m_{s_-}+2m_\psi\quad\mbox{mod $4\mathbb{Z}$}.\label{mpsispmconstraint1}\end{align} Take the highest weights $\boldsymbol\varepsilon_+^0=(1,\ldots,1)$ and $\boldsymbol\varepsilon_-^0=(1,\ldots,-1)$ for instance. $\boldsymbol\varepsilon_+^0\cdot\boldsymbol\Theta=\boldsymbol\varepsilon_-^0\cdot\boldsymbol\Theta+2\Theta^n$ imples $m_{s_+}(\boldsymbol\varepsilon_+^0)=m_{s_-}(\boldsymbol\varepsilon_+^0)+2m_\psi^n$. Lastly the fusion rules \begin{align}s_\pm\times s_\pm\left\{\begin{array}{*{20}l}1,&\mbox{for $n$ even}\\\psi,&\mbox{for $n$ odd}\end{array}\right.\end{align} requires the GEV's to obey \begin{align}\left\{\begin{array}{*{20}l}(-1)^{m_{s_\pm}}=1&\mbox{for $n$ even}\\(-1)^{m_{s_\pm}}=(-1)^{m_\psi}&\mbox{for $n$ odd}\end{array}\right.\label{mpsispmconstraint2}\end{align} for similar reasons.

The GN model therefore has four ground states when $r=2n>2$. They are specified by the quantum numbers (i) $m_{s_+}=0,1,2,3$ modulo 4 when $n$ is odd, or (ii) $m_{s_+}=0,2$ and $m_{s_-}=0,2$ modulo 4 when $n$ is even. The rest are fixed by \eqref{mpsispmconstraint1} and \eqref{mpsispmconstraint2}. Quasiparticle excitations are trapped between domain walls or kinks separating distinct ground states\cite{Witten78,ShankarWitten78,FendleySaleur01}. For example, the vertex operator $V_{s_+}^R(x_0)=e^{i\boldsymbol\varepsilon^0_+\cdot\widetilde{\boldsymbol\phi}_R(x_0)/2}$ of an even spinor field creates a jump in the GEV \eqref{fbackVEV} \begin{align}\left\langle V_{s_+}^R(x_0)^\dagger e^{i2\Theta^j(x)}V_{s_+}^R(x_0)\right\rangle
=(-1)^{m'_\psi+\theta(x_0-x)}\end{align} because of the Baker-Hausdorff-Campbell formula and the commutation relation from \eqref{ETcomm0} \begin{align}\left[2\Theta^j(x),\boldsymbol\varepsilon^0_+\cdot\widetilde{\boldsymbol\phi}_R(x_0)/2\right]=i\pi\left(\theta(x_0-x)-n+j-1\right)\end{align} for $\theta$ the unit step function $\theta(s)=0$ when $s\leq0$, or 1 when $s>0$, and $m'_\psi=m_\psi+n-j+1$. In general, the primary fields $V^R_{s_\pm}=e^{i\boldsymbol\varepsilon\cdot\widetilde{\boldsymbol\phi}_R}$ and $c_R^j=e^{i\widetilde{\phi}^j_R}$ corresponds to the domain walls of $m_{s_\pm}$: \begin{align}\left\langle V_{s_\pm}^R(x_0)^\dagger e^{i\boldsymbol\varepsilon^0_\pm\cdot\boldsymbol\Theta(x)}V_{s_\pm}^R(x_0)\right\rangle&=e^{\frac{i\pi}{2}\left(m'_{s_\pm}+n\theta(x_0-x)\right)}\nonumber\\\left\langle V_{s_\mp}^R(x_0)^\dagger e^{i\boldsymbol\varepsilon^0_\pm\cdot\boldsymbol\Theta(x)}V_{s_\mp}^R(x_0)\right\rangle&=e^{\frac{i\pi}{2}\left(m'_{s_\pm}+(n-2)\theta(x_0-x)\right)}\nonumber\\\left\langle c_R^j(x_0)^\dagger e^{i\boldsymbol\varepsilon^0_\pm\cdot\boldsymbol\Theta(x)}c_R^j(x_0)\right\rangle&=e^{\frac{i\pi}{2}\left(m'_{s_\pm}+2\theta(x_0-x)\right)}.\end{align}

Now we move on to the odd $r=2n+1>1$ case. First we pair the first $2n$ Majorana fermions into $n$ Dirac ones and bosonize them similar to the previous even $r$ case. This leaves a single unpaired Majorana fermion $\psi^r_{R/L}$. Dropping terms that only renormalizes velocities, the GN model \eqref{GN} takes the partially bosonized form \begin{align}\mathcal{H}_{\mathrm{GN}}&\sim -u\sum_{\boldsymbol\alpha\in\Delta_{so(2n)}}\cos\left(\boldsymbol\alpha\cdot2\boldsymbol\Theta\right)\nonumber\\&\quad\quad-u\left[\sum_{j=1}^n\cos\left(2\Theta^j\right)\right]i\psi^r_R\psi^r_L\label{GNso(2n+1)}\end{align} where the first line is identical to the even $r$ case \eqref{GNso(2n+1)} and is responsible for gapping out first $2n$ Majorana channels. Projecting onto the lowest energy states and taking the GEV $\langle\cos(2\Theta^j)\rangle=(-1)^{m_\psi}$, the interacting Hamiltonian becomes \begin{align}\mathcal{H}_{\mathrm{GN}}\sim-2n(n-1)u-nu(-1)^{m_\psi}i\psi^r_R\psi^r_L\label{GNlasefermionsector}\end{align} which is identical to the continuum limit of the quantum Ising model with transverse field after a Jordan-Wigner transformation. The remaining Majorana channel $\psi^r_{R/L}$ is gapped by the single-body backscattering term. The sign of the mass gap $nu(-1)^{m_\psi}$ determines the phase of the Ising model. We take the convention so that a negative (or positive) mass with $m_\psi\equiv1$ (resp.~$m_\psi\equiv0$) corresponds to the order (resp.~disorder) phase. 

Like the previous case, the fermion backscattering amplitude \eqref{fbackVEV} is not the only ground state expectation value. From \eqref{so(2r+1)sigma} appendix~\ref{sec:kleinfactors2}, the Ising twist field of $so(2n+1)_1$ can be written as the product $V_\sigma=e^{i\boldsymbol\varepsilon\cdot\widetilde{\boldsymbol\phi}/2}\sigma^r$, where $\boldsymbol\varepsilon=(\varepsilon_1,\ldots,\varepsilon_n)$ for $\varepsilon_j=\pm1$, and $\sigma^r_{R/L}=\sigma^{2n+1}_{R/L}$ is the twist field along the last Majorana channel. There are three possible GEV for the backscattering \begin{align}\left\langle V_\sigma^R(x)V_\sigma^L(x)^\dagger\right\rangle&=\left\langle e^{i\boldsymbol\varepsilon\cdot\boldsymbol\Theta(x)}\sigma^r_R(x)\sigma^r_L(x)\right\rangle\label{VEVsigma}\\&\sim\left\{\begin{array}{*{20}l}0&\mbox{for the disorder phase}\\\pm1&\mbox{for the order phase}\end{array}\right..\nonumber\end{align} Here we choose the convention so that $\sigma_R\sigma_L$ takes the role of the spin operator $\boldsymbol\sigma$ in the Ising model and its non-trivial GEV's in the order phase specify two ground states $|\uparrow\rangle$ and $|\downarrow\rangle$. 

Again, quasiparticle excitations are trapped between domain walls separating distinct ground states\cite{Witten78,ShankarWitten78,FendleySaleur01}. For example a twist field $V_\sigma^R$ (or $V_\sigma^L$) sits between the order to disorder phase boundary where the quantum number $m_\psi$ flips from 1 to 0, or equivalently the fermion mass gap in \eqref{GNlasefermionsector} changes sign. This is because the twist field $V^R_\sigma(x_0)$ introduces a flip in boundary condition $\psi_R(x_0+)=-\psi_R(x_0-)$ and corresponds to a change of sign in front of the fermion backscattering $i\psi_R\psi_L$. Alternatively, this can also be understood by identifying $V_\sigma$ as a Jackiw-Rebbi soliton\cite{JackiwRebbi76} or a zero energy Majorana bound state between a trivial and topological superconductor\cite{Kitaevchain} in 1D. 

Next a $\uparrow-\downarrow$ domain wall of opposite signs of the GEV \eqref{VEVsigma} in the order phase 
traps an excitation in the fermion sector $\psi$. This can be seen by equating the order Ising phase to a 1D topological superconductor\cite{Kitaevchain}, where the two Ising ground states corresponds to the even and odd fermion parity states among the pair of boundary Majorana zero modes. Adding (or subtracting) a fermion therefore flips the parity as well as the GEV in \eqref{VEVsigma}. We notice this domain wall interpretation of excitations is consistent with the non-Abelian fusion rule \begin{align}\sigma\times\sigma=1+\psi.\end{align} The trivial fusion channel corresponds to the annihilation of a domain wall pair such as \begin{align}|\underbrace{\ldots\uparrow\uparrow}_{\mathrm{order}}\underbrace{\leftarrow\leftarrow}_{\mathrm{disorder}}\underbrace{\uparrow\uparrow\ldots}_{\mathrm{order}}\rangle\xrightarrow{\mathrm{fusion}}|\ldots\uparrow\uparrow\ldots\rangle\end{align} while the fermion fusion channel corresponds to joining the pair of ``order - disorder" domain walls into a kink \begin{align}|\underbrace{\ldots\uparrow\uparrow}_{\mathrm{order}}\underbrace{\leftarrow\leftarrow}_{\mathrm{disorder}}\underbrace{\downarrow\downarrow\ldots}_{\mathrm{order}}\rangle\xrightarrow{\mathrm{fusion}}|\ldots\uparrow\uparrow\downarrow\downarrow\ldots\rangle.\end{align}

\subsubsection{The special case: \texorpdfstring{$so(4)_1=su(2)_1\times su(2)_1$}{so(4)=su(2)xsu(2)}}\label{sec:gappingN=4}
The case when $r=2$ requires special attention. The $O(2)$ GN model \eqref{GN} is a gapless Luttinger liquid because its bosonized form \eqref{GNso(2n)} contains no sine-Gordon terms and the rest only renormalizes velocities and the Luttinger parameter. As a result the fractionalization (or conformal embedding) $so(4)_1\supseteq so(2)_1\times so(2)_1$ of wires with $N=4$ Majorana channels does not lead to a gapped theory. Instead we turn to an alternative fractionalization $so(4)_1=su(2)_1^+\times su(2)_1^-$ that only applies for $N=4$.

The four Majorana $\psi^a_y$ along each wire can be paired into Dirac channels $c_y^1=(\psi^1_y+i\psi^2_y)/\sqrt{2}=e^{i\widetilde{\phi}^1_y}$ and $c_y^2=(\psi^3_y+i\psi^4_y)/\sqrt{2}=e^{i\widetilde{\phi}^2_y}$. It would be more convenient if we express the bosons in the new basis using the simple roots of $so(4)$: $\widetilde{\phi}^1=\phi^1-\phi^2$ and $\widetilde{\phi}^2=\phi^1+\phi^2$. Unlike when $r>2$, these bosons decouple in the Lagrangian density \eqref{L0} \begin{align}\mathcal{L}_0=\frac{1}{2\pi}\sum_{y=-\infty}^\infty(-1)^y\sum_{J=1}^22\partial_x\phi_y^J\partial_t\phi_y^J.\end{align} This is equivalent to the fact that the Cartan matrix $K_{so(4)}=\mathrm{diag}(2,2)$ is diagonal so that the Lie algebra splits into the product $su(2)^+\times su(2)^-$ of isoclinic rotations, each with Cartan matrix $K_{su(2)}=2$.

The $su(2)_1$ current generators are given by $S_{\mathsf{z}}^I(z)=i\sqrt{2}\partial\phi^I(z)$ and $S_\pm^I(z)=(S_{\mathsf{x}}^I\pm iS_{\mathsf{y}}^I)/\sqrt{2}=e^{i2\phi^I(z)}$, and they satisfy the OPE \begin{align}S^I_{\mathsf{i}}(z)S^I_{\mathsf{j}}(w)=\frac{\delta_{{\mathsf{i}}{\mathsf{j}}}}{(z-w)^2}+\frac{i\sqrt{2}\varepsilon_{{\mathsf{i}}{\mathsf{j}}{\mathsf{k}}}}{z-w}S^I_{\mathsf{k}}(w)+\ldots\label{su(2)1Jrelation}\end{align} for $I=1,2=+,-$. The $su(2)_1^+$ sector is completely decoupled from the $su(2)_1^-$ one as the OPE $S_{\mathsf{i}}^1(z)S^2_{\mathsf{j}}(w)$ is non-singular. They completely decomposes all low energy degrees of freedom as the energy momentum tensor splits into \begin{align}T_{so(4)_1}&=-\frac{1}{2}\sum_{j=1}^2\partial\widetilde{\phi}^j(z)\partial\widetilde{\phi}^j(z)\\&=-\sum_{J=1}^2\partial\phi^J(z)\partial\phi^J(z)=T_{su(2)_1^+}+T_{su(2)_1^-}.\nonumber\end{align} The gapping Hamiltonian is \begin{align}\mathcal{H}_{\mathrm{int}}&=u\sum_{y=-\infty}^\infty{\bf S}^2_y\cdot{\bf S}^1_{y+1}\label{su(2)gapppingH}\\&=2u\sum_{y=-\infty}^\infty\partial_x\phi_y^2\partial_x\phi_{y+1}^1-2\cos\left(4\Theta_{y+1/2}\right),\nonumber\\4\Theta_{y+1/2}&=2\phi^1_{y+1}-2\phi^2_y\\&=\widetilde{\phi}^1_{y+1}+\widetilde{\phi}^2_{y+1}+\widetilde{\phi}^1_y-\widetilde{\phi}^2_y.\nonumber\end{align} The first kinetic term of the interacting Hamiltonian only renormalizes velocities and the Luttinger parameter. The second sine-Gordon term involves four-fermion interactions and is responsible for the energy gap as it back-scatters the $su(2)_1^-$ sector on the $y^{\mathrm{th}}$ wire to the $su(2)^+_1$ sector on the $(y+1)^{\mathrm{th}}$ one. It pins the ground state expectation value (GEV) \begin{align}\left\langle e^{i2\Theta_{y+1/2}(x)}\right\rangle=(-1)^{m_s}\end{align} which characterizes the two distinct ground states. Like the previous cases, quasiparticle excitations are kinks in the GEV. The fundamental excitation can be created by the vertex operator $V_s=e^{i\phi^1_{y+1}}$, which is the semionic primary field in the $su(2)_1^+$ sector along the $(y+1)^{\mathrm{th}}$ wire.


\subsection{Gapping odd Majorana cones}\label{sec:odd}
We now move on to the case when there are $N=2r+1\geq3$ chiral Majorana channels on each wire in the coupled Majorana wire model \eqref{NMajcone} (of figure~\ref{fig:couplewires}). It corresponds to an odd number of Majorana cones on the surface of a 3D topological superconductor. The chiral degrees of freedom along each wire are described by a $so(N)_1$ WZW theory, which is going to be fractionalized into the pair $\mathcal{G}_N^+\times\mathcal{G}_N^-$ according to \eqref{GNfractionalization}. The $\mathcal{G}_N^-$ sector along the $y^{\mathrm{th}}$ wire will then be back-scattered onto the $\mathcal{G}_N^+$ sector along the $(y+1)^{\mathrm{th}}$ one by the current-current interaction \eqref{bcint}, which will introduce an energy gap.

Unlike the even $N$ case where $so(N)_1$ can simply be split into a pair of $so(N/2)_1$'s, here the decomposition is less trivial but leads to more exotic surface topological order. We begin with the particular case where 9 Majorana channels can be bipartite into \begin{align}so(9)_1\supseteq so(3)_3\times so(3)_3\label{so(9)fractionalization}\end{align} essentially by noticing that the tensor product $SO(3)\otimes SO(3)$ sits inside $SO(9)$. The two $so(3)_3$ WZW sectors carry decoupled current generators. They can then be back-scattered using the current-current interaction \eqref{bcint} onto adjacent wires in opposite directions (also see figure~\ref{fig:couplewires} and \ref{fig:fractionalization}).

For a general odd $N\geq9$, one can decompose the Majorana channels into $N=9+(N-9)$. The first 9 channels can be fractionalized by \eqref{so(9)fractionalization}, which we will discuss in detail below, and the remaining even number of channels can be split using the previous method, namely $so(N-9)_1=so\left(\frac{N-9}{2}\right)_1\times so\left(\frac{N-9}{2}\right)_1$. In the case when $N$ is smaller than 9, one can add $9-N$ number of {\em non-chiral} Majorana channels to each wire. These additional degrees of freedom can be interpreted as surface reconstruction as they do not violate fermion doubling\cite{NielsenNinomiya83} and are not required to live on the boundary of a topological bulk. Now each wire consists of 9 right (or left) propagating Majorana channels and $9-N$ left (resp.~right) propagating ones. We still refer the remaining even channels by $so(N-9)_1$ except now the negative $N-9$ signals the reverse propagating direction of these Majorana's.

The $so(9)_1$ and $so(N-9)_1$ sectors can then be bipartitioned independently. The fractionalization of a general odd number of Majorana channels is summarized by the sequence \begin{align}so(N)_1\supseteq so(9)_1\times so(N-9)_1\supseteq\mathcal{G}_N^+\times\mathcal{G}_N^-\label{oddNdecomposition}\end{align} for $\mathcal{G}_N^\pm=so(3)_3\times so\left(\frac{N-9}{2}\right)_1$. The ``+" and ``$-$" sectors can now be back-scattered independently using \eqref{bcint} onto adjacent wires in opposite directions. This removes all low energy degrees of freedom and opens up an energy gap.

\subsubsection{The conformal embedding \texorpdfstring{$so(9)_1\supseteq so(3)_3^+\times so(3)_3^-$}{so(9)~so(3)xso(3)}}\label{sec:conformalembedding}
As a matrix Lie algebra, $so(3)$ is generated by the three anti-symmetric matrices $\boldsymbol\Sigma=(\Sigma_{\mathsf{x}},\Sigma_{\mathsf{y}},\Sigma_{\mathsf{z}})$ \begin{align}\Sigma_{\mathsf{x}}=\left(\begin{smallmatrix}0&0&0\\0&0&1\\0&-1&0\end{smallmatrix}\right),\quad\Sigma_{\mathsf{y}}=\left(\begin{smallmatrix}0&0&1\\0&0&0\\-1&0&0\end{smallmatrix}\right),\quad\Sigma_{\mathsf{z}}=\left(\begin{smallmatrix}0&1&0\\-1&0&0\\0&0&0\end{smallmatrix}\right).\nonumber\end{align} They can be embedded into $so(9)$ by tensoring with $\openone_3$, the $3\times3$ identity matrix, on the left or right \begin{align}\boldsymbol\Sigma^+=\boldsymbol\Sigma\otimes\openone_3,\quad\boldsymbol\Sigma^-=\openone_3\otimes\boldsymbol\Sigma.\end{align} We denote $so(3)^\pm=\mathrm{span}\{\Sigma_{\mathsf{x}}^\pm,\Sigma_{\mathsf{y}}^\pm,\Sigma_{\mathsf{z}}^\pm\}$ to be the two mutually commuting subalgebras in $so(9)$.

Recall the free field representation \eqref{so(N)current} of the $so(9)_1$ WZW current generators $J^\beta=i\psi^at^\beta_{ab}\psi^b/2$ for $t^\beta$ an antisymmetric $9\times9$ matrix, the $so(3)_3^\pm$ current generators are given by the substitution of $t^\beta$: \begin{align}{\bf J}_{so(3)_3^\pm}(z)=\frac{i}{2}\psi^a(z)\boldsymbol\Sigma^\pm_{ab}\psi^b(z)\label{so(3)3current}\end{align} for $z=e^{\tau+ix}$ and  ${\bf J}=(J_{\mathsf{x}},J_{\mathsf{y}},J_{\mathsf{z}})$. Written explicitly, \begin{align}J^+_{\mathsf{x}}=i(\psi^{23}+\psi^{56}+\psi^{89}),\quad J^-_{\mathsf{x}}=i(\psi^{47}+\psi^{58}+\psi^{69})\nonumber\\J^+_{\mathsf{y}}=i(\psi^{13}+\psi^{46}+\psi^{79}),\quad J^-_{\mathsf{y}}=i(\psi^{17}+\psi^{28}+\psi^{39})\nonumber\\J^+_{\mathsf{z}}=i(\psi^{12}+\psi^{45}+\psi^{78}),\quad J^-_{\mathsf{z}}=i(\psi^{14}+\psi^{25}+\psi^{36})\nonumber\end{align} for $\psi^{ab}=\psi^a\psi^b$. Using Wick's theorem and the OPE $\psi^a(z)\psi^b(w)=\delta^{ab}/(z-w)+\ldots$, it is straightforward to deduce the $so(3)_3$ WZW current relations \begin{align}J^\pm_{\mathsf{i}}(z)J^\pm_{\mathsf{j}}(w)=\frac{3\delta_{{\mathsf{i}}{\mathsf{j}}}}{(z-w)^2}+\frac{i\varepsilon_{\mathsf{i}\mathsf{j}\mathsf{k}}}{z-w}J^\pm_{\mathsf{k}}(w)+\ldots\label{so(3)3Jrelation}\end{align} and $J^\pm_{\mathsf{i}}(z)J^\mp_{\mathsf{j}}(w)$ is non-singular, for ${\mathsf{i}},{\mathsf{j}}={\mathsf{x}},{\mathsf{y}},{\mathsf{z}}$ and $\varepsilon_{\mathsf{i}\mathsf{j}\mathsf{k}}$ the antisymmetric tensor.

The $so(3)_3$ current relations \eqref{so(3)3Jrelation} differs from the $so(3)_1$ ones \eqref{so(N)1Jrelation} by the coefficient 3 of the most singular term. This sets the level of the affine Lie algebra. The $so(3)_3$ WZW theory is identical to $su(2)_6$ by noticing that the structure factor of $su(2)$ is $f_{\mathsf{i}\mathsf{j}\mathsf{k}}=\sqrt{2}\varepsilon_{\mathsf{i}\mathsf{j}\mathsf{k}}$ (see \eqref{su(2)1Jrelation} and Ref.\onlinecite{bigyellowbook}). The $su(2)$ current generators thus need to be normalized by ${\bf S}_{su(2)_6^\pm}=\sqrt{2}{\bf J}_{so(3)_3^\pm}$ so that \begin{align}S^\pm_{\mathsf{i}}(z)S^\pm_{\mathsf{j}}(w)=\frac{6\delta_{{\mathsf{i}}{\mathsf{j}}}}{(z-w)^2}+\frac{i\sqrt{2}\varepsilon_{\mathsf{i}\mathsf{j}\mathsf{k}}}{z-w}S^\pm_{\mathsf{k}}(w)+\ldots\label{su(2)6Jrelation}\end{align} where the coefficient 6 of the most singular term sets the level of the $su(2)_6$ affine Lie algebra.

The Sugawara energy momentum tensors are the normal ordered product \begin{align}T_{so(3)_3^\pm}(z)=\frac{1}{8}{\bf J}_{so(3)_3^\pm}(z)\cdot{\bf J}_{so(3)_3^\pm}(z).
\end{align} Written explicitly in the fermion representation \eqref{so(3)3current} and using the normal ordered product \begin{align}\psi^a(z)\psi^b(z)\psi^a(z)\psi^b(z)=\psi^a(z)\partial\psi^a(z)+\psi^b(z)\partial\psi^b(z)\end{align} the energy momentum tensor takes the form \begin{align}T_{so(3)_3^\pm}(z)&=-\frac{1}{4}\sum_{a=1}^9\psi^a(z)\partial\psi^a(z)\mp\frac{1}{4}\mathcal{O}_\psi(z)\label{Tso(3)3}\\\mathcal{O}_\psi(z)&=\psi^{1245}+\psi^{1278}+\psi^{4578}+\psi^{1346}+\psi^{1379}\nonumber\\&\quad\quad+\psi^{4679}+\psi^{2356}+\psi^{2389}+\psi^{5689}\label{Opsi}\end{align} for $\psi^{abcd}=\psi^a(z)\psi^b(z)\psi^c(z)\psi^d(z)$. The four-fermion terms in $\mathcal{O}_\psi$ cancel when combining the ``$\pm$" sectors, and therefore the energy momentum tensor \eqref{SugawaraT} completely decomposes \begin{align}T_{so(9)_1}=-\frac{1}{2}\sum_{a=1}^9\psi^a\partial\psi^a=T_{so(3)_3^+}+T_{so(3)_3^-}.\label{Tso(9)=Tso(3)3+Tso(3)3}\end{align} Moreover, as the OPE between ${\bf J}_{so(3)_3^+}$ and ${\bf J}_{so(3)_3^-}$ is non-singular, so is the OPE between $T_{so(3)_3^+}$ and $T_{so(3)_3^-}$. Each sector carries half the total central charge of 9 Majorana channels \begin{align}c_{so(3)_3^\pm}=9/4.\end{align}

The primary fields of $so(3)_3=su(2)_6$ are characterized by half-integral ``angular momenta" $s=0,1/2,\ldots,3$.\cite{bigyellowbook}  Each primary field ${\bf V}_s=(V_s^{-s},V_s^{-s+1},\ldots,V_s^s)$ irreducibly represents the WZW algebra \begin{align}S_{\mathsf{i}}(z)V^m_s(w)=\frac{1}{z-w}\sum_{m'=-s}^s\left(S_{\mathsf{i}}^s\right)^m_{m'}V^{m'}_s(w)+\ldots\end{align} for $\mathsf{i}=\mathsf{x},\mathsf{y},\mathsf{z}$ and $S_{\mathsf{i}}^s$ the $su(2)$ generators in the spin-$s$ matrix representation. We label the seven primary fields by greek letters ${\bf V}_s=1,\alpha_\pm,\gamma_\pm,\beta,f$, each has conformal dimension $h_s=s(s+1)/8$ (see table~\ref{tab:so(3)3primaryfields}). In particular $1={\bf V}_0$ is the vacuum and $f={\bf V}_3$ is Abelian and fermionic with spin $3/2$. 

\begin{table}[htbp]
\centering
\begin{tabular}{l|lllllll}
${\bf V}_s$&1&$\alpha_+$&$\gamma_+$&$\beta$&$\gamma_-$&$\alpha_-$&$f$\\\hline
$s$&0&$1/2$&1&$3/2$&2&$5/2$&3\\
$h_s$&0&$3/32$&$1/4$&$15/32$&$3/4$&$35/32$&$3/2$\\
$d_s$&1&$\sqrt{2+\sqrt{2}}$&$1+\sqrt{2}$&$\sqrt{4+2\sqrt{2}}$&$1+\sqrt{2}$&$\sqrt{2+\sqrt{2}}$&1
\end{tabular}
\caption{The ``angular momenta" $s$, conformal dimensions $h_s$ and quantum dimensions $d_s$ of primary fields ${\bf V}_s$ of $so(3)_3=su(2)_6$.}\label{tab:so(3)3primaryfields}
\end{table}

The rest of the primary fields are non-Abelian. They obey multi-channel fusion rules \begin{align}{\bf V}_{s_1}\times{\bf V}_{s_2}=\sum_sN_{s_1s_2}^s{\bf V}_s\end{align} where the fusion matrix element $N_{s_1s_2}^s=0,1$ is determined by the Verlinde formula\cite{Verlinde88} \begin{align}N_{s_1s_2}^s=\sum_{s'}\frac{\mathcal{S}_{s_1s'}\mathcal{S}_{s_2s'}\mathcal{S}_{ss'}}{\mathcal{S}_{0s'}}\label{Verlindeformula}\end{align} and the modular $S$-matrix\cite{bigyellowbook} \begin{align}\mathcal{S}_{s_1s_2}=\frac{1}{2}\sin\left[\frac{\pi(2s_1+1)(2s_2+1)}{8}\right]\label{SO(3)3Smatrix}\end{align} which is symmetric and orthogonal. Explicitly, the fusion rules are given by \begin{gather}f\times f=1,\quad f\times\gamma_\pm=\gamma_\mp,\quad f\times\alpha_\pm=\alpha_\mp,\quad f\times\beta=\beta\nonumber\\\gamma_\pm\times\gamma_\pm=1+\gamma_++\gamma_-,\quad\alpha_\pm\times\alpha_\pm=1+\gamma_+\nonumber\\\beta\times\beta=1+\gamma_++\gamma_-+f\label{SO(3)3fusion}\\\alpha_\pm\times\gamma_\pm=\alpha_++\beta,\quad\beta\times\gamma_\pm=\alpha_++\alpha_-+\beta\nonumber\\\alpha_\pm\times\beta=\gamma_++\gamma_-\nonumber\end{gather} The quantum dimension $d_s$ of the primary field ${\bf V}_s$ is defined to be the largest eigenvalue of the fusion matrix $N_s=\left(N_{ss_1}^{s_2}\right)$. It coincides with the modular $S$ matrix element $\mathcal{S}_{0s}/\mathcal{S}_{00}$ and respects fusion rules so that \begin{align}d_{s_1}d_{s_2}=\sum_sN_{s_1s_2}^sd_s.\end{align} They are listed in table~\ref{tab:so(3)3primaryfields}.

\subsubsection{\texorpdfstring{$\mathbb{Z}_6$}{Z6} parafermions}\label{sec:Z6parafermions}
We first study the simplest odd case when there are 9 Majorana cones mimicked by the coupled Majorana wire model \eqref{NMajcone} with 9 chiral Majorana channels per wire. Now that we have bipartite the degrees of freedom according to the two $so(3)_3^\pm$ WZW current algebras in \eqref{so(3)3current}, they can be backscattered independently to adjacent wires in opposite directions (see eq.\eqref{bcint} and figure~\ref{fig:couplewires}). As the $so(3)_3^+$ sector completely decomposes from the $so(3)_3^-$ one, the current backscattering ${\bf J}_{so(3)_3^-}^{y-1}\cdot{\bf J}_{so(3)_3^+}^y$ between the $(y-1)^{\mathrm{th}}$ and $y^{\mathrm{th}}$ wire does not compete with the next pair ${\bf J}_{so(3)_3^-}^y\cdot{\bf J}_{so(3)_3^+}^{y+1}$.

The current-current interaction consists of four-fermion terms and is marginally relevant. This can be seen from the RG equation \eqref{RGeqn} using the operator product expansion $({\bf J}^y\cdot{\bf J}^{y+1})^2\sim+{\bf J}^y\cdot{\bf J}^{y+1}$. (Recall the time reversal symmetric convention \eqref{TRsymmcurrentconvention} and that ${\bf J}^y{\bf J}^y\sim i(-1)^y{\bf J}^y$.) To see that the interaction indeed opens up an excitation energy gap, it suffices to focus on a single pair of wires with the Hamiltonian \begin{align}\mathcal{H}_{\mathrm{int}}=u{\bf J}_{so(3)_3^-}^R\cdot{\bf J}_{so(3)_3^+}^L\label{Jso(3)3Hint}\end{align} where $R/L$ labels the counter-propagating directions along wire $y$ and $y+1$.

First we further decompose the $so(3)_3$ WZW theory by the coset construction\cite{bigyellowbook} \begin{align}so(3)_3=u(1)_6\times``\mathbb{Z}_6",\quad``\mathbb{Z}_6"=\frac{so(3)_3}{so(2)_3}=\frac{su(2)_6}{u(1)_6}\end{align} where $``\mathbb{Z}_6"$ refers to the $\mathbb{Z}_6$ parafermion CFT model by Zamolodchikov and Fateev\cite{FateevZamolodchikov82,ZamolodchikovFateev85}. This is done by noticing that $SO(3)$ (or equivalently $SU(2)$) contains the Abelian subgroup $SO(2)$ (resp.~$U(1)$) of rotations about the $\mathsf{z}$-axis, and on the CFT level, the $so(2)_3$ WZW sub-theory of $so(3)_3$ (resp. $u(1)_6\subseteq su(2)_6$) can be bosonized and single-out. To do this we first group three pairs of Majorana fermions into three Dirac fermions on each chiral sector \begin{align}c_R^1&=\frac{\psi^1_R+i\psi^4_R}{\sqrt{2}},& c_R^2&=\frac{\psi^2_R+i\psi^5_R}{\sqrt{2}},& c_R^3&=\frac{\psi^3_R+i\psi^6_R}{\sqrt{2}}\nonumber\\c_L^1&=\frac{\psi^1_L+i\psi^2_L}{\sqrt{2}},& c_L^2&=\frac{\psi^4_R+i\psi^5_L}{\sqrt{2}},& c_L^3&=\frac{\psi^7_L+i\psi^8_L}{\sqrt{2}}\nonumber\end{align} and bosonize \begin{align}c_{R/L}^j\sim\frac{1}{\sqrt{l}_0}\exp\left(i\widetilde{\phi}^j_{R/L}\right)\label{so(3)3bosonization}\end{align} for $j=1,2,3$. 
The $so(2)_3$ subalgebra in the $R$ and $L$ sectors are generated by the $J_{\mathsf{z}}^-$ and $J_{\mathsf{z}}^+$ currents operators in \eqref{so(3)3current} \begin{align}J_{\mathsf{z}}^R=-3i\partial\phi_R^\rho,\quad J_{\mathsf{z}}^L=3i\partial\phi_L^\rho\end{align} where the boson field of the ``charge" sector is the average \begin{align}\phi_{R/L}^\rho=\frac{\widetilde{\phi}_{R/L}^1+\widetilde{\phi}_{R/L}^2+\widetilde{\phi}_{R/L}^3}{3}.\end{align} The ``neutral" sector is carried by the three boson fields \begin{align}\phi_{R/L}^{\sigma,j}=\widetilde{\phi}^j_{R/L}-\phi_{R/L}^\rho\end{align} which are not independent as $\phi^{\sigma,1}+\phi^{\sigma,2}+\phi^{\sigma,3}=0$.

It is straightforward to check that the ``charge" and the ``neutral" sectors completely decouple from each other. For instance, the Lagrangian density decomposes \begin{align}\mathcal{L}_{R/L}&=\frac{(-1)^{R/L}}{2\pi}\sum_{j=1}^3\partial_x\widetilde{\phi}_{R/L}^j\partial_t\widetilde{\phi}_{R/L}^j\label{Lagrangianchargeneutral}\\&=\frac{(-1)^{R/L}}{2\pi}\left[3\partial_x\phi_{R/L}^\rho\partial_t\phi_{R/L}^\rho+\sum_{j=1}^3\partial_x\phi_{R/L}^{\sigma,j}\partial_t\phi_{R/L}^{\sigma,j}\right]\nonumber\end{align} where the remaining fermions $\psi_R^{7,8,9},\psi_L^{3,6,9}$ are suppressed, and $(-1)^R=1$, $(-1)^L=-1$.

The Lagrangian density \eqref{Lagrangianchargeneutral} involves more degrees of freedom in $so(9)_1^{R/L}$ than just $so(3)_3^{R,-}$ or $so(3)_3^{L,+}$. Therefore, a priori, it is not obvious that this $\rho-\sigma$ decomposition is a splitting of $so(3)_3$, and in fact it is not. Only the charge sector $\phi^\rho_{R/L}$ is entirely belonging to $so(3)_3^{R,-}$ or $so(3)_3^{L,+}$. To show this, we go back to the energy-momentum tensor $T_{so(3)_3^\pm}$ in \eqref{Tso(3)3}, say for $R$ movers. \begin{align}T_{so(3)_3^{R,\pm}}(z)&=\frac{1}{2}T_{so(9)_1^R}(z)\mp\frac{1}{4}\mathcal{O}_\psi(z)\end{align} where the total energy-momentum tensor in partially bosonized basis is \begin{align}T_{so(9)_1^R}&=-\frac{1}{2}\Bigg[3\partial\phi_{R}^\rho\partial\phi_{R}^\rho+\sum_{j=1}^3\partial\phi_{R}^{\sigma,j}\partial\phi_{R}^{\sigma,j}\nonumber\\&\quad\quad+\psi_R^7\partial\psi_R^7+\psi_R^8\partial\psi_R^8+\psi_R^9\partial\psi_R^9\Bigg]\label{Tso(3)3bosonized}\end{align} and the operator $\mathcal{O}_\psi$ defined in \eqref{Opsi} is now \begin{align}\mathcal{O}_\psi&=-3\partial\phi^\rho_R\partial\phi^\rho_R+\frac{1}{2}\sum_{j=1}^3\partial\phi_R^{\sigma,j}\partial\phi_R^{\sigma,j}\label{Opsibosonized}\\&-2i\left[\cos\left(\phi^{\sigma,1}_R-\phi^{\sigma,2}_R\right)\psi_R^{78}+\cos\left(\phi^{\sigma,1}_R-\phi^{\sigma,3}_R\right)\psi_R^{97}\right.\nonumber\\&\quad\quad\left.+\cos\left(\phi^{\sigma,2}_R-\phi^{\sigma,3}_R\right)\psi_R^{89}\right].\nonumber\end{align} Eq.\eqref{Opsibosonized} is deduced by substituting the fermions by the boson fields \eqref{so(3)3bosonization}, whose OPE can be found in (\ref{so(3)3bosonOPE1},\ref{so(3)3bosonOPE2},\ref{so(3)3bosonOPE3}) in appendix~\ref{sec:appZ6parafermion}. For instance, the factor of $i$ in \eqref{Opsibosonized} is a result of mutually non-commuting $\phi^{\sigma,j}$. More importantly, $\phi^\rho$, $\phi^\sigma$ and $\psi^{7,8,9}$ are completely decoupled. As the ``charge" sector $\phi^\rho_R$ only appears in $T_{so(3)_3^{R,-}}$, it belongs entirely in $so(3)_3^{R,-}$. Similarly $\phi^\rho_L$ belongs entirely in $so(3)_3^{L,+}$. The ``$\mathbb{Z}_6$" energy-momentum is defined by subtracting the decoupled ``charge" sector from $so(3)_3$. \begin{align}T_{so(2)_3^R}&=\frac{1}{6}J_{\mathsf{z}}J_{\mathsf{z}}=-\frac{1}{2}3\partial\phi_\rho\partial\phi_\rho\label{Tso(2)3}\\T^R_{\mathbb{Z}_6}&=T_{so(3)_3^{R,-}}-T_{so(2)_3^R}\label{TZ6}\\&=-\frac{1}{4}\sum_{a=7}^9\psi^a_R\partial\psi^a_R-\frac{1}{8}\sum_{j=1}^3\partial\phi^{\sigma,j}_R\partial\phi^{\sigma,j}_R\nonumber\\&\quad-\frac{i}{2}\left[\cos\left(\phi^{\sigma,1}_R-\phi^{\sigma,2}_R\right)\psi_R^{78}+\cos\left(\phi^{\sigma,1}_R-\phi^{\sigma,3}_R\right)\psi_R^{97}\right.\nonumber\\&\quad\quad\left.+\cos\left(\phi^{\sigma,2}_R-\phi^{\sigma,3}_R\right)\psi_R^{89}\right]\nonumber\end{align} and similarly for the $L$ movers.

The remaining current operators $J_\pm=(J_{\mathsf{x}}\pm iJ_{\mathsf{y}})/\sqrt{2}$ of $so(3)_3^-$ in the $R$ sector and $so(3)_3^+$ in the $L$ sector (see eq.\eqref{so(3)3current}) now split into ``charge" and ``netrual" parafermion components \begin{align}J^{R/L}_\pm=\mp\sqrt{3}e^{\mp i\phi^\rho_{R/L}}\Psi^\mp_{R/L}\label{Jso(3)3=Vso(2)3Z6}\end{align} where the $\mathbb{Z}_6$ parafermions are given by the combinations \begin{align}\Psi_R&=\frac{1}{\sqrt{3}}\left(e^{i\phi^{\sigma,1}_R}\psi_R^7+e^{i\phi^{\sigma,2}_R}\psi_R^8+e^{i\phi^{\sigma,3}_R}\psi_R^9\right)\label{Z6parafermiondefinition}\\\Psi_L&=\frac{1}{\sqrt{3}}\left(e^{i\phi^{\sigma,1}_L}\psi_L^3+e^{i\phi^{\sigma,2}_L}\psi_L^6+e^{i\phi^{\sigma,3}_L}\psi_L^9\right)\nonumber\end{align} for $\Psi_{R/L}^+=\Psi_{R/L}$ and $\Psi_{R/L}^-=\Psi_{R/L}^\dagger$. Unlike the $\phi^\sigma$'s, here the ``neutral" $\mathbb{Z}_6$ parafermions $\Psi_{R/L}$ belongs entirely in $so(3)_3^{R,-}$ or $so(3)_3^{L,+}$. This is because ${\bf J}^{R/L}$ and $\phi^\rho_{R/L}$ both completely sit inside the $so(3)_3$'s as seen above. Otherwise one can verified this by computing the OPE with the energy-momentum tensor \eqref{Tso(3)3bosonized} explicitly \begin{align}T_{so(3)_3^{R,-}}(z)\Psi_R(w)&=\frac{5/6}{(z-w)^2}\Psi_R(w)+\frac{\partial\Psi_R(w)}{z-w}+\ldots\nonumber\\T_{so(3)_3^{R,-}}(z)e^{\pm i\phi^\rho_R(w)}&=\frac{1/6}{(z-w)^2}e^{\pm i\phi^\rho_R(w)}+\frac{\partial e^{\pm i\phi^\rho_R(w)}}{z-w}+\ldots\end{align} and $T_{so(3)_3^{R,_+}}(z)\Psi_R(w)$ and $T_{so(3)_3^{R,+}}(z)e^{\pm i\phi^\rho_R(w)}$ are both non-singular. Similar OPE hold for the $L$ sector. The primary fields \eqref{Z6parafermiondefinition} generate the rest of the $\mathbb{Z}_6$ parafermions (see \eqref{appZ6parafermions} in appendix~\ref{sec:appZ6parafermion}) and they obey the known $\mathbb{Z}_6$ structure by Zamolodchikov and Fateev\cite{ZamolodchikovFateev85}. 

\subsubsection{Gapping potential}\label{sec:gappingpotential}
Now that we have further decomposed the $so(3)_3^\pm$ currents in each wire into $so(2)_3=U(1)_6$ and $\mathbb{Z}_6$ parafermion components (see eq.\eqref{Jso(3)3=Vso(2)3Z6}), the current-current backscattering interaction \eqref{Jso(3)3Hint} between a pair of wires takes the form of \begin{align}\mathcal{H}_{\mathrm{int}}&=9u\partial_x\phi^\rho_R\partial_x\phi^\rho_L+3u\left[e^{i(\phi^\rho_L-\phi^\rho_R)}\Psi_R^\dagger\Psi_L+h.c.\right].\label{parafermionHint}\end{align} The first term only renormalizes the velocity of the boson in the $so(2)_3$ sector. The second term is responsible for openning an excitation energy gap. It extracts a $\mathbb{Z}_6$ parafermion $\Psi$ and a quasiparticle $e^{i\phi^\rho}$ from the $so(3)_3^+$ sector on the $y^{\mathrm{th}}$ wire and backscatter them onto the $so(3)_3^-$ sector along the $(y+1)^{th}$ wire. This freezes all low energy degrees of freedom and the ground state is characterized by the $\mathbb{Z}_6$ expectation value (GEV) \begin{align}\left\langle\Psi_R^\dagger(x)\Psi_L(x)\right\rangle\sim-e^{i\left\langle\phi^\rho_R(x)-\phi^\rho_L(x)\right\rangle}=e^{2\pi im/6}\label{Z6GEV}\end{align} for $m$ an integer.

Like the $O(N)$ Gross-Neveu model we discussed in section~\ref{sec:GNmodel}, quasiparticle excitations here also manifest as kinks or domain walls between segments with different GEV's. The primary fields $\alpha_\pm,\gamma_\pm,\beta$ of the chiral $so(3)_3$ WZW theory in table~\ref{tab:so(3)3primaryfields} decompose into components in the ``$\mathbb{Z}_6$" and $so(2)_3$ sectors. \begin{gather}\alpha_+=[\sigma_1]\times[e^{i\phi^\rho/2}],\quad\alpha_-=[\sigma_5]\times[e^{-i\phi^\rho/2}]\nonumber\\\gamma_+=[\sigma_2]\times[e^{i\phi^\rho}],\quad\gamma_-=[\sigma_4]\times[e^{-i\phi^\rho}]\nonumber\\\beta=[\sigma_3]\times[e^{i3\phi^\rho/2}]\end{gather} where $\sigma_l$ are primary fields in the chiral $\mathbb{Z}_6$ parafermion theory so that $\sigma_l^R\sigma_l^L$ take the roles of the order parameters of the $\mathbb{Z}_6$ model\cite{FateevZamolodchikov82,ZamolodchikovFateev85}. They satisfy the exchange relations \begin{align}
\Psi(x)\sigma_l(x')&=\sigma_l(x')\Psi(x)e^{-2\pi i\frac{l}{6}\theta(x-x')}\label{Z6exchangerelation}\end{align} for $R$ sector, and similar relations hold for the $L$ sector with the $\mathbb{Z}_6$ phases conjugated. Therefore adding the operators $\alpha_\pm(x),\gamma_\pm(x),\beta(x)$ to the ground state create kinks of different hights in the GEV \eqref{Z6GEV} \begin{gather}\left\langle\alpha_\pm^\dagger(x_0)\Psi^\dagger_R(x)\Psi_L(x)\alpha_\pm(x_0)\right\rangle\sim e^{\frac{\pi i}{3}(m\pm\theta(x-x_0))}\nonumber\\\left\langle\gamma_\pm^\dagger(x_0)\Psi^\dagger_R(x)\Psi_L(x)\gamma_\pm(x_0)\right\rangle\sim e^{\frac{\pi i}{3}(m\pm2\theta(x-x_0))}\nonumber\\\left\langle\beta^\dagger(x_0)\Psi^\dagger_R(x)\Psi_L(x)\beta(x_0)\right\rangle\sim e^{\frac{\pi i}{3}(m+3\theta(x-x_0))}\end{gather} where $\theta(s)=(s/|s|+1)/2$ is the unit step function.

The fermionic supersector $f$ in $so(3)_3$ (see table~\ref{tab:so(3)3primaryfields}) consists of operators that admit free field representations. Again we focus on the the $so(3)_3^{R,-}$ sector. The operators \begin{gather}V_f^0=\Psi^3,\quad V_f^{\pm1}=e^{\mp i\phi^\rho}\Psi^{\mp2}\nonumber\\V_f^{\pm2}=e^{\mp2i\phi^\rho}\Psi^\mp,\quad V_f^{\pm3}=e^{\mp3i\phi^\rho}\nonumber\end{gather} span a $s=3$ representation of the affine $so(3)_3$ Lie algebra, where $\Psi^{-m}=\Psi^{6-m}$ are the $\mathbb{Z}_6$ parafermions satisfying the OPE $\Psi^m(z)\Psi^{m'}(w)\sim(z-w)^{-mm'/3}\Psi^{m+m'}$ (see appendix~\ref{sec:appZ6parafermion} for explicit definitions). From \eqref{Z6exchangerelation}, they create a kink to the order parameter $\langle b\rangle=\langle\beta_R(x)\beta_L(x)\rangle$ \begin{align}\left\langle{\bf V}_f^R(x_0)^\dagger\beta_R(x)\beta_L(x){\bf V}_f^R(x_0)\right\rangle=\langle b\rangle(-1)^{\theta(x-x_0)}\end{align} in the order phase.

The gapping potential can now be generalized to an arbitrary odd number of Majorana channels per wire. Using the decomposition \eqref{oddNdecomposition}, the $N$ Majorana channels are first split into $9+(N-9)$. The first 9 channels are fractionalized into $so(3)_3^+\times so(3)_3^-$ while the remaining $N-9$ can be split into $so(\frac{N-9}{2})_1^+\times so(\frac{N-9}{2})_1^-$ because $N-9$ is even. The interwire current backscattering \eqref{bcint} takes the form \begin{align}\mathcal{H}_{\mathrm{int}}&=u\sum_{y=-\infty}^\infty{\bf J}^y_{so(3)_3^-}\cdot{\bf J}^{y+1}_{so(3)_3^+}+{\bf J}^y_{so\left(\frac{N-9}{2}\right)_1^-}\cdot{\bf J}^{y+1}_{so\left(\frac{N-9}{2}\right)_1^+}\end{align} where different terms act on completely decoupled degrees of freedom. They also gap out {\em all} low energy degrees freedom as the energy-momentum tensor of the CFT along each wire decomposes \begin{align}T_{so(N)_1}&=T_{so(9)_1}+T_{so(N-9)_1}\\&=T_{so(3)_3^+}+T_{so(3)_3^-}+T_{so\left(\frac{N-9}{2}\right)_1^+}+T_{so\left(\frac{N-9}{2}\right)_1^-}\nonumber\end{align} using \eqref{Tso(9)=Tso(3)3+Tso(3)3} and the fact that \begin{align}T_{so(m+n)_1}=-\frac{1}{2}\sum_{a=1}^{m+n}\psi^a\partial\psi^a=T_{so(m)_1}+T_{so(n)_1}.\end{align}

\subsection{Gapping by fractional quantum Hall stripes}

\begin{figure}[htbp]
(a)\centering\includegraphics[width=0.38\textwidth]{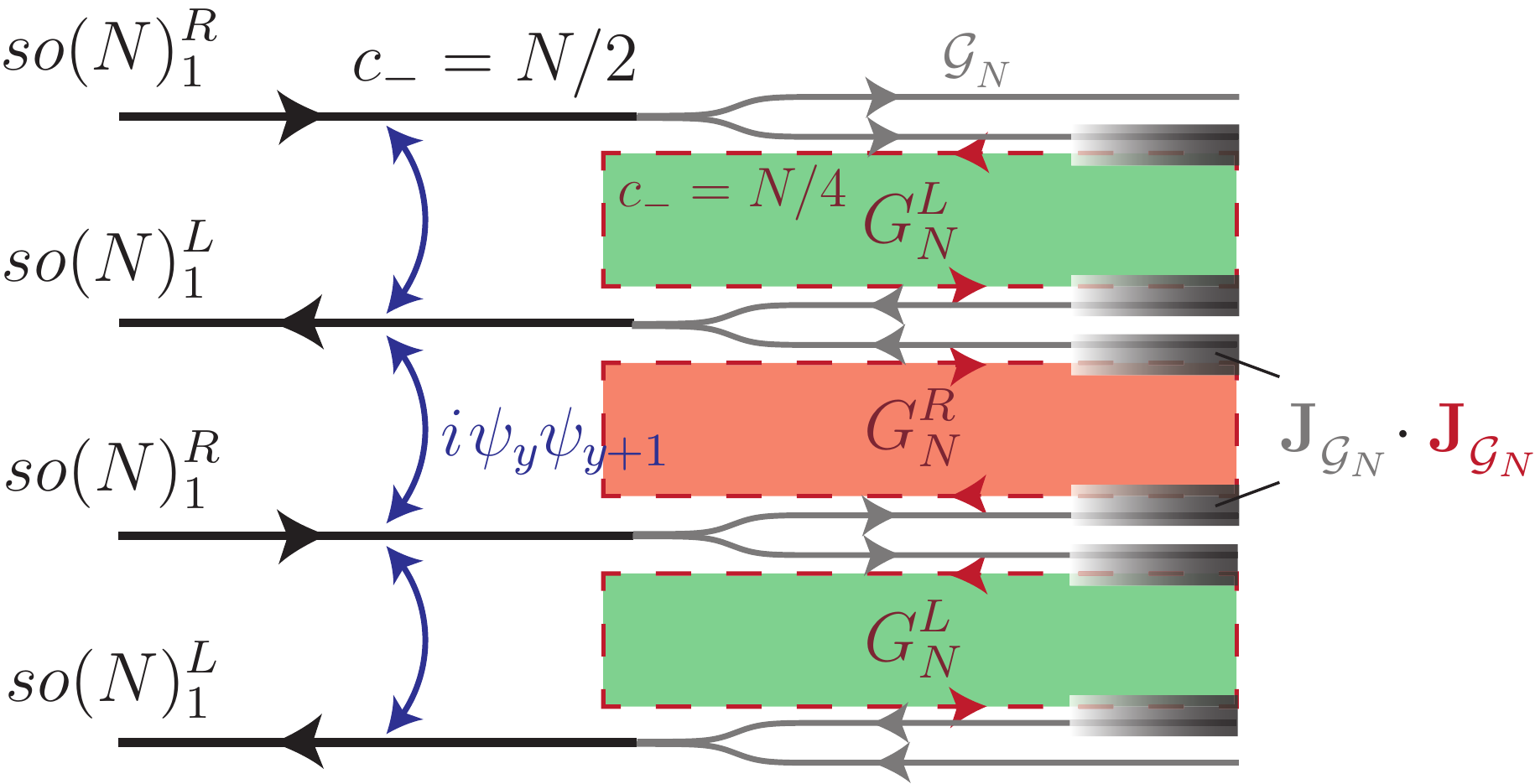}
(b)\centering\includegraphics[width=0.4\textwidth]{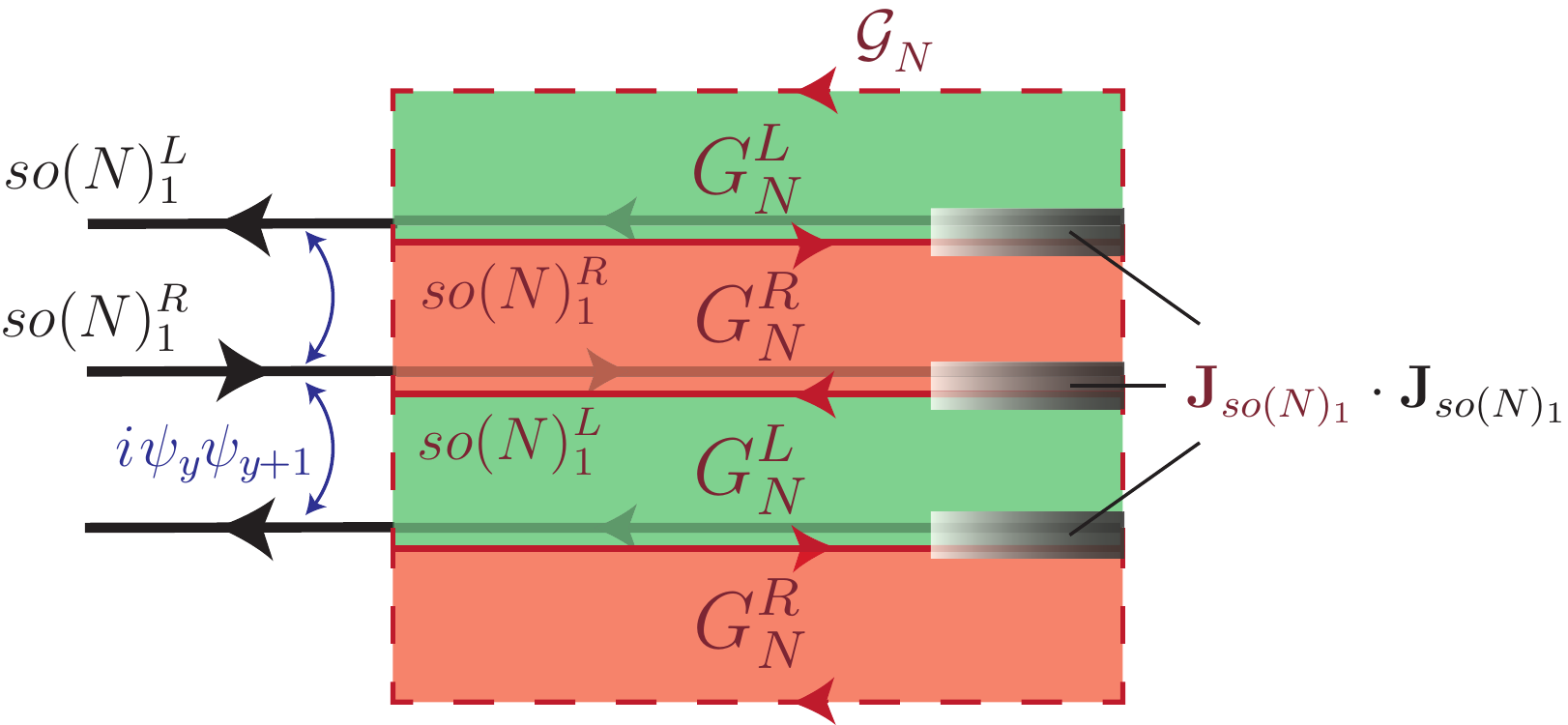}
\caption{Gapping $N$ surface Majorana cones by inserting $(2+1)$D $G_N$ stripe state and removing edge modes by current-current backscattering interaction.}\label{fig:couplewires2}
\end{figure}

Previously, we designed interwire interactions that gap all Majorana modes without breaking time reversal symmetry. Here we provide an alternative where each chiral Majorana wire is gapped by backscattering onto the edges of two topological stripes sandwiching the wire (see figure~\ref{fig:couplewires2}). The topological stripes could be fractional quantum Hall states for instance. Similar construction has been proposed to describe surface states of topological insulators\cite{MrossEssinAlicea15}.

First we consider inserting between each pairs of Majorana wire a $(2+1)$D topological state. It supports chiral boundary modes which move in a reverse direction to its neighboring Majorana wire. As adjacent wires have opposite propagation directions, the chiralities of the topological states also alternates. This alternating topological stripe state can be regarded as a surface reconstruction of the 3D topological superconductor. It preserves the antiferrormagnetic time reversal symmetry \eqref{TR}, which relates adjacent topological stripes by reversing their chirality. Unlike the coupled Majorana wire mode, the topological stripe state itself is a pure $(2+1)$D time reversal symmetric system and is not supported by a $(3+1)$D bulk. It has a gapless energy spectrum that is identical to $N$ surface Majorana cones and is carried by the interface modes between stripes (see figure~\ref{fig:couplewires2}(b)). However the topological stripe state also carry non-trivial anyonic excitations between wires. This distinguishes it from the coupled Majorana wire model and allows it to exist non-holographically in a pure $(2+1)$D setting.

The Majorana modes along the chiral wires then can be backscattered onto the boundaries or interfaces of the topological stripes by current-current couplings. In order for the boundary or interface modes to exactly cancel the Majorana modes along each wire, the topological stripes must carry specific topological orders. We take a $G_N$ topological state (see eq.\eqref{GNdefintro}) 
so that its boundary carries a $\mathcal{G}_N$ Kac-Moody current, for $\mathcal{G}_N$ the affine Lie algebra of $G_N$ defined in \eqref{GNfractionalization}. $G_N^R$ and $G_N^L$ denote stripes with opposite chiralities. The $(2+1)$D $G_N$ topological state itself can be constructed using a coupled wire construction similar to that in Ref.\onlinecite{TeoKaneCouplewires,KhanTeoHughesfuture2} and will not be discussed here. 

There are two ways the Majorana modes can be backscattered onto the topological stripes. The first is shown in figure~\ref{fig:couplewires2}(a). The $N$ Majorana modes along each chiral wire is bipartite into a pair of WZW theories $\mathcal{G}_N^+\times\mathcal{G}_N^-$ according to \eqref{fractioneq}. Each WZW theory is identical to the CFT along the boundary of an neighboring topological stripe but propagates in an opposite direction. It can be then be gapped out by the current-current backscattering \begin{align}\mathcal{H}_{\mathrm{int}}=u{\bf J}_{\mathcal{G}_N}^{\mathrm{wire}}\cdot{\bf J}_{\mathcal{G}_N}^{\mathrm{stripe}}.\end{align}

Alternatively, one could first glue the topological stripes together (see figure~\ref{fig:couplewires2}(b)) so that the line interface sandwiched between adjacent $G_N^R$ and $G_N^L$ states hosts a chiral $so(N)_1$ CFT. The stripes can then be put on top of the Majorana wire array so that each interface is sitting on top of a wire with opposite chirality. The current-current backscattering \begin{align}\mathcal{H}_{\mathrm{int}}=u{\bf J}_{so(N)_1}^{\mathrm{wire}}\cdot{\bf J}_{so(N)_1}^{\mathrm{interface}}\end{align} between each Majorana wire and stripe interface gaps out all low energy degrees of freedom.


\section{Surface topological order}\label{sec:topologicalorder}
In the previous section, we described how a coupled Majorana wire model, which mimics the surface Majorana modes of a 3D bulk topological superconductor (TSC), can be gapped by interwire current-current backscattering interaction without breaking time reversal (TR) symmetry. In this section, we pay more attention to the topological order and the anyon types\cite{Wilczekbook,Fradkinbook,Wenbook} of gapped excitations. The ground states are time reversal symmetric and there are no non-vanishing order parameters that breaks time reversal spontaneously. There is a finite ground state degeneracy that does not depend on system size. This signifies a non-trivial topological order\cite{Wentopologicalorder89,Wentopologicalorder90,WenNiu90}. 

\begin{figure}[htbp]
\centering
\includegraphics[width=0.45\textwidth]{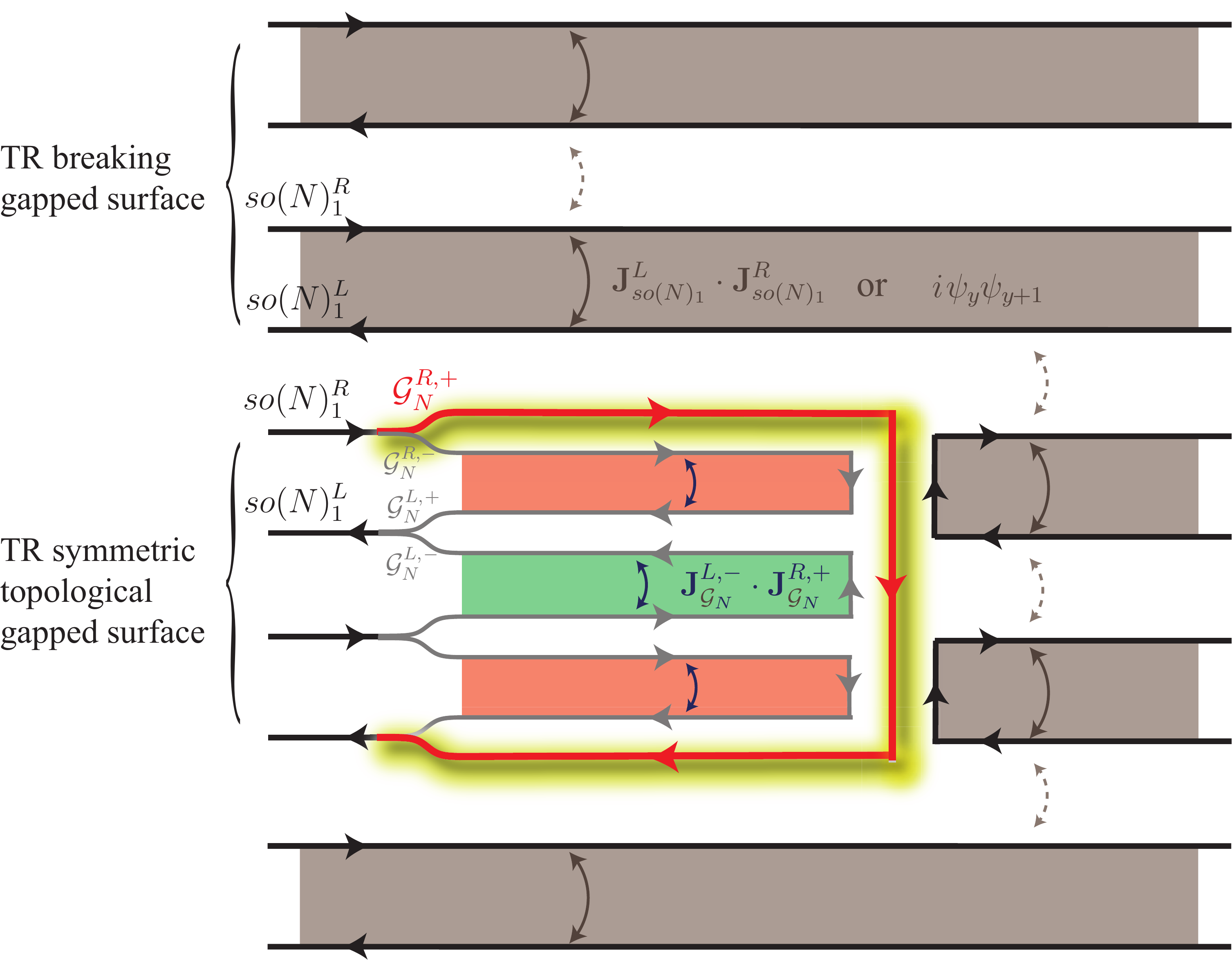}
\caption{Chiral interface (highlighted line) between a time reversal breaking gapped region and a TR symmetric topologically ordered gapped region.}\label{fig:edgemode}
\end{figure}

\begin{figure}[htbp]
\centering
\includegraphics[width=0.3\textwidth]{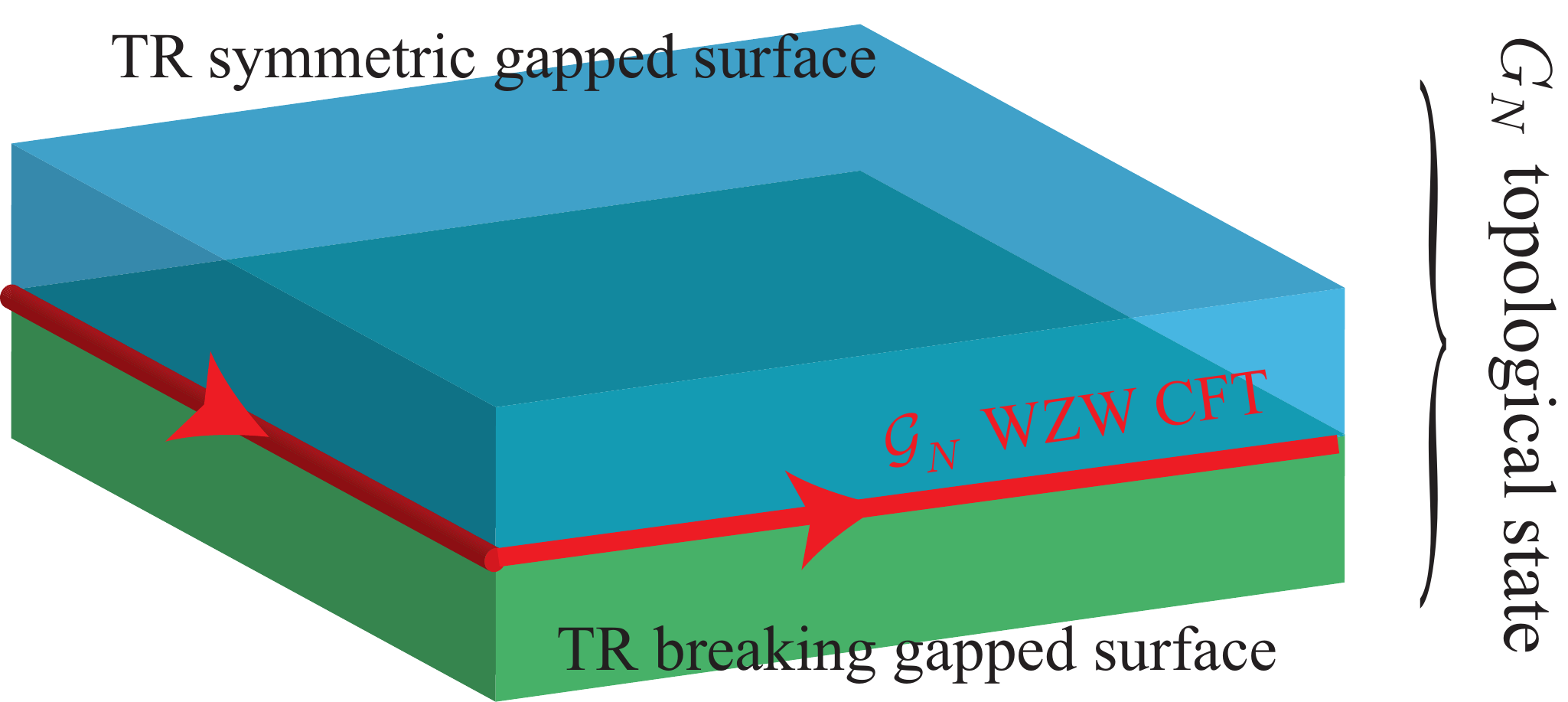}
\caption{The $G_N$ topological order of a quasi-2D slab with time reversal symmetric gapped top surface and time reversal breaking gapped bottom surface}\label{fig:slab}
\end{figure}

The surface topological order can be inferred from bulk-boundary correspondence\cite{Wenedgereview,MooreRead,ReadMoore,Kitaev06}. There is a one-to-one correspondence between the primary fields of the CFT along the $(1+1)$D gapless boundary and the anyon types in the $(2+1)$D gapped topological bulk. The conformal scaling dimension or spin $h=h_R-h_L$ of a primary field corresponds to the exchange statistical phase $\theta=e^{2\pi ih}$ of the corresponding anyon. The fusion rules of primary fields are identical to that of the anyons. And the modular $S$-matrix of the CFT at the boundary equals the braiding $S$-matrix\cite{Kitaev06} \begin{align}\mathcal{S}_{{\bf a}{\bf b}}=\frac{1}{\mathcal{D}}\sum_{\bf c}d_{\bf c}N_{{\bf a}{\bf b}}^{\bf c}\frac{\theta_{\bf c}}{\theta_{\bf a}\theta_{\bf b}}\label{braidingS}\end{align} in the bulk, where the non-negative integers $N_{{\bf a}{\bf b}}^{\bf c}$ are the degeneracies of the fusion rules \begin{align}{\bf a}\times{\bf b}=\sum_{\bf c}N_{{\bf a}{\bf b}}^{\bf c}{\bf c}\end{align} between anyons, and the total quantum dimension $\mathcal{D}=\sqrt{\sum_{\bf a}d^2_{\bf a}}$ quantifies topological entanglement\cite{KitaevPreskill06} and can be evaluated by knowing the quantum dimensions $d_{\bf a}\geq1$ of each anyon ${\bf a}$ by solving the fusion identities \begin{align}d_{\bf a}d_{\bf b}=\sum_{\bf c}N_{{\bf a}{\bf b}}^{\bf c}d_{\bf c}.\end{align}

On the surface of a topological superconductor, where there are no boundaries, the $(2+1)$D topological order corresponds to a $(1+1)$D interface that separate the time reversal symmetric topologically ordered domain and a time reversal breaking domain. This interface hosts a chiral gapless modes (see figure~\ref{fig:edgemode}). This geometry can be wrapped onto the surface of a slab where the TR symmetric and breaking domains occupy the top and bottom surface of a 3D bulk (see figure~\ref{fig:slab}). The quasi-2D system has an energy gap except along its boundary which is previously the interface that carries the $\mathcal{G}_N$ WZW CFT. The bulk-boundary correspondence then determines a bulk $G_N$ topological order on the quasi-2D slab.

Wires in the trivial TR-breaking domain are gapped by non-uniform current backscattering \begin{align}\mathcal{H}_{\mathrm{TR-breaking}}=&\sum_y\Delta{\bf J}_{so(N)_1}^{2y-1}\cdot{\bf J}_{so(N)_1}^{2y}\nonumber\\&\;\;\;+\delta{\bf J}_{so(N)_1}^{2y}\cdot{\bf J}_{so(N)_1}^{2y+1}\label{HTRbreaking}\end{align} or single-body fermion backscattering perturbation \begin{align}\mathcal{H}_{\mathrm{TR-breaking}}=\sum_yi\Delta\boldsymbol{\psi}^T_{2y-1}\boldsymbol{\psi}_{2y}+i\delta\boldsymbol{\psi}^T_{2y}\boldsymbol{\psi}_{2y+1}\label{HTRbreaking2}\end{align} to the coupled Majorana wire model \eqref{NMajcone}, for $\Delta>\delta$ and $\boldsymbol\psi_y=(\psi_y^1,\ldots,\psi_y^N)$. This violates the antiferrormagnetic time reversal symmetry \eqref{TR} and leads to a gapped surface with trivial topological order.
This TR breaking half-plane is put side by side against a TR symmetric gapped half-plane, where the $N$ Majorana channels per wire is {\em fractionalized} into $so(N)_1\supseteq\mathcal{G}_N^+\times\mathcal{G}_N^-$, for $\mathcal{G}_N$ previously defined in \eqref{GNfractionalization}. Each $\mathcal{G}_N$ sector is then paired with the adjacent one on the next wire and are gapped by current-current backscattering ${\bf J}_{\mathcal{G}_N^-}\cdot{\bf J}_{\mathcal{G}_N^+}$. The interface between the TR-symmetric and TR-breaking regions leaves behind one single unpaired fractional $\mathcal{G}_N$ channel. This can be regarded as a 2D analogue of the fractional boundary modes in the the Haldane integral spin chain\cite{Haldanespinchain1,Haldanespinchain2} and the AKLT spin chain\cite{AKLT}. 


As eluded in the introduction, when the coupled wire model involves only current-current backscattering interaction, it is a boson model where the bosonic current operators, rather than Majorana fermions, are treated as fundamental local objects. It is therefore more natural for us to use the current backscattering Hamiltonian \eqref{HTRbreaking} instead of the fermionic single-body one \eqref{HTRbreaking2} to introduce a time reversal breaking gap. In this case, $\pi$-fluxes are deconfined anyonic excitations realized as $\pi$-kinks along a stripe where there is no energy cost in separating a flux-antiflux pair. If the fermionic TR-breaking Hamiltonian \eqref{HTRbreaking2} were used instead, $\pi$-fluxes would be confined on the bottom layer and Majorana fermions would become local. We however will mostly be focusing on the former bosonic case, although the fermionic scenario may be more realistic in a superconducting medium.

The bulk-interface correspondence depends on the orientation of the time reversal breaking order. In eq.\eqref{HTRbreaking}, if the backscattering tunneling strengths are reversed so that $\delta>\Delta$, figure~\ref{fig:edgemode} will need to be shifted by $y\to y+1$  and all propagating directions will need to be inverted. As a result, the interface CFT will also be reversed to its time reversal partner $\mathcal{G}_N\to\overline{\mathcal{G}_N}$. This will flip the spins of all primary fields $h_{\bf a}\to h_{\overline{\bf a}}=-h_{\bf a}$ and conjugates all exchange phases $\theta_{\bf a}\to\theta_{\overline{\bf a}}=\theta^\ast_{\bf a}$.

An interface with a particular orientation therefore corresponds to a time reversal breaking topological order. This is also apparent in the slab geometry in figure~\ref{fig:slab} where the TR breaking order on the bottom surface can have opposite orientations. Unlike the conventional case on the surface of a topological superconductor where time reversal is local, here time reversal involves a half translation $y\to y+1$ and relates a stripe gapped by ${\bf J}^-_y\cdot{\bf J}^+_{y+1}$ to its neighbor ${\bf J}^-_{y+1}\cdot{\bf J}^+_{y+2}$. As anyonic excitations are realized as kinks or domain walls that separate distinct ground states along a stripe, time reversal non-locally translates anyons on an even stripe (green) to an odd one (red) or vice versa (see figure~\ref{fig:edgemode}). However an interface with a particular orientation can only correspond to anyons on stripes with a particular parity. For example the bulk-interface correspondence in figure~\ref{fig:edgemode} singles out anyons on even stripes gapped by ${\bf J}^-_{2y}\cdot{\bf J}^+_{2y+1}$. There is therefore no reason to expect the anyon theory would be closed under time reversal.

\subsection{Summary of anyon contents}
\begin{table}[htbp]
\begin{tabular}{l|llll|lll}
&\multicolumn{4}{c|}{$r$ even}&\multicolumn{3}{c}{$r$ odd}\\
${\bf x}$&1&$\psi$&$s_+$&$s_-$&1&$\psi$&$\sigma$\\\hline
$d_{\bf x}$&1&1&1&1&1&1&$\sqrt{2}$\\
$\theta_{\bf x}$&1&$-1$&$e^{\pi ir/8}$&$e^{\pi ir/8}$&1&$-1$&$e^{\pi ir/8}$\\
\end{tabular}
\caption{The exchange phase $\theta_{\bf x}=e^{2\pi ih_{\bf x}}$ and quantum dimensions of anyons {\bf x} in a $(2+1)$D $SO(r)_1$ topological phase.}\label{tab:toporderSO(r)}
\end{table}
The interface carries chiral gapless degrees of freedom, which are captured by the $\mathcal{G}_N$ WZW theory whose primary fields corresponds to the anyon content of the TR symmetry gapped surface. For even $N=2r$, the surface carries a \begin{align}G_N=SO(r)_1\label{GN=sor1}\end{align} topological order summarized in table~\ref{tab:toporderSO(r)}. Its anyonic excitations obey the abelian fusion rules  \begin{gather}\psi\times\psi=1,\quad s_\pm\times\psi=s_\mp\label{so(2n)fusion}\\s_\pm\times s_\pm=\left\{\begin{array}{*{20}c}1,&\mbox{for $r\equiv0$ mod 4}\\\psi,&\mbox{for $r\equiv2$ mod 4}\end{array}\right.\nonumber\end{gather} for $r$ even, or the Ising fusion rules \begin{gather}\psi\times\psi=1,\quad\psi\times\sigma=\sigma,\quad\sigma\times\sigma=1+\psi\label{Isingfusion}\end{gather} for $r$ odd. Eq.\eqref{so(2n)fusion} and \eqref{Isingfusion} follows directly from the fusion properties of the primary fields in the $so(r)_1$ Kac-Moody algebra (see section~\ref{sec:SO(N)} and appendix~\ref{sec:kleinfactors} and \ref{sec:kleinfactors2}). The exchange phase (also known as topological spin) $\theta_{\bf x}=e^{2\pi ih_{\bf x}}$ can be read off from the conformal dimension $h_{\bf x}$ of the primary field $V_{\bf x}$ in $so(r)_1$ that corresponds to the anyon type ${\bf x}$. Again we extend $r$ to negative integers by defining $SO(-r)_1=\overline{SO(r)_1}$ to be the time reversal conjugate of the $SO(r)_1$ topological state.

\begin{table}[htbp]
\centering
\begin{tabular}{l|lllllll}
${\bf x}$&1&$\alpha_+$&$\gamma_+$&$\beta$&$\gamma_-$&$\alpha_-$&$f$\\\hline
$d_{\bf x}$&1&$\sqrt{2+\sqrt{2}}$&$1+\sqrt{2}$&$\sqrt{4+2\sqrt{2}}$&$1+\sqrt{2}$&$\sqrt{2+\sqrt{2}}$&1\\
$\theta_{\bf x}$&1&$e^{\pi i\frac{3+2r}{16}}$&$e^{i\pi/2}$&$e^{\pi i\frac{15+2r}{16}}$&$e^{-i\pi /2}$&$e^{\pi i\frac{3+2r}{16}}$&$-1$\\
\multicolumn{8}{c}{$r$ even}
\end{tabular}
\begin{tabular}{l|lllllll}
${\bf x}$&1&$\alpha_+$&$\gamma_+$&$\beta$&$\gamma_-$&$\alpha_-$&$f$\\\hline
$d_{\bf x}$&1&$\sqrt{2+\sqrt{2}}$&$1+\sqrt{2}$&$\sqrt{4+2\sqrt{2}}$&$1+\sqrt{2}$&$\sqrt{2+\sqrt{2}}$&1\\
$\theta_{\bf x}$&1&$e^{\pi i\frac{15+2r}{16}}$&$e^{i\pi/2}$&$e^{\pi i\frac{3+2r}{16}}$&$e^{-i\pi /2}$&$e^{\pi i\frac{15+2r}{16}}$&$-1$\\
\multicolumn{8}{c}{$r$ odd}
\end{tabular}
\caption{The exchange phase $\theta_{\bf x}=e^{2\pi ih_{\bf x}}$ and quantum dimensions of anyons {\bf x} in a $(2+1)$D $SO(3)_3\boxtimes_bSO(r)_1$ topological phase.}\label{tab:toporderSO(3)3SO(r)1}
\end{table}
For odd $N=9+2r$, the $\mathcal{G}_N$ WZW CFT at the interface corresponds the TR symmetric gapped surface that carries a topological order given by the {\em relative} tensor product \begin{align}G_N=SO(3)_3\boxtimes_bSO(r)_1\label{GN=so3xsor}\end{align} where the fermion pair $b=\psi_{SO(3)_3}\times\psi_{SO(r)_1}$ is condensed. The concept of anyon condensation\cite{BaisSlingerlandCondensation} will be demonstrated more explicitly later in section~\ref{32class}. The topological state carries seven anyon types and are summarized in table~\ref{tab:toporderSO(3)3SO(r)1}. For instance, the anyon structure matches the primary field content of the $so(3)_3$ WZW theory (see table~\ref{tab:so(3)3primaryfields}) when $r=0$. The quasiparticle fusion rules of $G_N$ are similar to the $so(3)_3$ ones in \eqref{SO(3)3fusion} \begin{gather}f\times f=1,\quad f\times\gamma_\pm=\gamma_\mp,\quad f\times\alpha_\pm=\alpha_\mp,\quad f\times\beta=\beta\nonumber\\\gamma_\pm\times\gamma_\pm=1+\gamma_++\gamma_-,\quad\alpha_\pm\times\beta=\gamma_++\gamma_-\label{GNfusion1}\\\beta\times\beta=1+\gamma_++\gamma_-+f,\quad\beta\times\gamma_\pm=\alpha_++\alpha_-+\beta\nonumber\end{gather} except the following modifications that dependent on $r=(N-9)/2$. \begin{gather}\alpha_\pm\times\alpha_\pm=\left\{\begin{array}{*{20}c}1+\gamma_+,&\mbox{for $r\equiv0$ mod 4}\\f+\gamma_+,&\mbox{for $r\equiv1$ mod 4}\\f+\gamma_-,&\mbox{for $r\equiv2$ mod 4}\\1+\gamma_-,&\mbox{for $r\equiv3$ mod 4}\end{array}\right.\label{GNfusion2}\\
\alpha_\pm\times\gamma_\pm=\left\{\begin{array}{*{20}c}\alpha_++\beta,&\mbox{for $r$ even}\\\alpha_-+\beta,&\mbox{for $r$ odd}\end{array}\right.\nonumber\end{gather} This quasiparticle spin and fusion structure will be shown later in section~\ref{32class}. The braiding $S$-matrices of the $G_N$ states are summarized in appendix~\ref{sec:GNapp}.

The $G_N$ sequence extends the sixteenfold periodic anyon structure\cite{Kitaev06,khan2014,TeoHughesFradkin15} $SO(r+16)_1\cong SO(r)_1$ to a periodic class of thirty two topological states \begin{align}G_{N+32}\cong G_N.\end{align} This seemingly contradicts the sixteenfold prediction of topologically ordered surface states from Ref.\onlinecite{LukaszChenVishwanath,MetlitskiFidkowskiChenVishwanath14,WangSenthil14,Senthil2014,Kapustin2014e,Qi_AxionTFT,Witten15}. This is due to the non-local nature of the ``antiferromagnetic" time reversal symmetry in the coupled Majorana wire model. On the other hand, in general there are multiple possible gapping potentials that leads to distinct topological order. For instance, we will show in a subsequent section that for $N=16$, there is an extended $E_8$ symmetry or an alternative conformal embedding that would allow a different set of gapping terms but would forbid all electronic quasiparticle excitations. 

The thirty two topological states here follow a $\mathbb{Z}_{32}$ tensor product algebraic structure \begin{align}G_{N_1}\boxtimes_bG_{N_2}\cong G_{N_1+N_2}\label{Z32tensorstructure}\end{align} where certain maximal set of mutually local bosons from $G_{N_1}$ and $G_{N_2}$ are pair condensed in the relative tensor product. We will discuss this in more detail below.

\subsection{The 32-fold tensor product structure}\label{32class}
We first explain the relative tensor product that defines the $G_N$ topological state in eq.\eqref{GN=so3xsor}. We begin with the tensor product state $SO(3)_3\otimes SO(r)_1$ which consists of decoupled $SO(3)_3=SU(2)_6$ and $SO(r)_1$ topological states. The primary fields of the $su(2)_6$ WZW CFT are labeled by seven half-integral ``spins" $s={\bf 0},{\bf 1/2},{\bf 1},{\bf 3/2},{\bf 2},{\bf 5/2},{\bf 3}$ and are summarized in table~\ref{tab:so(3)3primaryfields} and eq.\eqref{SO(3)3fusion}. These correspond to the anyon structure of the $(2+1)$D $SO(3)_3$ topological state. The topological order of $SO(r)_1$ is well-known\cite{Kitaev06} and was summarized earlier in this section. For instance, ``spin" ${\bf 3}$ corresponds to the BdG fermion quasiparticle $f$, and the half-integral ``spins" ${\bf 1/2}$, ${\bf 3/2}$ and ${\bf 5/2}$ are $\pi$-fluxes that give a $-1$ monodromy phase of an orbiting fermion.

In the coupled Majorana wire model where there are $N=9+2r$ Majorana channels per wire, the gapping term explicitly seperates the first 9 and final $2r$ channels and the current backscattering potential does not mix these two sectors. This model would therefore give a decouple $SO(3)_3\otimes SO(r)_1$ topological state. However, there could be additional local time reversal symmetric terms, such as intrawire forward scattering $i\psi^R_a\psi^R_b$ and $i\psi^L_a\psi^L_b$, that mixes the two sectors and condenses the fermion pair $b=f_{SO(3)_3}\otimes\psi_{SO(r)_1}$. In fact, fermion pair condensation is natural in a superconducting medium where the ground state consists of Cooper pairs. The condensation of the bosonic anyon $b$ results in the confinement of certain quasiparticles that have non-trivially monodromy around it.\cite{BaisSlingerlandCondensation} These includes all the $\pi$ fluxes ${\bf 1/2}$, ${\bf 3/2}$ and ${\bf 5/2}$ in the $SO(3)_3$ sector, $s_\pm$ (or $\sigma$) in $SO(r)_1$ for $r$ even (resp.~odd), as well as the tensor product ${\bf 1/2}\otimes\psi$, ${\bf 3/2}\otimes\psi$, ${\bf 5/2}\otimes\psi$, ${\bf 1}\otimes s_\pm$, ${\bf 2}\otimes s_\pm$ and ${\bf 3}\otimes s_\pm$ (or ${\bf 1}\otimes\sigma$, ${\bf 2}\otimes\sigma$ and ${\bf 3}\otimes\sigma$). The remaining anyons are local with respect to the boson $b$ and survive the condensation, but certain pairs are identified if they differ only by the boson condensate, ${\bf a}\times b\equiv{\bf a}$.  This includes ${\bf 3}\equiv\psi$, ${\bf 1}\otimes\psi\equiv{\bf 2}$, ${\bf 2}\otimes\psi\equiv{\bf 1}$, ${\bf 1/2}\otimes s_\pm\equiv{\bf 5/2}\otimes s_\mp$ and ${\bf 3/2}\otimes s_+\equiv{\bf 3/2}\otimes s_-$ for even $r$, or ${\bf 1/2}\otimes\sigma\equiv{\bf 5/2}\otimes\sigma$ for $r$ odd. Special care has to be taken for the tensor product ${\bf 3/2}\otimes\sigma$ when $r$ is odd. After condensation, the fusion rule of a pair of ${\bf 3/2}\otimes\sigma$ becomes \begin{align}({\bf 3/2}\otimes\sigma)\times({\bf 3/2}\otimes\sigma)&=({\bf 0}+{\bf 1}+{\bf 2}+{\bf 3})\otimes(1+\psi)\nonumber\\&\equiv{\bf 0}+{\bf 0}+{\bf 1}+{\bf 1}+{\bf 2}+{\bf 2}+{\bf 3}+{\bf 3}\end{align} which has two vacuum fusion channels and indicates that ${\bf 3/2}\otimes\sigma$ cannot be a simple object. This leads to the decomposition \begin{align}{\bf 3/2}\otimes\sigma\equiv\alpha_++\alpha_-\label{3/2sigmadecomposition}\end{align} where $\alpha_\pm$ are simple anyons with identical exchange statistics but opposite fermion parity $\alpha_\pm\times f=\alpha_\mp$ and obey the fusion rules \eqref{GNfusion2}.

\begin{table}[htbp]
\centering
\begin{tabular}{l|lllllll}
&1&$\alpha_+$&$\gamma_+$&$\beta$&$\gamma_-$&$\alpha_-$&$f$\\\hline
$r$ even &${\bf 0}$&${\bf 1/2}\otimes s_+$&${\bf 1}$&${\bf 3/2}\otimes s_\pm$&${\bf 2}$&${\bf 5/2}\otimes s_+$&${\bf 3}$\\
$r$ odd &${\bf 0}$&$({\bf 3/2}\otimes\sigma)_+$&${\bf 1}$&${\bf 1/2}\otimes\sigma$&${\bf 2}$&$({\bf 3/2}\otimes\sigma)_-$&${\bf 3}$
\end{tabular}
\caption{Identification of the seven anyon types in table~\ref{tab:toporderSO(3)3SO(r)1} as tensor products.}\label{tab:anyonsSO(3)3SO(r)1}
\end{table}

We summarize the identification of the seven anyon types in $G_N=SO(3)_3\boxtimes_bSO(r)_1$ as tensor products in table~\ref{tab:anyonsSO(3)3SO(r)1}. This explains the exchange statistics and quantum dimensions of the quasiparticles in table~\ref{tab:toporderSO(3)3SO(r)1} \begin{align}\theta_{{\bf a}\otimes{\bf b}}=\theta_{\bf a}\theta_{\bf b},\quad d_{{\bf a}\otimes{\bf b}}=d_{\bf a}d_{\bf b}\end{align} with the exception of the non-simple object ${\bf 3/2}\otimes\sigma$ in \eqref{3/2sigmadecomposition} where each component $\alpha_\pm$ carries half of its dimension. The fusion rules in \eqref{GNfusion1} and \eqref{GNfusion2} are explained by the tensor product \begin{align}({\bf a}_1\otimes{\bf b}_1)\times({\bf a}_2\otimes{\bf b}_2)=({\bf a}_1\times{\bf a}_2)\otimes({\bf b}_1\times{\bf b}_2)\end{align} except in the odd $r$ cases where again the non-simple object ${\bf 3/2}\otimes\sigma=\alpha_++\alpha_-$ requires special attention.

The fusion rules \eqref{GNfusion2} of $\alpha_\pm$ in the odd $r$ cases are fixed by modular invariance. The braiding $S$-matrix \eqref{braidingS} is determined by fusion rules and quasiparticle exchange statistics. On the other hand fusion rules can also be determined by the $S$-matrix using the Verlinde formula \eqref{Verlindeformula}.\cite{Verlinde88} Moreover one can define the $T$-matrix according to the quasiparticle exchange statistics \begin{align}T_{{\bf a}{\bf b}}=\delta_{{\bf a}{\bf b}}\theta_{\bf a}\end{align} which corresponds to the modular $T$-transformation in the CFT along the boundary. As a consequence they satisfies the $SL(2;\mathbb{Z})$ algebraic relation\cite{Kitaev06} \begin{align}\left(\mathcal{S}T^\dagger\right)^3=e^{-2\pi ic_-/8}\mathcal{S}^2\label{modularity}\end{align} where $c_-=c_R-c_L$ is the chiral central charge of the corresponding CFT along the boundary \begin{align}c_-(G_N)=c_-(so(3)_3)+c_-(so(r)_1)=\frac{9}{4}+\frac{r}{2}=\frac{N}{4}.\end{align} These put a very restrictive constraint on the allowed topological field theory and fix the fusion rules \eqref{GNfusion2} for $\alpha_\pm$ when $r$ is odd. The braiding $S$ matrices can be found in appendix~\ref{sec:GNapp}.

The relative tensor product structure of the sixteenfold $SO(r)_1$ sequence itself can also be understood using anyon condensation \begin{align}SO(r_1)_1\boxtimes_bSO(r_2)_1\cong SO(r_1+r_2)_1\label{SO(r)tensorproduct}\end{align} where the fermion pair $\psi_1\otimes\psi_2$ is condensed. This can be verified by a similar condensation procedure as the one presented above. For instance, if $r_1$ and $r_2$ are both odd, the tensor product $\sigma_1\otimes\sigma_2$ will become non-simple after condensation and decompose into a pair of abelian $\pi$-fluxes, $s_++s_-$, with identical exchange statistics but opposite fermion parities $s_\pm\times\psi=s_\mp$ and are related by an anyonic symmetry\cite{khan2014,TeoHughesFradkin15}.

Next we move on to explaining the general relative tensor product structure \eqref{Z32tensorstructure} of the 32-fold $G_N$ states. Eq.\eqref{SO(r)tensorproduct} describes the cases when both $N_1$ and $N_2$ are even, i.e.~$G_{2r_1}\boxtimes_bG_{2r_2}\cong G_{2r_1+2r_2}$. A similar anyon condensation procedure that defined the relative tensor product $SO(3)_3\boxtimes_bSO(r)_1$ above would prove that \begin{align}G_N\boxtimes_bSO(r)_1\cong G_{N+2r}\end{align} for $N$ odd, where the fermion pair $b=f_{G_N}\otimes\psi_{SO(r)_1}$ is condensed. 

When both $N_1=9+2r_1$ and $N_2=9+2r_2$ are odd, each of the two $G_{N_i}=SO(3)_3\boxtimes_bSO(r_i)_1$ theories contains seven anyon types $1,\alpha^i_\pm,\gamma^i_\pm,\beta^i,f^i$. The tensor product state $G_{N_1}\otimes G_{N_2}$ contains three non-trivial bosons \begin{align}b=\{b_0,b_+,b_-\}=\left\{f^1\otimes f^2,\gamma^1_+\otimes\gamma^2_-,\gamma^1_-\otimes\gamma^2_+\right\}\label{GNGNbosons}\end{align} as $\gamma_\pm$ have conjugate exchange phases $\theta_{\gamma_\pm}=\pm i$. Moreover, these bosons are mutually local. Firstly, $b_0$ have trivial monodromy around $b_\pm$ as $\gamma_\pm$ are local with respect to the fermion $f$. Secondly, as there are bosonic fusion channels $b_\pm\times b_\pm=1+b_++b_-+\ldots$ and $b_\pm\times b_\mp=b_0+b_++b_-+\ldots$, $b_\pm$ are local among themselves because their mutual monodromy phases are trivial. We first condensed the Abelian fermion pair $b_0=f^1\otimes f^2$. The resulting theory contains the following set of (non-confined) anyon types \begin{align}G_{N_1}\boxtimes_{b_0}G_{N_2}=\left\langle\begin{array}{*{20}c}1,f,\gamma_\pm^1,\gamma_\pm^2,\gamma_+^1\gamma_+^2,\gamma_+^1\gamma_-^2,\\\alpha_+^1\alpha_+^2,\alpha_+^1\alpha_-^2,\alpha^1_+\beta^2,\beta^1\alpha^2_+,\beta^1\beta^2\end{array}\right\rangle\label{GNGNb0anyons}\end{align} where some anyon types are identified by the $b_0$ condensate, such as $f\equiv f^1\equiv f^2$ and $\gamma_-^1\gamma_-^2=\gamma_+^1\gamma_+^2\times b_0$, and are therefore not listed. Next we condense the non-Abelian boson $b_+=\gamma_+^1\gamma_-^2$, which is already equated with $b_-=b_+\times b_0$. The general condensation procedure of a non-Abelian boson was proposed by Bais and Slingerland in Ref.\onlinecite{BaisSlingerlandCondensation}. In the present case, it begins with the fusion theory $\mathcal{F}$ of $G_{N_1}\boxtimes_{b_0}G_{N_2}$ that only encodes the associative fusion content but neglects the braiding structure of the anyons. As the boson $b_+$ is condensed, it decomposes as $b_+=\gamma_+^1\gamma_-^2=1+\ldots$, which now contains the vacuum channel 1. This reduces the fusion theory $\mathcal{F}$ into a new fusion theory $\mathcal{F}'$, where the certain anyons in \eqref{GNGNb0anyons} become non-simple objects and decompose into simpler components while others are identified by the boson condensate. This new fusion category $\mathcal{F}'$ contains the non-confined anyons in the resulting state as well as confined non-point-like objects.

We start with the first line of anyons in \eqref{GNGNb0anyons}, which are all local with respect to the fermion $f$. The semion $\gamma_+^1$ is self-conjugate as $\gamma_+^1\times\gamma_+^1=1+\gamma_+^1+\gamma_-^1$. However $\gamma_-^2$ is now also an antiparticle of $\gamma_+^1$ since $\gamma_+^1\times\gamma_-^2=b_+=1+\ldots$ also contains the vacuum channel. The uniqueness of antipartner guarantees the identifications \begin{align}\gamma_+\equiv\gamma_+^1\equiv\gamma_-^2,\quad\gamma_-\equiv\gamma_-^1\equiv\gamma_+^2\label{gammaidentification}\end{align} which obey the usual fusion rules $\gamma_\pm\times\gamma_\pm=1+\gamma_++\gamma_-$ and $f\times\gamma_\pm=\gamma_\mp$. This in turn determines the decomposition of the non-Abelian boson \begin{align}b_+=\gamma_+^1\gamma_-^2\equiv\gamma_+\times\gamma_+=1+\gamma_++\gamma_-\label{b+identification}\end{align} which is consistent with the boson quantum dimension $d_{b_+}=d_\gamma^2=1+2d_\gamma$. Moreover the non-Abelian fermion also decomposes \begin{align}\gamma_+^1\gamma_+^2\equiv\gamma_+\times\gamma_-=f+\gamma_++\gamma_-.\label{NAfermionidentification}\end{align}

Next we move on to the second line of anyons in \eqref{GNGNb0anyons}, which are $\pi$ fluxes with respect to the fermion $f$. From the original fusion rules \eqref{GNfusion1}, \eqref{GNfusion2} and the identification \eqref{gammaidentification}, \eqref{b+identification} and \eqref{NAfermionidentification}, the $\pi$ fluxes satisfy the fusion rules \begin{align}&(\alpha_+^1\alpha_+^2)\times(\alpha_+^1\alpha_+^2)\nonumber\\&=\left\{\begin{array}{*{20}l}1+f+2\gamma_++2\gamma_-,&\mbox{for $r_1+r_2$ even}\\1+1+\gamma_++\gamma_-+2\gamma_\pm,&\mbox{for $r_1+r_2\equiv3$ mod 4}\\f+f+\gamma_++3\gamma_-,&\mbox{for $r_1+r_2\equiv1$ mod 4}\end{array}\right.\label{aaaa}\end{align}\begin{align}(\alpha^1\beta^2)\times(\alpha^1\beta^2)&=1+1+f+f+4\gamma_++4\gamma_-\label{abab}\\(\beta^1\beta^2)\times(\beta^1\beta^2)&=4(1+f+2\gamma_++2\gamma_-)\\(\alpha^1_+\alpha^2_+)\times(\alpha^1_+\beta^2)&=1+f+3\gamma_++3\gamma_-\label{aaab}\\(\alpha^1_+\alpha^2_+)\times(\beta^1\beta^2)&=1+1+f+f+4\gamma_++4\gamma_-\end{align} for $N_1=9+2r_1$ and $N_2=9+2r_2$. 

These show $\alpha^1\beta^2$ and $\beta^1\beta^2$ must be non-simple because their corresponding fusion rules contain multiple vacuum channels. The decomposition of $\beta^1\beta^2$ is simplest and applies to all $r_1$, $r_2$ \begin{align}\beta^1\beta^2=\alpha_+^1\alpha_+^2+\alpha_+^1\alpha_-^2\label{bbdecomposition}\end{align} where $\alpha_+^1\alpha_-^2=\alpha_+^1\alpha_+^2\times f$. For instance, it is straightforward to check that this decomposition is consistent with the fusion rules. $\alpha^1_+\beta^2$ and $\alpha^1_-\beta^2$ are clearly identified as they differ only by the Abelian boson $b_0=f^1f^2$. We therefore will simply denote them as $\alpha^1\beta^2$. Moreover, one can show that $\alpha^1\beta^2$ and $\beta^1\alpha^2$ are also identified after the condensation of the non-Abelian boson $\gamma_+^1\gamma_-^2=1+\gamma_++\gamma_-$ in \eqref{b+identification}. This can be verify by equating the fusion equations $(\alpha^1\beta^2)\times(\gamma_+^1\gamma_-^2)=(\alpha^1\beta^2)\times(1+\gamma_++\gamma_-)$. The decomposition of $\alpha^1\beta^2\equiv\beta^1\alpha^2$ depends on the parity of $r_1+r_2$. 

When $r_1+r_2$ is even, the pair fusion rule for $\alpha_+^1\alpha_+^2$ allows it to be simple since there is a unique vacuum channel. Moreover as the pair fusion rule is unaltered by the addition of a fermion $f$, it is identical to $(\alpha_+^1\alpha_+^2)\times(\alpha^1_+\alpha^2_-)$. This shows $\alpha^1_\pm\alpha^2_-$ conjugates and therefore identifies with $\alpha^1_\pm\alpha^2_+$, which is self-conjugate. \begin{align}\alpha^1\alpha^2\equiv\alpha^1_\pm\alpha^2_\pm\equiv\alpha^1_\pm\alpha^2_\mp.\end{align} In this case, $\alpha^1\beta^2$ is decomposed into \begin{align}\alpha^1\beta^2=\sigma+\alpha^1\alpha^2\label{abdecomposition1}\end{align} where we introduce the Ising anyon $\sigma$ that obey \begin{gather}\sigma\times\sigma=1+f,\quad\sigma\times f=\sigma\label{redfusioneven1}\\\sigma\times\alpha^1\alpha^2=\gamma_++\gamma_-,\quad\sigma\times\gamma_\pm=\alpha^1\alpha^2.\nonumber\end{gather} The decomposition \eqref{abdecomposition1} is consistent with the fusion rules \eqref{aaab} and \eqref{abab}. The reduced fusion category after condensing the boson \eqref{b+identification} is therefore generated by the following simple objects \begin{align}\mathcal{F}'_{\mathrm{even}}=\left\langle1,f,\sigma,\gamma_\pm,\alpha^1\alpha^2\right\rangle\label{F'even}\end{align} when $r_1+r_2$ is even. It has the fusion rules \eqref{redfusioneven1} together with $\gamma_\pm\times\alpha^1\alpha^2=\sigma+2\alpha^1\alpha^2$.

When $r_1+r_2$ is odd, we need to further separate into two cases. When $r_1+r_2\equiv3$ mod 4, the fusion rule of a pair of $\alpha^1_+\alpha^2_+$ in \eqref{aaaa} forbids it to be simple. It decomposes into \begin{align}\alpha^1_+\alpha^2_+=s_++\gamma_+\quad\mbox{or}\quad s_++\gamma_-\label{aadecomposition}\end{align} where $s_\pm$ are Abelian anyons that satisfy the fusion rules \begin{align}s_\pm\times s_\pm=1,\quad s_\pm\times f=s_\mp,\quad s_+\times\gamma_\pm=\gamma_\pm\end{align} and the fermion parity $\gamma_\pm$ in \eqref{aadecomposition} depends on $(r_1,r_2)\equiv(0,3)$ or $(1,2)$ mod 4 but is unimportant for the current discussion. The decomposition \eqref{aadecomposition} is consistent with the fusion rule \eqref{aaaa}. In this case, the fusion rules $(\alpha^1_+\alpha^2_+)\times(\alpha^1\beta^2)$ in \eqref{aaab} requires a different decomposition of $\alpha^1\beta^2$ than \eqref{abdecomposition1}. \begin{align}\alpha^1\beta^2=\gamma_++\gamma_-.\label{abdecomposition2}\end{align} The reduced fusion category after condensing the boson \eqref{b+identification} is therefore generated by the following simple objects \begin{align}\mathcal{F}'_3=\left\langle1,f,s_\pm,\gamma_\pm\right\rangle\label{F'3}\end{align} when $r_1+r_2\equiv3$ mod 4.

When $r_1+r_2\equiv1$ mod 4, the fusion rule \eqref{aaaa} again forbids $\alpha^1_+\alpha^2_+$ to be simple. Moreover as the vacuum channel is absent, it is no longer self-conjugate but instead is conjugate with $\alpha^1_+\alpha^2_-$ since it has opposite fermion parity and $(\alpha^1_+\alpha^2_+)\times(\alpha^1_+\alpha^2_-)=1+1+3\gamma_++\gamma_-$. We decompose \begin{align}\alpha^1_+\alpha^2_+=s_++g_+\label{aadecomposition2}\end{align} where $s_\pm$ are Abelian anyons and $g_\pm$ are non-Abelian objects that satisfy \begin{align}s_\pm\times s_\pm=f,\quad s_\pm\times f=s_\mp,\quad g_\pm=\gamma_+\times s_\pm.\end{align} The decomposition of $\alpha^1\beta^2$ also needs to be modified \begin{align}\alpha^1\beta^2=g_++g_-.\label{abdecomposition3}\end{align} One can check that these decompositions are consistent with the original fusion rules. The reduced fusion category after condensing the boson \eqref{b+identification} is therefore generated by the following simple objects \begin{align}\mathcal{F}'_1=\left\langle1,f,s_\pm,\gamma_\pm,g_\pm\right\rangle\label{F'1}\end{align} when $r_1+r_2\equiv1$ mod 4.

Not all objects in the reduced fusion theories $\mathcal{F}'_{\mathrm{even}}$, $\mathcal{F}'_1$ and $\mathcal{F}'_3$ in \eqref{F'even}, \eqref{F'1} and \eqref{F'3} are non-confined anyons in the new topological states. Some may be non-local with respect to the boson $b_+$ \eqref{b+identification} and are therefore not point-like objects when $b_+$ is condensed. They are equipped with a physical string or branch cut that extends. The anyon theory, which encodes both fusion and braiding information, after condensation excludes these confined extended objects. To determine which objects in the reduced fusion categories $\mathcal{F}'$ are non-confined anyons, we look at the possible monodromy around the condensed boson $b_+$. Suppose ${\bf a}_1\otimes{\bf a}_2$ and ${\bf b}_1\otimes{\bf b}_2$ are anyons in the tensor product state $G_{N_1}\boxtimes_{b_0}G_{N_2}$ \eqref{GNGNb0anyons} that are related by the fusion rule $b_+\times({\bf a}_1\otimes{\bf a}_2)={\bf b}_1\otimes{\bf b}_2+\ldots$, the monodromy under this fixed fusion channel is\cite{BaisSlingerlandCondensation}  \begin{align}\vcenter{\hbox{\includegraphics[width=0.5in]{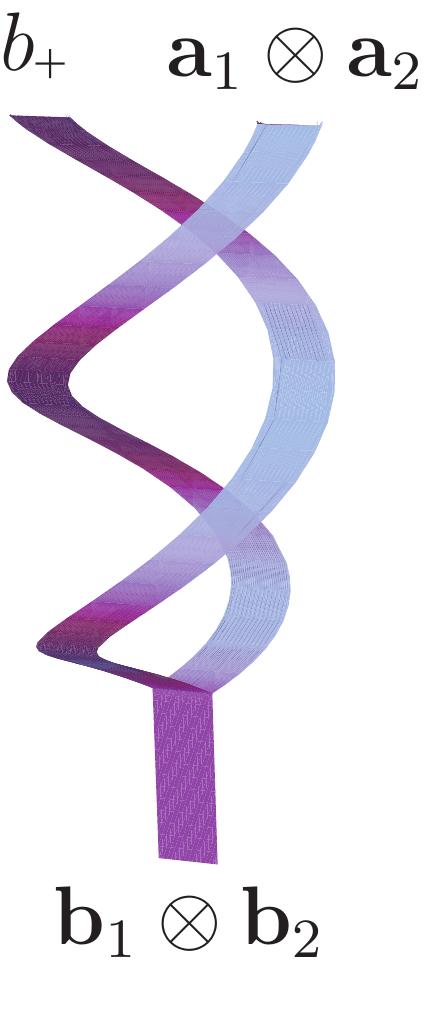}}}=\vcenter{\hbox{\includegraphics[width=0.5in]{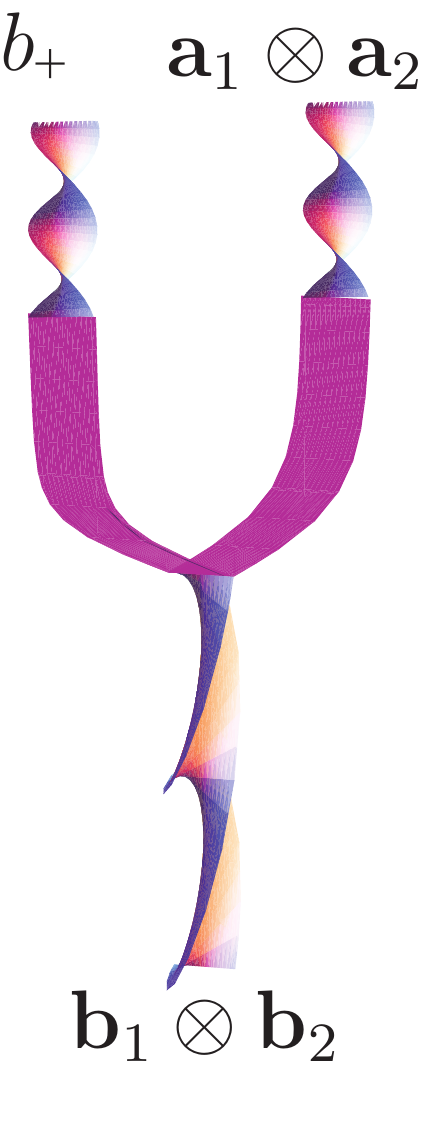}}}=\frac{\theta_{{\bf b}_1\otimes{\bf b}_2}}{\theta_{b_+}\theta_{{\bf a}_1\otimes{\bf a}_2}}=\frac{\theta_{{\bf b}_1\otimes{\bf b}_2}}{\theta_{{\bf a}_1\otimes{\bf a}_2}}\label{ribbon}\end{align} as $b_+$ is a boson with $\theta_{b_+}=1$. In other words trivial monodromy simply reqires the invariance of exchange statistics upon an addition of the boson. 

Given any simple object ${\bf x}$ in the reduced fusion category $\mathcal{F}'$ in \eqref{F'even}, \eqref{F'1} or \eqref{F'3}, it may be {\em lifted} to multiple anyons in the tensor product state $G_{N_1}\boxtimes_{b_0}G_{N_2}$ in \eqref{GNGNb0anyons} in the sense that it belongs in distinct decompositions ${\bf a}_1\otimes{\bf a}_2={\bf x}+\ldots$ and ${\bf b}_1\otimes{\bf b}_2={\bf x}+\ldots$. For instance, $\gamma_\pm$ are components of the boson $\gamma^1_+\gamma^2_-=1+\gamma_++\gamma_-$ as well as the fermion $\gamma^1_+\gamma^2_+=f+\gamma_++\gamma_-$ (see \eqref{b+identification} and \eqref{NAfermionidentification}). If ${\bf x}$ is an object not confined by the boson condensation, then its exchange statistics should be independent from the choices of lift \begin{align}\theta_{\bf x}=\theta_{{\bf a}_1\otimes{\bf a}_2}=\theta_{{\bf b}_1\otimes{\bf b}_2}\end{align} since the monodromy \eqref{ribbon} should be trivial. Otherwise, the object ${\bf x}$ has to be non-point-like and extended as it does not have well defined statistics. For example since $\gamma_\pm$ belongs to the decomposition of a non-Abelian boson and fermion, they have to be confined objects after condensation.

The relative tensor product $G_{N_1}\boxtimes_bG_{N_2}$ with the condensation of the set of bosons $b$ \eqref{GNGNbosons} contains non-confined anyons in the reduced fusion categories $\mathcal{F}'_{\mathrm{even}}$, $\mathcal{F}'_1$ and $\mathcal{F}'_3$ in \eqref{F'even}, \eqref{F'1} and \eqref{F'3}. For example when $r_1+r_2$ is even, the simple object $\alpha^1\alpha^2$ in \eqref{F'even} is confined and is not an anyon because it can be lifted into $\alpha^1\beta^2$ and $\beta^1\beta^2$, which have distinct statistics, in \eqref{abdecomposition1} and \eqref{bbdecomposition}. When $r_1+r_2\equiv1$ mod 4, the simple objects $g_\pm$ are also confined because they belong in $\alpha^1\beta^2$ and $\alpha^1_+\alpha^2_\pm$, which have different spins, in \eqref{abdecomposition3} and \eqref{aadecomposition2}. This shows $G_{N_1}\boxtimes_bG_{N_2}$ is generated by the non-confined anyons \begin{align}G_{N_1}\boxtimes_bG_{N_2}=\left\{\begin{array}{*{20}c}\left\langle1,f,\sigma\right\rangle,&\mbox{for $r_1+r_2$ even}\\\left\langle1,f,s_\pm\right\rangle,&\mbox{for $r_1+r_2$ odd}\end{array}\right.\end{align} The exchange statistics of $\sigma$ and $s_\pm$ are determined by that of their lifts. For instance,  \begin{align}\theta_\sigma=\theta_{\alpha^1\beta^2}=\theta_\alpha\theta_\beta=e^{\pi i\frac{9+r_1+r_2}{8}}=e^{\pi i(N_1+N_2)/16}\end{align} using table~\ref{tab:toporderSO(3)3SO(r)1} when $r_1+r_2$ is even. This shows \begin{align}G_{N_1}\boxtimes_bG_{N_2}=SO\left(\frac{N_1+N_2}{2}\right)_1\end{align} when both $N_1$ and $N_2$ are odd and concludes the 32-fold tensor product algebraic structure of the $G_N$-series.

\section{Other possibilities}\label{sec:otherpossibilities}

In the previous sections, we proposed time reversal symmetric interactions that gap the coupled Majorana wire model and lead to a $G_N$ topological order (see eq.\eqref{GN=sor1} and \eqref{GN=so3xsor}). The interwire current-current backscattering interactions depend on a particular fractionalization, $so(N)_1\supseteq\mathcal{G}_N\times\mathcal{G}_N$, of the $N$ Majorana channels per wire. However, in special cases, we have already seen that alternative decompositions exist and correspond to different gapping interactions and topological orders. For example, at the beginning of section~\ref{sec:gapping}, we showed when there are even Majorana channels per wire, the model could simply be gapped by a single-body backscattering potential (see \eqref{HbcN=2}) and have trivial topological order. This is consistent with the $\mathbb{Z}_2$ classification of gapless Majorana modes protected by the ``antiferromagnetic" time reversal symmetry \eqref{TR}. Another example was given in section~\ref{sec:gappingN=4} for the special case when there are $N=4$ Majorana channels per wire where the decomposition needs to be changed into $so(4)_1\supseteq su(2)_1\times su(2)_1$. The resulting gapped state carries the $SU(2)_1$ semion topological order instead of $G_4=SO(2)_1$.

Moreover the sixteenfold classification of topological superconductors (TSC) with the presence of interaction\cite{LukaszChenVishwanath, MetlitskiFidkowskiChenVishwanath14, WangSenthil14, Senthil2014, Kapustin2014e, Qi_AxionTFT, Witten15} suggests the 32-fold $G_N$-series could have redundancies. On the other hand, the $\mathbb{Z}_{16}$ classification of TSC is based on the canonical local time reversal symmetry, which is fundamentally different from the non-local ``antiferrormagnetic" time reversal considered in this manuscript. The $\mathbb{Z}_{32}$ structure of surface topological order could be an artifact of such unconventional time reversal symmetry. Nonetheless, here in section~\ref{sec:N=16} and \ref{sec:altconfemb}, we discuss altenative gapping interactions when $N=16$ that removes all electronic quasiparticles. 

\subsection{Consequence of the emergent \texorpdfstring{$E_8$}{E8} when \texorpdfstring{$N=16$}{N=16}}\label{sec:N=16}

We design alternative interwire backscattering terms in the coupled wire model \eqref{NMajcone} with $N=16$ Majorana channels per wire. They open a time reversal symmetric energy gap among 16 surface Majorana cones with the same chirality. In general, these terms can also apply when the number of chiral Majorana channel per wire is larger than 16 by acting on a subset of channels.  
We begin with the bosonized description presented previously in section~\ref{sec:bosonization}, where each wire consists of an 8-component chiral $U(1)$ boson $\widetilde{\boldsymbol\phi}=(\tilde\phi^1,\ldots,\tilde\phi^8)$ that bosonizes the complex fermions $c_j=(\psi_{2j-1}+i\psi_{2j})/\sqrt{2}=\exp(i\tilde\phi^j)$. 
This theory is special because it carries non-trivial bosonic primary fields, which can condense. For example the two spinor representations $s_\pm$ correspond to {\em bosonic} primary fields of $so(16)_1$ with conformal dimension $h_{s_\pm}=1$ (see eq.\eqref{spinsigmas}). In particular we will focus on the even sector $s_+$. It consists of vertex operators \begin{align}V^{\bf \boldsymbol\varepsilon}_{s_+}
=e^{i\boldsymbol\varepsilon\cdot\widetilde{\boldsymbol\phi}/2},\quad\boldsymbol\varepsilon=(\varepsilon_1,\ldots,\varepsilon_8)\label{E8roots}\end{align} (see eq.\eqref{sprimaryso(2r)}) 
for $\varepsilon_j=\pm1$ with $\varepsilon_1\ldots\varepsilon_8=+1$. 
The $128=2^7$ number of combinations naturally matches with the dimension of the even spinor representation of $so(16)$ (see appendix~\ref{sec:so(N)app}). These $V^{\boldsymbol\varepsilon}_{s_+}$ are related to each other through the OPE with the raising and lowering operators $E^{\boldsymbol\alpha}=e^{i\boldsymbol\alpha\cdot\widetilde{\boldsymbol\phi}}=e^{i(\pm\tilde\phi^i\pm\tilde\phi^j)}$ of $so(16)_1$ (see \eqref{Ealpharoot} in appendix~\ref{sec:kleinfactors}). The 128 lattice vectors $\boldsymbol\varepsilon/2$ extend the 112 roots $\boldsymbol\alpha$ of $so(16)$ to the root lattice of the exceptional simple Lie algebra $E_8$ with size 240.\cite{bigyellowbook} The unit dimensional vertex operators $V^{\boldsymbol\varepsilon}_{s_+}$ themselves can be regarded as raising and lowering operators that enlarge the $so(16)_1$ current algebra to $E_8$ at level 1. This extends the current algebra of each wire \begin{align}so(16)_1\subseteq(E_8)_1\end{align} and is intimately related to the fact that the surface state can be gapped out without leaving electronic quasiparticles which are non-local with respect to the boson $s_+$. 

The gapping strategy is to condense primary fields in the bosonic sector $s_+$ between adjacent wires. This is facilitated by interwire backscattering interactions that bipartite the emergent $E_8$ symmetry. \begin{align}E_8\supseteq\widetilde{so(8)^+_1}\times\widetilde{so(8)^-_1}\end{align} However, these $\widetilde{so(8)_1}$ subalgebras are distinct from the ones in the decomposition $so(16)_1\supseteq so(8)_1\times so(8)_1$. In particular, we will see that they do not support electronic primary fields $c_j=e^{i\tilde\phi^j}$. Out of 128 $\boldsymbol\varepsilon$ lattice vectors in \eqref{E8roots}, there is a (non-unique) maximal set of 8 orthonormal vectors $\boldsymbol\varepsilon_{(1)},\ldots,\boldsymbol\varepsilon_{(8)}$ \begin{align}\frac{1}{2}\boldsymbol\varepsilon_{(m)}\cdot\frac{1}{2}\boldsymbol\varepsilon_{(n)}=2\delta_{mn}.\end{align} We choose the set containing the highest weight vector $\boldsymbol\varepsilon_{(1)}=(1,1,1,1,1,1,1,1)$: 
\begin{align}\left(\begin{array}{*{20}c}|&&|\\\boldsymbol\varepsilon_{(1)}&\ldots&\boldsymbol\varepsilon_{(8)}\\|&&|\end{array}\right)=
\left(\begin{smallmatrix}
 1 & 1 & 1 & 1 & 1 & 1 & 1 & 1 \\
 1 & 1 & 1 & 1 & -1 & -1 & -1 & -1 \\
 1 & 1 & -1 & -1 & 1 & 1 & -1 & -1 \\
 1 & 1 & -1 & -1 & -1 & -1 & 1 & 1 \\
 1 & -1 & 1 & -1 & 1 & -1 & 1 & -1 \\
 1 & -1 & 1 & -1 & -1 & 1 & -1 & 1 \\
 1 & -1 & -1 & 1 & 1 & -1 & -1 & 1 \\
 1 & -1 & -1 & 1 & -1 & 1 & 1 & -1
\end{smallmatrix}\right).\end{align} From \eqref{ETcomm0}, they give 8 mutually commuting bosons $\boldsymbol\varepsilon_{(n)}\cdot\boldsymbol\phi_y/2$ per wire \begin{align}&\left[\frac{1}{2}\boldsymbol\varepsilon_{(m)}\cdot\boldsymbol\phi_y(x,t),\frac{1}{2}\boldsymbol\varepsilon_{(n)}\cdot\boldsymbol\phi_{y'}(x',t)\right]
\nonumber\\&=2\pi i\delta_{mn}(-1)^y\delta_{yy'}\mbox{sgn}(x'-x)\end{align} up to a constant integral multiple of $2\pi i$.

We separate the 8 vectors into two groups $\mathcal{S}^+=\{\boldsymbol\varepsilon_{(1)},\boldsymbol\varepsilon_{(2)},\boldsymbol\varepsilon_{(3)},\boldsymbol\varepsilon_{(4)}\}$ and $\mathcal{S}^-=\{\boldsymbol\varepsilon_{(5)},\boldsymbol\varepsilon_{(6)},\boldsymbol\varepsilon_{(7)},\boldsymbol\varepsilon_{(8)}\}$. They defines the two $\widetilde{so(8)^\pm_1}$ subalgebras in $E_8$, whose roots lie in the root lattice of $E_8$ orthogonal to $\mathcal{S}^\mp$ respectively. One could pick the simple roots \begin{align}\widetilde{\boldsymbol\alpha}^+_1=\boldsymbol\varepsilon_{(1)}/2,\;\widetilde{\boldsymbol\alpha}^+_2={\bf e}_1+{\bf e}_2,\;\widetilde{\boldsymbol\alpha}^+_2={\bf e}_3+{\bf e}_4,\;\widetilde{\boldsymbol\alpha}^+_4={\bf e}_5+{\bf e}_6\nonumber\\\widetilde{\boldsymbol\alpha}^-_1=\boldsymbol\varepsilon_{(5)}/2,\;\widetilde{\boldsymbol\alpha}^-_2={\bf e}_2-{\bf e}_1,\;\widetilde{\boldsymbol\alpha}^-_2={\bf e}_4-{\bf e}_3,\;\widetilde{\boldsymbol\alpha}^-_4={\bf e}_6-{\bf e}_5\nonumber\end{align} so that their inner product recover the Cartan matrix of $so(8)$ \begin{align}\widetilde{\boldsymbol\alpha}^\pm_I\cdot\widetilde{\boldsymbol\alpha}^\pm_J=K_{IJ},\quad K=\left(\begin{smallmatrix}2&-1&-1&-1\\-1&2&0&0\\-1&0&2&0\\-1&0&0&2\end{smallmatrix}\right)\end{align} while opposite sectors decouple $\widetilde{\boldsymbol\alpha}^\pm_I\cdot\widetilde{\boldsymbol\alpha}^\mp_J=0$.

The new gapping potential is constructed by backscattering the two decoupled $\widetilde{so(8)^\pm_1}$ currents to adjacent wires in opposite directions. \begin{align}\mathcal{H}_{\mathrm{int}}=u\sum_{y=-\infty}^\infty{\bf J}_{\widetilde{so(8)^-_1}}^y\cdot{\bf J}_{\widetilde{so(8)^+_1}}^{y+1}\label{E8int}\end{align} However not every terms can be written down as 4-fermion interactions. In particular $\mathcal{H}_{\mathrm{int}}$ contains interwire $s_+$ quasiparticle backscattering \begin{align}V^{\boldsymbol\varepsilon}_yV^{-\boldsymbol\varepsilon'}_{y+1}+h.c.\sim\cos\left(\sum_{j=1}^8\frac{\varepsilon_j}{2}\tilde\phi^j_y-\frac{\varepsilon'_j}{2}\tilde\phi^j_{y+1}\right)\end{align} for $\varepsilon_j,\varepsilon'_j=\pm1$, that condenses pairs of $s_+$'s along adjacent wires and confines all electronic excitations. The $\widetilde{so(8)_1^\pm}$ WZW CFT carries three emergent fermionic primary fields \begin{align}\widetilde{c}^\pm_p=\exp\left[\frac{i}{2}\left(\tilde\phi^{2p-1}\pm\tilde\phi^{2p}-\tilde\phi^7\mp\tilde\phi^8\right)\right]\end{align} for $p=1,2,3$. All of which have neutral electric charge and even fermion parity with respect to the original electronic operators $c_j=e^{i\tilde\phi^j}$. This is because the $\widetilde{c}^\pm_p$'s are invariant under the $U(1)$ gauge transformation $\tilde\phi^j\to\tilde\phi^j+\varphi$. As a result, the interaction \eqref{E8int} corresponds to a gapped $\widetilde{SO(8)_1}$ topological order but contains no electron-like anyon excitations. Lastly we notice that this matches with the surface topological order of a type-II topological paramagnet.\cite{WangPotterSenthil13,Senthil2014}

\subsection{Alternative conformal embeddings}\label{sec:altconfemb}
The fractionalization $so(9)_1\supseteq so(3)_3\otimes so(3)_3$ in \ref{sec:conformalembedding} is the corner stone for the construction of symmetric gapping interactions when there is an odd number of Majorana species. However, this is not the unique decomposition. In general when the number of Majorana channels is a whole square, the wire can be bipartitioned into $so(n^2)_1\supseteq so(n)_n\otimes so(n)_n$.\cite{bigyellowbook,BaisEnglertTaorminaZizzi87,SchellekensWarner86,BaisBouwknegt87} 

For instance, this provides yet another alternative when $N=16$ where each wire is fractionalized into a pair of $so(4)_4=su(2)_4\times su(2)_4$. The $so(4)_4^\pm$ current operators can be constructed in a similar fashion as those in the $so(3)_3^\pm$ case, ${\bf J}=\frac{i}{2}\boldsymbol\Sigma^\pm_{ab}\psi^a\psi^b$ for $\boldsymbol\Sigma^+=\boldsymbol\Sigma\otimes\openone_4$ and $\boldsymbol\Sigma^-=\openone_4\otimes\boldsymbol\Sigma$ where $\boldsymbol\Sigma$ are antisymmetric $4\times 4$ matrices generating $so(4)$. After introducing the current-current backscattering interactions ${\bf J}_{so(4)_4^-}^y\cdot{\bf J}_{so(4)_4^+}^{y+1}$, the surface would carry a $SO(4)_4=SU(2)_4\times SU(2)_4$ topological order. Each $SU(2)_4$ theory contains five anyon types $j={\bf 0},{\bf 1/2},{\bf 1},{\bf 3/2},{\bf 2}$ with spins $h_j=j(j+1)/6$. The $SO(4)_4$ topological state does not carry fermionic excitations, and therefore, like the previous example in \ref{sec:N=16}, this gapping potential also removes all electronic quasiparticle excitations. 

The gapped symmetric states for $N$ odd are not unique either. For example, the decomposition $so(25)_1\supseteq so(5)_5\otimes so(5)_5$ leads to a surface $SO(5)_5$ topological order which is inequivalent to $G_{25}=SO(3)_3\boxtimes_bSO(8)_1$.

\section{Conclusion and discussion}\label{sec:conclusion}
We constructed a coupled Majorana wire model in $(2+1)$D that imitates the massless Majorana modes on the surface of a topological superconductor. This model had a non-local ``antiferromagnetic" time reversal symmetry and consequently was $\mathbb{Z}_2$ classified -- rather than $\mathbb{Z}$ in the class DIII TSC case -- under the single-body framework. Despite the difference, this model adequetly described the surface behavior of a TSC when the number $N$ of Majorana species was odd, and it was worth studying and interesting in and of itself. 

We introduced the 4-fermion gapping potentials in section~\ref{sec:gapping}. They relied on the fractionalization or bipartition of the $so(N)_1$ current along each wire into a pair of $\mathcal{G}_N$ channels (see eq.\eqref{fractioneq} and \eqref{GNfractionalization}). The two fractional channels then were backscattered onto adjacent wires in opposite directions. This freezed all low energy degrees of freedom and opened an excitations energy gap without breaking time reversal symmetry. When $N=2r$ was even, each wire could simply be split into a pair of $\mathcal{G}_N=so(r)_1$ channels. The fractionalization was not as obvious when $N$ was odd. We first made use of the conformal embedding that decomposed nine Majorana's into two subsectors, $so(9)_1\supseteq so(3)_3\otimes so(3)_3$ (see section~\ref{sec:conformalembedding}). This division could be generalized by all odd cases by splitting a subset of 9 Majorana's into a pair of $so(3)_3$ and the remaining even number of Majorana's into a pair of $so(r)_1$. This could even be applied when $N$ is less then 9 because each wire could be reconstructed by adding an arbitrary number of helical Majorana modes with the same number of right and left movers. 

The surface $G_N$ topological ordered was inferred from the bulk-boundary correspondence (see eq.\eqref{GNdefintro}). These topological states followed a 32-fold periodicity $G_N\cong G_{N+32}$ and a relative tensor product structure $G_{N_1}\boxtimes_bG_{N_2}\cong G_{N_1+N_2}$. We presented the quasiparticle types as well as their fusion and braiding statistics properties. We explained the relative tensor product structure using the notion of anyon condenstion\cite{BaisSlingerlandCondensation}. On a more fundamental level, one should be able to deduce the topological order without the knowledge of the boundary by studying the modular properties of the degenerate bulk ground states under a compact torus geometry\cite{Wenbook}, or by directly looking at exchange and braiding behaviors of bulk excitations. In fact the coupled wire construction provided a fitting model for this purpose. Being an exactly solvable model, a ground state could be explicitly expressed as entangled superposition of tensor product ground states between each pair of wires. In the simplest case when the model is bosonizable, a ground state could be specified by the pinned angle variables of a collection of sine-Gordon potentials. The bulk excitations could be realized as kinks between a pair of wires and could be created by vertex operators. The virtue of a bulk description is that the action of time reversal on quasiparticle excitations could be examined explicitly, which we have not performed or addressed here. These issues are beyond the scope of this article and we refer a more detail discussion to subsequent works.

We noticed that there were alternative ways of fractionalization that led to different gapping interactions and consequently different topological orders. We saw in section~\ref{sec:gappingN=4} that $N=4$ was an exceptional case that requires the special bipartition $so(4)_1\supseteq su(2)_1\times su(2)_1$ instead of two copies of $so(2)_1$. We also saw in section~\ref{sec:otherpossibilities} that when $N=16$, the surface could be gapped by alternative interactions that corresponded to a $\widetilde{SO(8)_1}$ or $SO(4)_4$ topological order, none of which contained electronic quasiparticle excitations. Other conformal embeddings $so(n^2)_1\supseteq so(n)_n\otimes so(n)_n$ could give rise to multiple possibilities. Our 32-fold topological states, which only utilized $so(9)_1\supseteq so(3)_3\otimes so(3)_3$, therefore should belong into a wider universal framework. These should be addressed in future works.



{}


\appendix

\section{The \texorpdfstring{$so(N)$}{so(N)} Lie algebra and its representations}\label{sec:so(N)app}
The $so(N)$ Lie algebra are generated by real antisymmetric matrices $t^{(rs)}=\left(t^{(rs)}_{ab}\right)_{N\times N}$ with entries \begin{align}t^{(rs)}_{ab}=\delta^r_a\delta^s_b-\delta^r_b\delta^s_a\label{fundrep}\end{align} for $r,s=1,\ldots,N$. There are $N(N-1)/2$ linearly independent generators since $t^{(rs)}=-t^{(sr)}$ and $t^{(rr)}=0$. In the main text, we write the basis labels as $\beta=(rs)$, for $r<s$, for conciseness. The generators obey the commutator relation \begin{align}\left[t^{(rs)},t^{(pq)}\right]=\sum_{m<n}f_{(rs)(pq)(mn)}t^{(mn)}\label{soNalgrel}\end{align} where the structure constant is \begin{align}f_{(rs)(pq)(mn)}=&\delta_{mr}\delta_{nq}\delta_{sp}-\delta_{mr}\delta_{np}\delta_{sq}\nonumber\\&+\delta_{ms}\delta_{rq}\delta_{np}-\delta_{ms}\delta_{nq}\delta_{rp}.\end{align} 

The matrix representation \eqref{fundrep} is referred as the fundamental representation of $so(N)$ and is labeled by $\psi$. In general the generators of $so(N)$ can have different irreducible matrix representations $t_\lambda^{(rs)}=t^\beta_\lambda$ labeled by $\lambda$. Since the quadratic Casimir operator \begin{align}\hat{\mathcal{Q}}_\lambda=-\sum_\beta t^\beta_\lambda t^\beta_\lambda\label{secondCasimir}\end{align} commutes with all the generators, it must have a fixed eigenvalue $\mathcal{Q}_\lambda$ that (incompletely) characterizes the irreducible representation $\lambda$. For instance, the fundamental representation in \eqref{fundrep}, denoted by $\psi$, has quadratic Casimir value $\mathcal{Q}_\psi=N-1$. 

The spinor representation $\sigma$ of $so(N)$ makes use of the Clifford algebra\cite{spingeometrybook} $\{\gamma_a,\gamma_b\}=\gamma_a\gamma_b+\gamma_b\gamma_a=2\delta_{ab}$ where $\gamma_1,\ldots,\gamma_N$ are hermitian matrices of dimension $d=2^{N/2}$ for $N$ even or $d=2^{(N-1)/2}$ for $N$ odd. The $so(N)$ generators are represented as the quadratic combination \begin{align}t_\sigma^{(rs)}=\frac{1}{4}\sum_{ab}\gamma_at^{(rs)}_{ab}\gamma_b=\frac{1}{2}\gamma_r\gamma_s\end{align} and satisfy \eqref{soNalgrel}. When $N$ is even, the parity operator $(-1)^F=i^{N/2}\gamma_1\ldots\gamma_N$ commutes with all $t_\sigma^{(rs)}$ and the representation is decomposable into $\sigma=s_+\oplus s_-$, where $s_\pm$ are $2^{N/2-1}$-dimensional sectors with $(-1)^F=\pm1$. The $so(N)$ generators are then irreducibly represented by \begin{align}t^{(rs)}_{s_\pm}=P_\pm t^{(rs)}_\sigma P_\pm^\dagger\end{align} where $P_\pm$ are the projection operators onto the fixed parity subspaces. As $t^{(rs)}_\sigma t^{(rs)}_\sigma=-(1/4)\openone$, the quadratic Casimir values \eqref{secondCasimir} of spinor representations are \begin{align}\mathcal{Q}_\sigma=\frac{N(N-1)}{8},\quad\mathcal{Q}_{s_\pm}=\frac{N(N-1)}{8}.\label{spinorQ}\end{align}

The complexified $so(N)$ Lie algebra has an alternative set of {\em Cartan-Weyl} generators. It consists of a maximal set of commuting hermitian generators $H^1,\ldots,H^r$, and a finite set of raising of lowering operators $E^{\boldsymbol\alpha}=(E^{-\boldsymbol\alpha})^\dagger$, labeled by integral vectors $\boldsymbol\alpha=(\alpha^1,\ldots,\alpha^r)\in\Delta$ called roots. The root lattice is given by the set \begin{align}\Delta_{so(2r)}&=\left\{\pm{\bf e}_I\pm{\bf e}_J:1\leq I<J\leq r\right\}\nonumber\\\Delta_{so(2r+1)}&=\Delta_{so(2r)}\cup\left\{\pm{\bf e}_I:1\leq I\leq r\right\}\label{approots}\end{align} where ${\bf e}_I$ are unit basis vectors of $\mathbb{R}^r$. In particular, there are $r$ simple roots $\boldsymbol\alpha_1,\ldots,\boldsymbol\alpha_r$ that forms a basis for the root lattice. For $so(N)$ they can be chosen to be \begin{align}\boldsymbol\alpha_I=\left\{\begin{array}{*{20}l}{\bf e}_I-{\bf e}_{I+1},&\mbox{for $I=1,\ldots,r-1$}\\{\bf e}_r,&\mbox{for $I=r$ and $N$ odd}\\{\bf e}_{r-1}+{\bf e}_r,&\mbox{for $I=r$ and $N$ even}\end{array}\right..\label{simpleroots}\end{align} The set of roots $\Delta$ consists of integral combinations of the simple roots $\boldsymbol\alpha=\sum_{J=1}^rb^J\boldsymbol\alpha_J$ so that its length is $|\boldsymbol\alpha|=\sqrt{2}$, for even $N$, or $|\boldsymbol\alpha|=1$ or $\sqrt{2}$, for odd $N$.

The integer $r$ is the rank of the $so(N)$ Lie algebra and is determined by $N=2r$ for $N$ even or $N=2r+1$ for $N$ odd. These generators satisfy \begin{gather}\left[H^i,E^{\boldsymbol\alpha}\right]=\alpha^iE^{\boldsymbol\alpha},\quad\left[E^{\boldsymbol\alpha},E^{-\boldsymbol\alpha}\right]=\frac{2}{|\boldsymbol\alpha|^2}\sum_{i=1}^r\alpha^iH^i\label{CartanWeylrel}\\\left[E^{\boldsymbol\alpha},E^{\boldsymbol\beta}\right]\propto\left\{\begin{array}{*{20}c}E^{\boldsymbol\alpha+\boldsymbol\beta},&\mbox{if $\boldsymbol\alpha+\boldsymbol\beta\in\Delta$}\\0,&\mbox{if otherwise}\end{array}\right.,\quad\mbox{for $\boldsymbol\alpha\neq\boldsymbol\beta$}.\nonumber\end{gather} 

The Cartan matrix $K=(K_{IJ})_{r\times r}$ of the algebra is defined by the scalar product \begin{align}K_{IJ}=\frac{2\boldsymbol\alpha_I^T\boldsymbol\alpha_J}{|\boldsymbol\alpha_J|^2}=\sum_{i=1}^r\frac{2\alpha_I^i\alpha_J^i}{|\boldsymbol\alpha_J|^2}.\end{align} $so(2r)$ is {\em simply-laced} in the sense that all roots have identical length and the Cartan matrix is therefore symmetric \begin{align}K_{so(2r)}=\left(\begin{array}{*{20}c}2&-1&0&\ldots&0\\-1&2&\ddots&&\vdots\\0&\ddots&2&-1&-1\\\vdots&&-1&2&0\\0&\ldots&-1&0&2\end{array}\right).\label{Kso(2r)}\end{align} Sometimes it would be convenient to use the {\em Chevalley} basis so that the commuting generators are redefined \begin{align}h^I=\frac{2}{|\boldsymbol\alpha_I|^2}\sum_{i=1}^r\alpha_I^iH^i\end{align} so that the commutator relations \eqref{CartanWeylrel} becomes \begin{align}\left[h^I,E^{\pm\boldsymbol\alpha_J}\right]=\pm K_{IJ}E^{\pm\boldsymbol\alpha_J},\quad\left[E^{\boldsymbol\alpha_J},E^{-\boldsymbol\alpha_J}\right]=\delta^{IJ}h^J.\label{Chevalleyrel}\end{align}

\section{Bosonizing the \texorpdfstring{$so(2r)_1$}{so(2r)} current algebra}\label{sec:kleinfactors}
Here we review the bosonization\cite{bigyellowbook,witten1984,Fradkinbook} of a chiral wire with $N=2r$ Majorana fermions, and express the $so(2r)_1$ current operators in bosonized form. The $2r$ Majorana (real) fermions can be paired into $r$ Dirac (complex) fermions and bosonized into the normal ordered vertex operators \begin{align}c^j(z)=\frac{\psi^{2j-1}(z)+i\psi^{2j}(z)}{\sqrt{2}}\sim\exp\left(i\widetilde{\phi}^j(z)\right).\label{appcomplexfermion}\end{align} Here we focus on a single wire, say at an even $y$, so that all fields depend on the holomorphic parameter $z=e^{\tau+ix}$. The $r$-component boson $\widetilde{\boldsymbol\phi}=(\widetilde{\phi}^1,\ldots,\widetilde{\phi}^r)$ is governed by the Lagrangian density \begin{align}\mathcal{L}_0=\frac{1}{2\pi}\sum_{j=1}^r\partial_x\widetilde{\phi}^j\partial_t\widetilde{\phi}^j=\frac{1}{2\pi}\partial_x\widetilde{\boldsymbol\phi}\partial_t\widetilde{\boldsymbol\phi}\label{appbosonization}\end{align} and follows the algebraic relations \begin{align}\left[\widetilde{\phi}^j(x,t),\widetilde{\phi}^{j'}(x',t)\right]=i\pi\left[\delta^{jj'}\mathrm{sgn}(x'-x)+\mathrm{sgn}(j-j')\right]\label{appETcomm}\end{align} or equivalently the time-ordered correlation function \begin{align}\langle\widetilde{\phi}^j(z)\widetilde{\phi}^{j'}(w)\rangle=-\delta^{jj'}\log(z-w)+\frac{i\pi}{2}\mathrm{sgn}(j-j')\label{appbosonOPE}\end{align} for $\mathrm{sgn}(s)=s/|s|$ when $s\neq0$ and $\mathrm{sgn}(0)=0$. Operator product expansions between unordered vertex operators can be evaluated by $e^{A(z)}e^{B(w)}=e^{A(z)+B(w)+\langle A(z)B(w)\rangle}$, for $A,B$ linear combination of the bosons $\widetilde{\phi}^j$. 
For instance, the vertex operators in \eqref{appcomplexfermion} reproduce the product expansion of a pair of identical Dirac fermions \begin{align}c^j(z)\left(c^j(w)\right)^\dagger=\frac{1}{z-w}+i\partial\widetilde{\phi}^j(w)+\ldots\end{align} and the singular piece is dropped when the product is normal ordered in the limit $z\to w$. The non-singular sign factor $i\pi\mathrm{sgn}(j-j')$ ensures fermions with distinct flavors anticommutes \begin{align}c^j(z)c^{j'}(w)=-c^{j'}(w)c^j(z).\end{align} 

The $so(2r)_1$ currents in the Cartan-Weyl basis can now be bosonized \begin{align}H^j(z)&=c^j(z)c^j(z)^\dagger=i\partial_z\widetilde\phi^j(z)
\label{CartanWeylcurrent}\\E^{\boldsymbol\alpha}(z)&=\prod_{j=1}^rc^j(z)^{\alpha^j}=\exp\left(i\boldsymbol\alpha\cdot\widetilde{\boldsymbol\phi}(z)\right)\nonumber\end{align} where $\boldsymbol\alpha=(\alpha^1,\ldots,\alpha^r)\in\Delta$ are roots of $so(2r)$ (see \eqref{approots}) and the fermion products are normal ordered. For instance, $\boldsymbol\alpha$ has two and only two non-zero entries and $E^{\boldsymbol\alpha}$ must be of the form \begin{align}E^{\boldsymbol\alpha}(z)=c^i(z)^\pm c^j(z)^\pm=e^{i(\pm\widetilde{\phi}^i(z)\pm\widetilde{\phi}^j(z))}.\label{Ealpharoot}\end{align} Combining raising or lowering operators give \begin{align}E^{\boldsymbol\alpha}(z)E^{\boldsymbol\beta}(w)=i^{-\boldsymbol\alpha\cdot\boldsymbol\beta}\epsilon(\boldsymbol\alpha,\boldsymbol\beta)\frac{e^{i(\boldsymbol\alpha\cdot\widetilde{\boldsymbol\phi}(z)+\boldsymbol\beta\cdot\widetilde{\boldsymbol\phi}(w))}}{(z-w)^{-\boldsymbol\alpha\cdot\boldsymbol\beta}}\end{align} where the vertex operator here is again normal ordered and the 2-cocyle is given by the star product  \begin{align}\epsilon(\boldsymbol\alpha,\boldsymbol\beta)=(-1)^{\boldsymbol\alpha\ast\boldsymbol\beta}=(-1)^{\sum_{i>j}\alpha^i\beta^j}.\label{starproduct}\end{align} As $\sum_{i=1}^r\alpha^i$ is even for all roots, we have the following simplification when interchanging $\boldsymbol\alpha\leftrightarrow\boldsymbol\beta$ \begin{align}\epsilon(\boldsymbol\alpha,\boldsymbol\beta)\epsilon(\boldsymbol\beta,\boldsymbol\alpha)=(-1)^{\boldsymbol\alpha\cdot\boldsymbol\beta}.\label{2cocylecomm}\end{align} 
	
Using the boson OPE \eqref{appbosonOPE}, the product of the two vertex operators above is singular only when (i) $\boldsymbol\alpha=-\boldsymbol\beta$, or (ii) $\boldsymbol\alpha\cdot\boldsymbol\beta=-1$ in other words $\boldsymbol\alpha+\boldsymbol\beta\in\Delta$. To summarize, the Cartan-Weyl generators satisfy the product expansion \begin{align}H^i(z)H^j(w)&=\frac{\delta^{ij}}{(z-w)^2}-\partial\widetilde{\phi}^i(w)\partial\widetilde{\phi}^j(w)+\ldots\nonumber\\H^i(z)E^{\boldsymbol\alpha}(w)&=\frac{\alpha^i}{z-w}E^{\boldsymbol\alpha}(w)+\ldots
\nonumber\\E^{\boldsymbol\alpha}(z)E^{-\boldsymbol\alpha}(w)&=\frac{1}{(z-w)^2}+\sum_{i=1}^r\frac{\alpha^i}{z-w}H^i(w)\label{appcurrentrelations}\\&\quad\quad
-\frac{1}{2}\left(\boldsymbol\alpha\cdot\partial\widetilde{\boldsymbol\phi}(w)\right)^2+\ldots\nonumber\\E^{\boldsymbol\alpha}(z)E^{\boldsymbol\beta}(w)&=\frac{i\epsilon(\boldsymbol\alpha,\boldsymbol\beta)}{z-w}E^{\boldsymbol\alpha+\boldsymbol\beta}(w)+\ldots,\quad\mbox{if $\boldsymbol\alpha\cdot\boldsymbol\beta=-1$}.\nonumber
\end{align} 
For instance, the 2-cocyle coefficient $\epsilon(\boldsymbol\alpha,\boldsymbol\beta)$ ensures the OPE between $E^{\boldsymbol\alpha}(z)$ and $E^{\boldsymbol\beta}(w)$ commute as the sign in \eqref{2cocylecomm} when exchanging $\boldsymbol\alpha\leftrightarrow\boldsymbol\beta$ cancels that in $1/(z-w)$ when switching $z\leftrightarrow w$.

In certain derivations, especially when involving quasiparticle excitations, it may be more convenient to use the Chevalley basis. Here fields are expressed in terms of non-local bosons $\boldsymbol\phi=(\phi^1,\ldots,\phi^r)$, which are related to the original ones by the (non-unimodular) basis transformation \begin{align}\widetilde{\phi}^i=\sum_{I=1}^r\alpha^i_I\phi^I\label{phiphitilde}\end{align} using the simple roots $\boldsymbol\alpha_I=(\alpha_I^1,\ldots,\alpha_I^r)\in\mathbb{Z}^r$ (see \eqref{simpleroots} in appendix~\ref{sec:so(N)app}). The Lagrangian density \eqref{appbosonization} now becomes \begin{align}\mathcal{L}_0=\frac{1}{2\pi}\sum_{I,J=1}^rK_{IJ}\partial_x\phi^I\partial_t\phi^J\label{bosonization}\end{align} where $K=(K_{IJ})_{r\times r}=\boldsymbol\alpha_I\cdot\boldsymbol\alpha_J$ is the Cartan matrix of $so(2r)_1$ (see eq.\eqref{Kso(2r)}).

The current generators are rewritten in the Chevalley basis by \begin{align}h_I(z)&=\sum_{i=1}^r\alpha^i_IH^i(z)=i\sum_{J=1}^rK_{IJ}\partial_z\phi^J(z)\nonumber\\E^{\bf b}(z)&=E^{\boldsymbol\beta}(z)=\exp\left(i{\bf b}^TK\boldsymbol\phi^J(z)\right)\end{align} where  $\boldsymbol\beta=\sum_{J}b^J\boldsymbol\alpha_J$ are roots expressed in integral combinations of the simple ones, for ${\bf b}=(b^1,\ldots,b^r)\in\mathbb{Z}^r$. The Chevalley generators satisfy the modified current relations from \eqref{appcurrentrelations} \begin{align}h_I(z)h_J(w)&=\frac{K_{IJ}}{(z-w)^2}
+\ldots\nonumber\\h_I(z)E^{\bf b}(w)&=\frac{K_{IJ}b^J}{z-w}E^{\bf b}(w)+\ldots\\E^{\bf b}(z)E^{-{\bf b}}(w)&=\frac{1}{(z-w)^2}+\sum_{I=1}^r\frac{b^I}{z-w}h_I(w)+\ldots
\nonumber\\E^{{\bf b}_1}(z)E^{{\bf b}_2}(w)&=\frac{i\epsilon(\boldsymbol\beta_1,\boldsymbol\beta_2)}{z-w}E^{{\bf b}_1+{\bf b}_2}(w)+\ldots\nonumber\end{align} if ${\bf b}_1^TK{\bf b}_2=-1$.

The (normal ordered) energy-momentum tensor can be turned from the Sugawara form \eqref{SugawaraT} to the usual bosonic one \begin{align}T(z)&=\frac{1}{2(N-1)}\left[\sum_{i=1}^rH^i(z)H^i(z)+\sum_{\boldsymbol\alpha\in\Delta}E^{\boldsymbol\alpha}(z)E^{-\boldsymbol\alpha}(z)\right]\nonumber\\&=-\frac{1}{2}\partial\widetilde{\boldsymbol\phi}(z)\cdot\partial\widetilde{\boldsymbol\phi}(z)=-\frac{1}{2}\partial\boldsymbol\phi(z)\cdot K\partial\boldsymbol\phi(z).\label{appTtensor}\end{align}
Excitations in the CFT can be easily represented by vertex operators \begin{align}V^{\bf a}(z)=\exp\left(i{\bf a}\cdot\boldsymbol\phi(z)\right)=\exp\left(i{\bf a}_\vee\cdot\widetilde{\boldsymbol\phi}(z)\right)\label{appvertex}\end{align} labeled by integral lattice vectors ${\bf a}=(a_1,\ldots,a_r)$, or equivalently dual root lattice vectors ${\bf a}_\vee=(a_\vee^1,\ldots,a_\vee^r)$ with rational entries \begin{align}a_\vee^j=\sum_{IJ}a_I(K^{-1})^{IJ}\alpha_J^j.\end{align} The conformal dimension of $V^{\bf a}$ can be read off by the inner product \begin{align}h_{\bf a}&=\frac{1}{2}{\bf a}^TK^{-1}{\bf a}=\frac{1}{2}(K^{-1})^{IJ}a_Ia_J\nonumber\\&=\frac{1}{2}{\bf a}_\vee^T{\bf a}_\vee=\frac{1}{2}\delta_{ij}a_\vee^ia_\vee^j.\label{acondim}\end{align} This can be evaluated from definition \eqref{TVdim} using the energy-momentum tensor \eqref{appTtensor} and the OPE \begin{align}\partial_z\phi_I(z)\phi_J(w)=-(K^{-1})^{IJ}\log(z-w)+\ldots\end{align} which is equivalent to \eqref{appbosonOPE}.

Most vertex operators \eqref{appvertex} however are not WZW primary and do not represent the $so(2r)_1$ Kac-Moody algebra. The OPE with the current generators \begin{align}h_I(z)V^{\bf a}(w)&=\frac{a_I}{z-w}V^{\bf a}(w)+\ldots\nonumber\\E^{\bf b}(z)V^{\bf a}(w)&=c^{\bf b}_{\bf a}(z-w)^{{\bf a}\cdot{\bf b}}V^{{\bf a}+K{\bf b}}(w)+\ldots\label{appraisinglowering}\end{align} would match the requirement \eqref{currentrepOPE} for a primary field only when the exponent of the singular term is bounded below, i.e.~${\bf a}\cdot{\bf b}\geq-1$ for all roots $\boldsymbol\beta=\sum_Ib^I\boldsymbol\alpha_I$. Such lattice vectors ${\bf a}$ are called weights or {\em Dynkin labels} of $so(2r)$ at level 1. When the exponenet ${\bf a}\cdot{\bf b}$ in \eqref{appraisinglowering} is $-1$, the vertex operators $V^{\bf a}$ and $V^{{\bf a}+K{\bf b}}$ are related by the $SO(2r)_1$ symmetry and belong to the same primary field sector. For example the unit vector ${\bf a}={\bf e}_1$ is the highest weight that generates the fermion sector $\psi$. Applying lowering operators $E^{-{\bf b}}$ to $V^{{\bf e}_1}=c^1$ gives all $2r$ Dirac fermions \begin{align}{\bf V}_\psi=\mathrm{span}\left\{(c^j)^\pm=e^{\pm i\widetilde{\phi}^j}:j=1,\ldots,r\right\}\label{appfermionsector}\end{align} which in turn irreducibly represent the $so(2r)_1$ algebra (see \eqref{currentrepOPE}) according to the fundamental vector representation.

The unit vectors ${\bf a}={\bf e}_{r-1}$ and ${\bf e}_r$ generate the two spinor sectors $s_-$ and $s_+$ respectively. Each of them consists of $2^{r-1}$ twist fields \begin{align}{\bf V}_{s_\pm}&=\sigma^1\ldots\sigma^{2r}\label{sprimaryso(2r)}\\&=\mathrm{span}\left\{\exp\left(i\sum_{j=1}^r\frac{(-1)^{s_j}}{2}\widetilde{\phi}^j\right):\prod_{j=1}^r(-1)^{s_j}=\pm1\right\}.\nonumber\end{align} They irreducibly represent the $so(2r)_1$ algebra according to the even and odd spinor representations. These are the only primary fields of $so(2r)_1$ and their conformal dimensions are given by $h_\psi=1/2$ and $h_{s_\pm}=r/8$. 

The four primary fields $1,\psi,s_\pm$ obey a set of fusion rules, which are OPE keeping only primary fields. \begin{gather}s_\pm\times\psi=s_\mp\label{so(2r)fusion1}\\s_\pm\times s_\pm\left\{\begin{array}{*{20}l}1,&\mbox{for $r$ even}\\\psi,&\mbox{for $r$ odd}\end{array}\right.,\quad s_\pm\times s_\mp\left\{\begin{array}{*{20}l}\psi,&\mbox{for $r$ even}\\1,&\mbox{for $r$ odd}\end{array}\right..\label{so(2r)fusion2}\end{gather} For instance, the OPE \begin{align}V_{s_+}(z)c^r(w)^\dagger&=e^{i\frac{\widetilde{\phi}^1(z)+\ldots+\widetilde{\phi}^r(z)}{2}}e^{-i\widetilde{\phi}^r(w)}\nonumber\\&\propto(z-w)^{-\frac{1}{2}}e^{i\frac{\widetilde{\phi}^1(w)+\ldots+\widetilde{\phi}^r(w)-\widetilde{\phi}^r(w)}{2}}+\ldots\nonumber\\&=(z-w)^{-\frac{1}{2}}V_{s_-}(w)+\ldots
\end{align} shows $s_+\times\psi=s_-$, and \begin{align}e^{i\sum_j\widetilde{\phi}^j(z)/2}e^{-i\sum_j\widetilde{\phi}^j(w)/2}\propto(z-w)^{-\frac{r}{4}}+\ldots\end{align} shows $s_+\times s_+=1$ for $r$ even, or $s_+\times s_-=1$ for $r$ odd.

\section{Bosonizing the \texorpdfstring{$so(2r+1)_1$}{so(2r+1)} current algebra}\label{sec:kleinfactors2}
A chiral wire with $N=2r+1$ Majorana fermions can be partially bosonized by grouping $\psi^1,\ldots,\psi^{2r}$ in pairs to form $r$ Dirac fermions (see \eqref{appcomplexfermion}). This leaves a single Majorana $\psi^{2r+1}$ behind. In order for the fermions to obey the correct anticommutation relations, the bosonized complex fermions \eqref{appcomplexfermion} have to be modified by a Klein factor \begin{align}c^j(z)=(-1)^\Pi e^{i\widetilde{\phi}^j(z)}=e^{i\widetilde{\phi}^j(z)+i\pi\Pi}\end{align} where $(-1)^\Pi$ is the fermion parity operator that anticommutes with $\psi^{2r+1}$, and both $\Pi$ and $\psi_{2r+1}$ commute with the rest of the bosons $\widetilde{\phi}^j$. In a non-chiral system, $(-1)^\Pi$ can be chosen to be the combination $i\gamma_L\gamma_R$, for $\gamma_{L/R}$ the zero mode of $\psi^{2r+1}_{L/R}$. In the chiral case, it can be defined by $i\gamma\gamma_\infty$ using an additional Majorana zero mode $\gamma_\infty$ that completes the Cliffort algebra $\{\gamma,\gamma_\infty\}=0$.

The $so(2r+1)_1$ current algebra extends the $so(2r)_1$ algebra by the short roots with length 1 (see \eqref{approots}). It contains the $so(2r)_1$ generators $H^j=i\partial\widetilde{\phi}^j$ and $E^{\boldsymbol\alpha}=e^{i\boldsymbol\alpha\cdot\widetilde{\boldsymbol\phi}}$ (see \eqref{CartanWeylcurrent} in apendix~\ref{sec:kleinfactors}), for $\boldsymbol\alpha\in\Delta_{so(2r)}$ the long roots with length $|\boldsymbol\alpha|=\sqrt{2}$. The remaining raising and lowering operators with the short roots are represented by the normal ordered products \begin{align}E^{\pm{\bf e}_j}(z)=e^{\pm i\widetilde{\phi}^j(z)}\psi^{2r+1}(z).\end{align} In addition to \eqref{appcurrentrelations}, the Cartan-Weyl generators satisfy the current relations \begin{align}H^i(z)E^{\pm{\bf e}_j}(w)&=\frac{\pm\delta^{ij}}{z-w}E^{\pm{\bf e}_j}(w)+\ldots\nonumber\\E^{{\bf e}_j}(z)E^{-{\bf e}_j}(w)&=\frac{1}{(z-w)^2}+\frac{1}{z-w}H^j(w)\label{appcurrentrelations2}\\&\quad\quad-\frac{1}{2}\partial\widetilde{\phi}^j(w)\partial\widetilde{\phi}^j(w)\nonumber\\&\quad\quad-\psi^{2r+1}(w)\partial\psi^{2r+1}(w)+\ldots\nonumber\\E^{s_1{\bf e}_{j_1}}(z)E^{s_2{\bf e}_{j_2}}(w)&=\frac{i^{-s_1s_2}\epsilon({\bf e}_{j_1},{\bf e}_{j_2})}{z-w}E^{s_1{\bf e}_{j_1}+s_2{\bf e}_{j_2}}(w)\nonumber\\&\quad\quad\quad+\ldots\nonumber\end{align} for $j_1\neq j_2$ and $s_1,s_2=\pm1$. Moreover, when $\boldsymbol\alpha\cdot(\pm{\bf e}_j)=-1$, i.e.~$\boldsymbol\alpha\pm{\bf e}_j\in\Delta_{so(2r+1)}$, \begin{align}E^{\boldsymbol\alpha}(z)E^{\pm{\bf e}_j}(w)&=\frac{i\epsilon(\boldsymbol\alpha,{\bf e}_j)(-1)^{\sum_j\alpha^j/2}}{z-w}E^{\boldsymbol\alpha\pm{\bf e}_j}(w)+\ldots\nonumber\end{align} where $\epsilon({\bf m},{\bf n})=(-1)^{{\bf m}\ast{\bf n}}$ is defined in \eqref{starproduct}.

The (normal ordered) energy-momentum tensor can be turned from the Sugawara form \eqref{SugawaraT} to the usual bosonic and fermionic one \begin{align}T(z)&=\frac{1}{2(N-1)}\left[\sum_{i=1}^rH^i(z)H^i(z)+\sum_{\boldsymbol\alpha\in\Delta}E^{\boldsymbol\alpha}(z)E^{-\boldsymbol\alpha}(z)\right.\nonumber\\&\quad\quad\left.+\sum_{j=1}^rE^{{\bf e}_j}(z)E^{-{\bf e}_j}(z)+E^{-{\bf e}_j}(z)E^{{\bf e}_j}(z)\right]\nonumber\\&=-\frac{1}{2}\partial\widetilde{\boldsymbol\phi}(z)\cdot\partial\widetilde{\boldsymbol\phi}(z)-\frac{1}{2}\psi^{2r+1}(z)\partial\psi^{2r+1}(z).\end{align} There are only two non-trivial primary fields $\psi$ and $\sigma$. The fermion sector $\psi$ consists of the $2r$ Dirac fermions $c^j,(c^j)^\dagger$ in \eqref{appfermionsector} as well as the remaining Majorana fermion $\psi^{2r+1}$. The $\sigma$ sector consists of $2^r$ twist fields \begin{align}{\bf V}_\sigma&=\sigma^1\ldots\sigma^{2r+1}\label{so(2r+1)sigma}\\&=\mathrm{span}\left\{\exp\left(i\sum_{j=1}^r\frac{(-1)^{s_j}}{2}\widetilde{\phi}^j\right)\sigma^{2r+1}:s_j=0,1\right\}\nonumber\end{align} which represents $so(2r+1)_1$ according to the spinor representation. 
Their conformal dimensions are given by $h_\psi=1/2$ and $h_\sigma=(2r+1)/16$.

\section{\texorpdfstring{$\mathbb{Z}_6$}{Z6} parafermion model}\label{sec:appZ6parafermion}
Here we represent the $\mathbb{Z}_6$ parafermions using bosonized fields and Majorana fermions in the $so(9)_1$ CFT. We focus on a single Majorana wire containing 9 right moving real fermions. The CFT is fractionalized using the conformal embedding into $so(9)_1\supseteq so(3)_3^+\times so(3)_3^-$ (see section~\ref{sec:conformalembedding}). Each $so(3)_3$ sector is then further decomposed into $so(2)_3\times``\mathbb{Z}_6"$ using the coset construction $``\mathbb{Z}_6"=so(3)_3/so(2)_3$ (see section~\ref{sec:Z6parafermions}). We now provide a more detail description of the $\mathbb{Z}_6$ parafermion sector. We will focus on the one in $so(3)_3^-$.

First we pair six Majorana channels into three Dirac fermions and bosonize $c^1=(\psi^1+i\psi^4)/\sqrt{2}=e^{i\widetilde{\phi}^1}$, $c^2=(\psi^2+i\psi^5)/\sqrt{2}=e^{i\widetilde{\phi}^2}$ and $c^3=(\psi^3+i\psi^6)/\sqrt{2}=e^{i\widetilde{\phi}^3}$. The Lagrangian density of the boson fields are given in \eqref{Lagrangianchargeneutral}. Like the $so(N)_1$ case, extra care is required so that the Dirac fermions $c^j$ satisfies the appropriate mutual anticommutation relations. Here we use a slightly different but more convenient convention \begin{align}\left\langle\widetilde{\phi}^i(z)\widetilde{\phi}^j(w)\right\rangle&=-\delta^{ij}\log(z-w)+\frac{i\pi}{2}S^{ij}\label{so(3)3bosonOPE1}\\S^{ij}&=\left\{\begin{array}{*{20}l}0&\mbox{if $i=j$}\\1&\mbox{if $i-j\equiv1$ mod 3}\\-1&\mbox{if $i-j\equiv-1$ mod 3}\end{array}\right.\nonumber\end{align} so that the constant phases $S^{ij}$ have a threefold cyclic symmetry. The $so(2)_3$ sub-theory is generated by the ``charged" boson $\phi_\rho=(\widetilde{\phi}^1+\widetilde{\phi}^2+\widetilde{\phi}^3)/3$. It satisfies \begin{align}\left\langle\phi_\rho(z)\phi_\rho(w)\right\rangle=-\frac{1}{3}\log(z-w).\label{so(3)3bosonOPE2}\end{align} The remaining ``neutral" bosons $\phi_\sigma^j=\widetilde{\phi}^j-\phi_\rho$ are linearly dependent $\phi_\sigma^1+\phi_\sigma^2+\phi_\sigma^3=0$ and obey the OPE \begin{align}\left\langle\phi_\sigma^i(z)\phi_\sigma^j(w)\right\rangle=-\left(\delta^{ij}-\frac{1}{3}\right)\log(z-w)+\frac{i\pi}{2}S^{ij}.\label{so(3)3bosonOPE3}\end{align} The ``charge" and ``neutral" sector completely decoupled so that $\langle\phi_\rho(z)\phi_\sigma^j(w)\rangle=0$. Lastly, there are three remaining Majoranan fermions $\psi^{7,8,9}$ in the $so(9)_1$ theory. They completely decouple with $\phi_\sigma$ and $\phi_\rho$. Although the vertex $e^{i\phi_\rho}$ anticommutes with $\psi^{7,8,9}$, this has no effect on any of our derivations. More importantly the ``neutral" vertices $e^{i\phi_\sigma^j}$ commute with the remaining fermions.

In section~\ref{sec:Z6parafermions}, we defined the $\mathbb{Z}_6$ parafermion \eqref{Z6parafermiondefinition} \begin{align}\Psi=\frac{1}{\sqrt{3}}\left(e^{i\phi_\sigma^1}\psi^7+e^{i\phi_\sigma^2}\psi^8+e^{i\phi_\sigma^3}\psi^9\right)\end{align} which is part of the $so(3)_3^-$ current (see \eqref{Jso(3)3=Vso(2)3Z6}). It generates the rest of the $\mathbb{Z}_6$ parafermions \begin{align}\Psi^2&=\frac{1}{\sqrt{15}}\Bigg[\sum_{j=1}^3e^{i2\phi_\sigma^j}\nonumber\\&\quad\quad+2i\left(e^{-i\phi_\sigma^1}\psi^{89}+e^{-i\phi_\sigma^2}\psi^{97}+e^{-i\phi_\sigma^3}\psi^{78}\right)\Bigg]\nonumber\\\Psi^3&=\sqrt{\frac{2}{5}}\left[i\psi^{789}-\cos\left(\phi_\sigma^1-\phi_\sigma^2\right)\psi^9\right.\label{appZ6parafermions}\\&\quad\quad\left.-\cos\left(\phi_\sigma^2-\phi_\sigma^3\right)\psi^7-\cos\left(\phi_\sigma^3-\phi_\sigma^1\right)\psi^8\right]\nonumber\\\Psi^4&=\left(\Psi^2\right)^\dagger,\quad\Psi^5=\left(\Psi_1\right)^\dagger,\quad\Psi^0=\Psi^6=1\nonumber\end{align} where $\psi^{ab}=\psi^a\psi^b$ and $\psi^{abc}=\psi^a\psi^b\psi^c$. Their conformal dimensions \begin{align}h_{\Psi^m}=\frac{m(6-m)}{6}\end{align} as well as the fusion rules \begin{align}\Psi^m(z)\Psi^{m'}(w)&=\frac{c^{mm'}}{(z-w)^{mm'/3}}\Psi^{m+m'}(w)+\ldots\\\Psi^m(z)\Psi^{6-m}(w)&=\frac{1}{(z-w)^{2h_{\Psi^m}}}\nonumber\\&\;\;\;\;\times\left[1+\frac{2h_{\Psi^m}}{c_{\mathbb{Z}_6}}(z-w)^2T_{\mathbb{Z}_6}+\ldots\right]\nonumber\end{align} match with the known result by Zamolodchikov and Fateev\cite{ZamolodchikovFateev85}, for $T_{\mathbb{Z}_6}$ the energy-momentum tensor \eqref{TZ6} with central charge $c_{\mathbb{Z}_6}=5/4$ and \begin{align}c^{mm'}=\sqrt{\frac{(m+m')!(6-m)!(6-m')!}{m!m'!(6-m-m')!6!}}.\end{align}

\section{The \texorpdfstring{$S$}{S}-matrices of the \texorpdfstring{$G_N$}{GN} state}\label{sec:GNapp}
The surface topological orders of the time reversal symmetric gapped coupled wire model are described in section~\ref{sec:topologicalorder}. There are thirty two distinct topological states defined in eq.\eqref{GN=sor1} and \eqref{GN=so3xsor}, which we repeat here. \begin{align}G_N=\left\{\begin{array}{*{20}l}SO(r)_1,&\mbox{for $N=2r$}\\SO(3)_3\boxtimes_bSO(r)_1,&\mbox{for $N=9+2r$}\end{array}\right..\label{GNappintro}\end{align} In this appendix we summarize the modular properties of these states. In particular we present there braiding $S$-matrices \eqref{braidingS} \begin{align}\mathcal{S}_{{\bf a}{\bf b}}=\frac{1}{\mathcal{D}}\sum_{\bf c}d_{\bf c}N_{{\bf a}{\bf b}}^{\bf c}\frac{\theta_{\bf c}}{\theta_{\bf a}\theta_{\bf b}}\label{braidingSapp}\end{align} which are identical to the modular $S$-matrix\cite{bigyellowbook} of the $\mathcal{G}_N$ WZW CFT. The fusion matrices $N_{\bf ab}^{\bf c}$ that characterize fusion rules ${\bf a}\times{\bf b}=\sum_{\bf c}N_{\bf ab}^{\bf c}{\bf c}$ can in turned be determined by $S$-matrix throught the Verlinde formula\cite{Verlinde88} \eqref{Verlindeformula} \begin{align}N_{s_1s_2}^s=\sum_{s'}\frac{\mathcal{S}_{s_1s'}\mathcal{S}_{s_2s'}\mathcal{S}_{ss'}}{\mathcal{S}_{0s'}}.\label{Verlindeformulaapp}\end{align}

The $G_N$ state is Abelian and carries four anyon types $1,\psi,s_+,s_-$ when $N$ is a multiple of four. It is non-Abelian otherwise and carries three anyon types $1,\psi,\sigma$ when $N$ is 2 mod 4, or seven anyon types $1,\alpha_+\gamma_+,\beta,\gamma_-,\alpha_-,f$ when $N$ is odd. The quasiparticle exchange statistics $\theta_{\bf x}$ and quantum dimensions $d_{\bf x}$ are summarized in table~\ref{tab:toporderSO(r)} and \ref{tab:toporderSO(3)3SO(r)1}. The total quantum dimensions $\mathcal{D}=\sqrt{\sum_{\bf x}d_{\bf x}^2}$ are given by \begin{align}\mathcal{D}_{G_N}=\left\{\begin{array}{*{20}l}2&\mbox{for $N$ even}\\2\csc(\pi/8)&\mbox{for $N$ odd}\end{array}\right.\end{align} where $\csc(\pi/8)=\sqrt{4+2\sqrt{2}}$.

The $S$-matrices of $G_N$ for $N=2r$ even are well-known and are given by those of the $SO(r)_1$ states.\cite{Kitaev06,khan2014} \begin{align}\mathcal{S}_{G_N}&=\frac{1}{\mathcal{D}_{G_N}}\left(\begin{smallmatrix}1&1&1&1\\1&1&-1&-1\\1&-1&i^n&-i^n\\1&-1&-i^n&i^n\end{smallmatrix}\right),\quad\mbox{for $N=4n$},\\\mathcal{S}_{G_N}&=\frac{1}{\mathcal{D}_{G_N}}\left(\begin{smallmatrix}1&1&\sqrt{2}\\1&1&-\sqrt{2}\\\sqrt{2}&-\sqrt{2}&0\end{smallmatrix}\right),\quad\mbox{for $N=4n+2$}.\end{align} The $S$-matrices for the odd $N$ cases are modification of the $G_9=SO(3)_3$ prototype \eqref{SO(3)3Smatrix} \begin{align}\mathcal{S}^{SO(3)_3}_{s_1s_2}=\frac{1}{2}\sin\left[\frac{\pi(2s_1+1)(2s_2+1)}{8}\right]\label{SO(3)3Smatrixapp}\end{align} where $s_j=0,1/2,1,3/2,2,5/2,3$ label the seven anyon types $1,\alpha_+,\gamma_+,\beta,\gamma_-,\alpha_-,f$ (see table~\ref{tab:so(3)3primaryfields}). For $N=9+2r$ mod 32, the $S$-matrix of $G_N$ is given by \begin{align}\mathcal{S}_{G_N}=\mathcal{F}^rS^e(\lceil r/2\rceil)\mathcal{F}^{-r}\end{align} where $\lceil r/2\rceil\geq r/2$ is the smallest integral ceiling of $r/2$, $\mathcal{S}^e(n)$ is the $S$-matrix when $r=2n$ is even \begin{align}\mathcal{S}^e(n)_{s_1s_2}=i^{n(4s_1s_2)^2}\mathcal{S}^{SO(3)_3}_{s_1s_2}\end{align} and $\mathcal{F}$ is the operator that flips the fermion parity of $\alpha_+\leftrightarrow\alpha_-$ and $\gamma_+\leftrightarrow\gamma_-$ \begin{align}\mathcal{F}=\left(\begin{smallmatrix}1&&&&&&\\&&&&&1&\\&&&&1&&\\&&&1&&&\\&&1&&&&\\&1&&&&&\\&&&&&&1\end{smallmatrix}\right).\end{align}


%

\end{document}